\documentclass{aa} 

\usepackage{graphicx}
\usepackage{xcolor}
\usepackage{txfonts}
\usepackage{booktabs}
\usepackage{multirow}

\usepackage{makecell}

\usepackage{nicefrac}

\usepackage[colorlinks=true,linkcolor=blue,allcolors=blue]{hyperref}

\definecolor{mycolor}{rgb}{0.05, 0.65, 0.2}

\newcommand{\lya}{{\text{Ly}\ensuremath{\alpha}}}
\newcommand{\Halpha}{{\text{H}\ensuremath{\alpha}}}
\newcommand{\Hbeta}{{\text{H}\ensuremath{\beta}}}
\newcommand{\oiii}{{\text{[\ion{O}{iii}]}}}
\newcommand{\niv}{{\text{\ion{N}{iv}]}}}
\newcommand{\fesclya}{\ensuremath{f_{\rm esc}(\lya{})}}
\newcommand{\fesclyc}{\ensuremath{f_{\rm esc}(\rm LyC)}}
\newcommand{\fcov}{\ensuremath{f_{\rm cov}}}
\newcommand{\monster}{A2744-45924}
\newcommand{\jwst}{{JWST}}
\newcommand{\nircam}{NIRCam}
\newcommand{\muse}{MUSE}
\newcommand{\flambdaunits}{\ensuremath{\rm erg\,s^{-1}\,cm^{-2}\,{\AA{}}^{-1}}}
\newcommand{\contblue}{{\it cont. blue}}
\newcommand{\contred}{{\it cont. red}}

\defcitealias{naidu2024}{Naidu \& Matthee et al. 2024}

\usepackage{comment}

\begin{document}

   \title{A weak \lya{} halo for an extremely bright little red dot}
   \subtitle{Indications of enshrouded supermassive black hole growth}
    \titlerunning{A weak \lya{} halo for an extremely bright LRD}

   \author{
    Alberto~Torralba\inst{\ref{inst:ista}}\thanks{alberto.torralba@ista.ac.at}
    \and Jorryt~Matthee\inst{\ref{inst:ista}}
    \and Gabriele~Pezzulli\inst{\ref{inst:kapteyn}}
    \and Tanya~Urrutia\inst{\ref{inst:leibniz}}
    \and Max~Gronke\inst{\ref{inst:MPIA_garching}}
    \and Sara~Mascia\inst{\ref{inst:ista}}
    \and Francesco~D'Eugenio\inst{\ref{inst:kavli_cambr},\ref{inst:cavendish_cambr}}
    \and Claudia~Di~Cesare\inst{\ref{inst:ista}}
    \and Anna-Christina~Eilers\inst{\ref{inst:MIT_kavli}}
    \and Jenny~E.~Greene\inst{\ref{inst:princeton}}
    \and Edoardo~Iani\inst{\ref{inst:ista}}
    \and Yuzo~Ishikawa\inst{\ref{inst:MIT_kavli}}
    \and Ruari~Mackenzie\inst{\ref{inst:EPFL}}
    \and Rohan~P.~Naidu\inst{\ref{inst:MIT_kavli}}
    \and Benjamín~Navarrete\inst{\ref{inst:ista}}
    \and Gauri~Kotiwale\inst{\ref{inst:ista}}
    }

   \institute{
    Institute of Science and Technology Austria (ISTA), Am Campus 1, 3400 Klosterneuburg, Austria.\label{inst:ista}
    \and Kapteyn Astronomical Institute, University of Groningen, Landleven 12, NL-9747 AD Groningen, the Netherlands.\label{inst:kapteyn}
    \and Leibniz-Institut für Astrophysik, Potsdam, An der Sternwarte 16, Potsdam 14482, Germany.\label{inst:leibniz}
    \and Max Planck Institute for Astrophysics, Karl-Schwarzschild-Str. 1, 85748 Garching, Germany.\label{inst:MPIA_garching}
    \and  Kavli Institute for Cosmology, University of Cambridge, Madingley Road, Cambridge, CB3 0HA, United Kingdom\label{inst:kavli_cambr}
    \and  Cavendish Laboratory - Astrophysics Group, University of Cambridge, 19 JJ Thomson Avenue, Cambridge, CB3 0HE, United Kingdom\label{inst:cavendish_cambr}
    \and MIT Kavli Institute for Astrophysics and Space Research, Massachusetts Institute of Technology, Cambridge, MA 02139, USA.\label{inst:MIT_kavli}
    \and Department of Astrophysical Sciences, Princeton University, Princeton, NJ 08544, USA.\label{inst:princeton}
    \and Institute of Physics, Laboratory of Astrophysics, Ecole Polytechnique F\'ed\'erale de Lausanne (EPFL), Observatoire de Sauverny, 1290 Versoix, Switzerland\label{inst:EPFL}
    }

   \date{Accepted XXX. Received YYY; in original form ZZZ}

  \abstract{
    The abundant population of little red dots (LRDs), compact objects with red UV to optical colors and broad Balmer lines at high redshift, is revealing new insights into the properties of early active galactic nuclei (AGN). Perhaps the most surprising features of this population are the presence of Balmer absorption and ubiquitous strong Balmer breaks. Recent models link these features to an active supermassive black hole (SMBH) cocooned in very dense gas ($N_{\rm H}\sim10^{24}\,\rm cm^{-2}$). We present a stringent test of such models using VLT/MUSE observations of A2744-45924, the most luminous LRD known to date ($L_{\rm H\alpha}\approx10^{44}~\rm erg\,s^{-1}$), located behind the Abell-2744 lensing cluster at $z=4.464$ ($\mu=1.8$). We detect a moderately extended Ly$\alpha$ nebula ($h\approx5.7$ pkpc), spatially offset from the point-like H$\alpha$ seen by JWST by $\approx 1.6$~pkpc. The Ly$\alpha$ emission is narrow ($\rm FWHM=270\pm 15~km\,s^{-1}$), and faint ($\rm Ly\alpha=0.07H\alpha$) compared to Ly$\alpha$ nebulae typically observed around quasars of similar luminosity. We detect compact N\,{\sc iv}]$\lambda$1486 emission, spatially aligned with H$\alpha$, and a spatial shift in the far-UV continuum matching the Ly$\alpha$ offset. We discuss that H$\alpha$ and Ly$\alpha$ have distinct physical origins: H$\alpha$ originates from the AGN, while Ly$\alpha$ is powered by star formation.  In the environment of A2744-45924, we identified four extended Ly$\alpha$ halos ($\Delta z<0.02$, $\Delta r<100$~pkpc). Their Ly$\alpha$ luminosities match the expectations based on H$\alpha$ emission, and show no evidence for radiation from A2744-45924 affecting its surroundings. The lack of strong, compact, and broad Ly$\alpha$ and the absence of a luminous extended halo, suggest that the UV AGN light is obscured by dense gas cloaking the SMBH with a covering factor close to unity.
  }

    \keywords{ Galaxies: high-redshift --
        Galaxies: halos --
        Galaxies: active
        }

   \maketitle

\section{Introduction}\label{sec:introduction}

\jwst{}'s significantly increased sensitivity in near-infrared imaging and spectroscopy is offering a new perspective on galaxies in the distant Universe \citep[see][for an overview]{adamo2024}. One of the most debated discoveries since the first data arrived in 2022 is the nature of a loosely defined class of objects that became known as the little red dots \citep[LRDs;][]{matthee2024a}.
These objects are characterized by their compact morphologies and red rest-frame optical colors, often in combination with blue UV colors resulting in a particular V shape \citep[e.g.,][]{labbe2025, akins2024, kokorev2024a, barro2024}.
To date, $\sim1000$ LRDs have been photometrically identified over redshifts $z\approx2-9$ spanning luminosities $L_{\rm H\alpha} \sim 10^{41 - 44}$ erg s$^{-1}$, with number densities $\sim10^{-5}$ to $10^{-4}$ cMpc$^{-3}$ \citep[e.g.,][]{harikane2023,akins2024,kokorev2024a,greene2024,maiolino2024a,matthee2024a,kocevski2024, Lin25}, 1000 to 10\,000 times higher than those of UV-selected quasars at similar redshifts \citep[e.g.,][]{niida2020, matsuoka2023}. These numbers imply that they constitute a significant fraction of the galaxy population ($\sim1\%$). These high number densities and their moderate overdensities \citep[e.g.,][]{pizzati2024, matthee2024b,Arita25} suggest that the {majority} of LRDs reside in relatively low-mass galaxies in low-mass halos (M$_{\rm star}\sim10^8$ $M_{\odot}$, $M_{\rm halo}\sim10^{11}$ $M_{\odot}$).
In many cases the Balmer emission lines show significant broad components with widths $>1000$ km s$^{-1}$ \citep[e.g.,][]{kocevski2023, matthee2024a, taylor2024, lin2024}, which are indicative of AGN activity \citep[e.g.,][]{perez-gonzalez2024, kokorev2024b, leung2024, volonteri2025}. Other AGN indicators are the detection of high-ionization UV lines \citep{treiber2024} and \ion{Fe}{ii} emission \citep{labbe2024} and indications of time variability in the Balmer emission lines \citep{ji2025, furtak2025} and in the Balmer Break strength \citep{ji2025} in select LRDs. On the other hand, LRDs typically lack the IR emission associated with a hot torus \citep[e.g.,][]{williams2024, akins2024, Setton2025}, and they also show weak or no radio \citep[e.g.,][]{latif2025, perger2025, gloudemans2025, mazzolari2024} and X-ray emission \citep[e.g.,][]{yue2024, ananna2024, maiolino2025}.

A particularly intriguing feature observed in a significant subset of LRDs is the presence of strong narrow Balmer absorption in the \Halpha{} and \Hbeta{} lines \citep[e.g.,][]{matthee2024a, juodzbalis2024, taylor2024,deugenio2025}, along with a pronounced Balmer break appearing consistently at rest-frame wavelengths $\approx 3645$~\AA{} \citep[e.g.,][]{setton2024, wang2024}.
These two properties may be correlated and can be explained by very dense gas along the line of sight \citep[$n_{\rm H}\sim 10^{9}$~cm$^{-3}$ and column densities $N_{\rm H}\sim 10^{24}$~cm$^{-2}$; e.g.,][]{inayoshi2025, ji2025, naidu2025,Taylor25z9}, which could also explain the X-ray weakness \citep[e.g.,][]{kocevski2023} as being due to Compton thick absorption \citep[e.g.,][]{juodzbalis2024, maiolino2025}.
The abrupt and consistent Balmer break of many LRDs could otherwise only be explained by classical models with a very unusual and specific combination of young and old stellar populations and dust attenuation laws \citep[e.g.,][]{wang2024, labbe2024, labbe2025, ma2025}.
Moreover, the strongest Balmer breaks observed cannot be reproduced by stellar population models, even assuming extreme dust attenuation laws \citep{graaff2025, naidu2025}.

Recent studies have proposed models in which a very dense, turbulent, and dust-free gas completely cocoons the supermassive black hole (SMBH) of LRDs \citep[{\it Black Hole Star} models; BH*;][]{naidu2025, rusakov2025, graaff2025}.
In these models, the observed Balmer line profiles could have been broadened partly by scattering in the dense gas \citep{rusakov2025, naidu2025}. This would imply that the single epoch virial estimators of black hole (BH) mass  cannot be simply applied to the broad components of the emission lines, implying that the BH mass is overestimated; this has significant implications for the interpretation of the BH-stellar mass relation \citep[cf.][]{pacucci2024}.
However, it remains an open question whether the dense gas fully obscures the central SMBH (covering factor $\fcov \sim 1$) or whether the absorption arises from an irregular distribution of clouds along the line of sight ($\fcov < 1$). 
Assessing the covering factor is crucial to understanding SMBH growth at high redshift, offering insights into the rapid early assembly of black holes in the very early Universe. Determining \fcov{} is also essential for evaluating the impact of AGN on their surroundings, particularly whether they produce enough ionizing photons to significantly contribute to cosmic reionization.

A detailed study of the UV spectrum of LRDs is key to determining \fcov{}.
In particular, the strength, spatial distribution, and spectral profile of the Lyman-alpha (\lya{}) emission line are highly sensitive to the presence and distribution of neutral gas \citep[e.g.,][]{neufeld1990, steidel2011}. If $\fcov{}\ll 1$, the escape fraction of \lya{} and ionizing photons must be high; therefore, we would expect to see large and luminous \lya{} nebulae around LRDs, similar to the ones seen around luminous quasars. On the contrary, if \fcov{} is close to unity, the extremely dense gas would strongly suppress \lya{} emission and the ionizing continuum, and the \lya{} halo would be weak.

In this work we present deep VLT/MUSE observations of \monster{}, the most luminous LRD known \citep{greene2024, labbe2024}; these observations  provide the first \lya{} halo study of an LRD and its environment. MUSE has demonstrated the existence of large \lya{} halos around quasars \citep[e.g.,][]{borisova2016}, which are powered by elevated photoionization levels in the vicinity of quasars. The surface brightness depends slightly on UV luminosity \citep[e.g.,][]{mackenzie2021} and Type II AGN tend to have more elongated halos \citep[e.g.,][]{brok2020}, exemplifying the use of extended Ly$\alpha$ nebulae to probe AGN covering factors and escaping ionizing luminosity. The Ly$\alpha$ surface brightness around quasars is significantly elevated compared to Ly$\alpha$ halos around star-forming galaxies \citep[e.g.,][]{leclercq2017}. The key questions that we aim to answer include the influence of the AGN on the circumgalactic medium (CGM) gas, and what  the covering factor of \monster{}'s central SMBH is.

This paper is structured as follows. In Sect.~\ref{sec:data} we describe the observational data from VLT/\muse{} and \jwst{}/NIRCam used throughout this work. In Sect.~\ref{sec:results} we describe the spatially resolved properties of the rest-frame UV spectrum of \monster{}. In Sect.~\ref{sec:large_scale_lya} we analyze the \lya{} emission of the galaxies in the environment of \monster{}. In Sect.~\ref{sec:discussion_what_powers_lya} we discuss the implications of our results for the interpretation of the origin of the \lya{} emission. In Sect.~\ref{sec:discussion:implications} we discuss  the possible interpretations of our results in the context of LRD models involving dense gas enshrouded SMBH growth. Finally, in Sect.~\ref{sec:summary} we summarize our main findings.

Throughout this work we use a $\Lambda$CDM cosmology as described by \citet{collaboration2020}, with $\Omega_\Lambda=0.69$, $\Omega_\text{M}=0.31$, and $H_0=67.7$ km\,s$^{-1}$\,Mpc$^{-1}$. All photometric magnitudes are given in the AB system \citep{oke1983}.

\section{Data}\label{sec:data}

\subsection{\jwst}\label{sec:data:jwst}

Among the LRD population, \monster{} stands out for its luminosity $L_{\rm H\alpha}\approx 10^{44}\,\mathrm{erg}\,\mathrm{s}^{-1}$ \citep{labbe2024}, which places it in the (obscured) quasar regime. Found behind the lensing cluster Abell-2744 \citep[A2744][]{abell1989} at $z\approx 4.464$, \monster{} is magnified by a factor $\mu = 1.8$, with $\mu_t = 1.684$ and $\mu_r=1.072$ being the transverse and radial magnification, respectively, relative to $\theta= 19.9^\circ$ \citep[][]{furtak2023, price2025}.

\monster{} was first identified in the Cycle 1 Treasury program \#2561 \citep[UNCOVER;][]{bezanson2024} DR1 images and selected as an LRD based on the compact morphology in the long wavelength and V-shaped SED. NIRSpec/PRISM spectroscopy confirmed the AGN nature through the detection of broad H$\alpha$ emission \citep{greene2024}.
\monster{} was also observed in the \jwst{} Cycle 2 survey \#3516 \citepalias[ALT;][object ID 66543]{naidu2024}. The ALT survey targeted the A2744 lensing cluster, taking imaging data over $\sim$30 arcmin$^2$ using the \nircam{} short-wavelength channel broadband filters F070W and F090W and deep grism observations in F356W. These grism observations covered the H$\alpha$ line with a resolution of $R\sim1600$, revealing broad non-Gaussian wings and narrow H$\alpha$ absorption \citep[Fig.~5 in][]{labbe2024}.

The NIRCam imaging revealed a compact yet resolved rest-frame UV morphology, along with a more extended and fainter component showing up in the far-UV \citep{labbe2024}. It shows broad Balmer lines, H$\alpha$ absorption, a Balmer break and various \ion{Fe}{ii} emission lines: all signs of AGN activity. Despite its rest-frame optical luminosity and unlike quasars, \monster{} is not strongly detected in the sub-millimeter (neither continuum nor [\ion{C}{ii}] emission) and the faintness in \jwst{}/MIRI data indicates the lack of hot dust emission \citep{Setton2025}. \cite{akins2025} identified faint [\ion{C}{i}] emission from neutral gas with ALMA, possibly spatially offset in a similar direction as the extended component in the far UV. Moreover, \citet{chen2025} studied the extended component of \monster{}, identifying it as a star-forming galaxy with stellar mass $\log_{10}(M/M_\odot)=8.17\pm0.08$ and blue rest-frame UV colors ($\beta_{\rm UV}=-2.12$). \monster{} is located in the largest galaxy overdensity in the A2744 field, as evidenced by the elevated number of \Halpha{} emitters in its vicinity \citep[$1 + \delta = 31 \pm 5$, within a radius of 1~cMpc;][]{matthee2024b}.

\subsection{MUSE}\label{sec:data:muse}

Here we present new VLT/MUSE observations of \monster{}. These data were obtained using the VLT/\muse{} integral field spectrograph \citep{bacon2010} as part of the ESO P114 program 114.27M6 (PI: Matthee). Observations were taken under clear conditions over the 10--31 October 2024 period. We used MUSE with the wide field mode which has a field of view of $1\arcmin\times1\arcmin$, nominal wavelength coverage (470--935 nm) and asssisted with ground-layer adaptive optics. Individual exposure times were 675s and we rotated the position angle by 90 degrees between every exposure. To improve the background subtraction, we used random dithering offsets between 0.2\arcsec{} and 1.5\arcsec{} between each exposure. The pointing was roughly centered on \monster{} and the starting PA was 30 degrees to maximize the number of sources with known spectroscopic redshifts from ALT \citepalias{naidu2024} in the pointing. The total exposure time was 18.9 ks.

Basic data reduction was carried out using version 2.8 of the MUSE pipeline \citep{weilbacher2020}. Each exposure was aligned to the JWST/F070W WCS and resampled onto a common grid. A superflat correction was applied to mitigate slicer stack transition artifacts \citep[see][]{bacon2023}, and pixels with fewer than 2000 valid wavelength layers were masked. Additional PCA-based sky subtraction was performed using ZAP \citep{soto2016}. The aligned exposures were then mean-combined voxel by voxel using a 6$\sigma$ clipping threshold. In the final cube, we applied a DC offset correction to bring empty sky regions to zero, and replaced the formal variance with an empirically estimated effective variance that accounts for nonpropagated covariances \citep{urrutia2019, weilbacher2020}.

The astrometric corrections and the effective PSF of the MUSE datacube were derived using the \jwst{} F070W image as a reference. This broadband filter includes the wavelength of \lya{} at $z = 4.464$. First, we used the \textsc{imphot} package\footnote{\url{https://github.com/musevlt/imphot}} to reproject the F070W image onto the same RA-DEC grid as the MUSE datacube. Next, we generated a pseudo-broadband image by convolving the MUSE datacube with the \jwst{} F070W filter’s transmission cube and aligned it to the \jwst{} counterpart using \textsc{imphot}. This process yielded small astrometric corrections of $\Delta({\rm RA},{\rm DEC}) = (-0.01656\arcsec, -0.00346\arcsec)$ and an effective PSF FWHM of $0.7396\arcsec$, assuming a Moffat profile with index $\beta = 2.5$. For this, we assumed that the effective PSF FWHM of the F070W image is significantly smaller than that of MUSE.
The spectral resolution at the \lya{} wavelength of \monster{} is $R\approx2700$ \citep{Bacon2017}.

When analyzing the Ly$\alpha$ emission from galaxies in the large scale environment of \monster{}, we also make use of deep public MUSE data from \cite{richard2021} which covers a $2.3\times2.3'$ region in the A2744 field centered on the most massive core of the cluster, with no overlap with our main cube. As the \jwst{} coverage is virtually identical, these data serve to increase the statistics.

\section{The spatially resolved properties of \monster{}}\label{sec:results}

\begin{table}
    \centering
    \caption{Collection of relevant properties of the rest-frame UV and \lya{} emission of \monster{}.}
    \label{tab:monster_properties}
    \begin{tabular}{llc}
         \toprule
         Property & Value & Notes\\
         \midrule
         RA & 00h 14m 20.34s & $^a$\\
         DEC & -30:20:37.06 & $^a$\\
         $z$ & 4.464 & $^{b}$\\
         $\lya$ & $20\pm 5$ & $^c$\\
         ${\niv{}\lambda 1483}$ & $0.8\pm 0.5$ & $^c$\\
         ${\niv{}\lambda 1486}$ & $2.5\pm 0.6$ & $^c$\\
         ${\text{\ion{O}{iii}]\ensuremath{\lambda}1661}}$ & $0.9\pm 0.5$ & $^c$\\
         ${\text{\ion{O}{iii}]\ensuremath{\lambda}1666}}$ & $1.9\pm 0.9$ & $^{c*}$\\
         $\Halpha$ & $310 \pm 16$ & $^{c\dag}$\\
         $\niv\lambda 1483 / \niv\lambda 1486$ & $0.32 \pm 0.21$ & $^d$\\
         EW$_0$(\lya{}) & $97 \pm 28$~\AA{} & $^e$\\
         EW$_0$($\niv{}\lambda$1483) & $2.8 \pm 1.7$~\AA{} & $^e$\\
         EW$_0$($\niv{}\lambda$1486) & $9 \pm 2$~\AA{} & $^e$\\
         FWHM$_\lya$ & $270 \pm 15$~km\,s$^{-1}$ & $^f$\\
         $\Delta v_{\lya{}, {\rm red}}$ & $183\pm 5$~km\,s$^{-1}$ & $^g$\\
         $h_{\rm halo}$ & $5.7 \pm 0.7$~pkpc & $^h$\\
         $M_{\rm UV, MUSE}$ & $-19.39^{+0.12}_{-0.10}$ & $^i$\\
    \bottomrule
    \end{tabular}
    \tablefoot{
    $^a$Right ascension and declination (J2000).
    $^b$Systemic redshift, measured from \Halpha{} in NIRCam grism data. 
    $^c$Emission line fluxes in units of $10^{-18}$\,erg\,s$^{-1}$\,cm$^{-2}$, corrected for magnification.
    $^d$Flux ratio of the \niv{}$\lambda\lambda$1483,1486 doublet.
    $^e$Rest-frame equivalent widths.
    $^f$FWHM of the \lya{} emission line.
    $^g$Velocity offset of the \lya{} red peak with respect to the \Halpha{} systemic redshift.
    $^h$Exponential scale length of the \lya{} halo.
    $^i$UV magnitude at 1500~\AA{}, measured from the MUSE 1D spectrum. 
    $^*$The flux of \ion{O}{iii}]$\lambda$1666 is affected by a skyline, and may be regarded as an upper limit.
    $^\dag$From \citet{labbe2024}, added here for completeness.
    }
\end{table}

Little red dots are compact objects in the rest-frame optical; however, \monster{} shows a complex multi-component morphology in the rest-frame UV (Fig.~\ref{fig:NIRCam_UV_model}). The gentle effective magnification of $\mu=1.8$ virtually increases the exposure time by a factor $\mu^2\approx 3$, making \monster{} ideal to study its different components to shed light into the relation of SMBHs in LRDs with their host galaxies.

\subsection{\nircam{} UV morphology}\label{sec:uv_morphology}

\begin{figure}
    \centering
    \includegraphics[width=\linewidth]{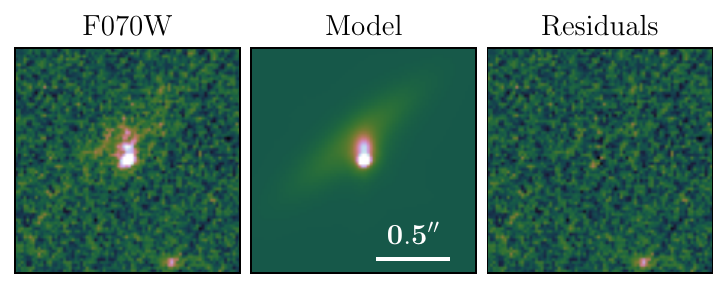}
    \includegraphics[width=\linewidth]{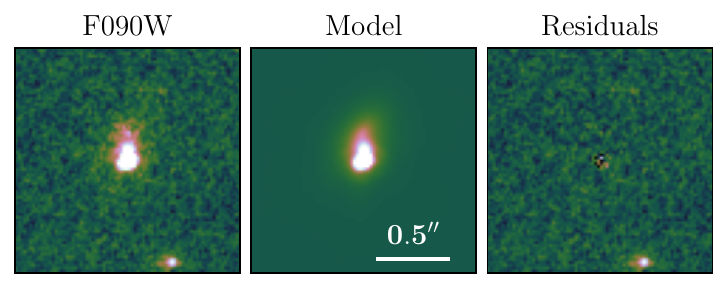}
    \caption{Morphological model of \nircam{} rest-frame UV of \monster{} (PSF FWHM $\approx 0.03\arcsec$). The left panels display the F070W and F090W stamps, corresponding to central rest-frame wavelengths of approximately 1290~\AA{} and 1651~\AA{}, respectively. The center panels show the best-fit models obtained with \textsc{imfit}, while the right panels present the residuals. The F070W image, probing the bluest part of the rest-frame UV spectrum of \monster{}, including \lya{}, reveals a prominent extended component. In contrast, this component is less clear in F090W. An alternative view of the models can be found in Fig.~\ref{fig:NIRCam_UV_model_contours}.}
    \label{fig:NIRCam_UV_model}
\end{figure}

We used the \jwst{}/NIRCam F070W and F090W images to fit the UV morphology of \monster{}. The purpose of this analysis is to establish a baseline for our MUSE IFU study rather than providing a detailed examination of the UV continuum morphology.  
At the redshift of \monster{} ($z=4.464$) F070W covers the rest-frame wavelengths 1140--1429~\AA{} (which include \lya{}), and F090W covers 1455--1839~\AA{}.

We used the \textsc{imfit}\footnote{\url{https://github.com/perwin/imfit}} package \citep{erwin2015} to fit the morphology of \monster{}. We tested multi-component models, combinations of 2D elliptical Gaussians and exponentials that minimize $\chi^2$, convolved with the PSF of the corresponding filter---the effective PSFs of the NIRCam images were computed as described in~\citet{weibel2024}. The fitted models are shown in Fig.~\ref{fig:NIRCam_UV_model}. Both F070W and F090W show clear multi-component morphologies.
The shape of \monster{} in F070W appears to be composed of a compact core plus a more extended and diffuse structure. The core component has two clumps: one well fitted by a Gaussian with $\sigma=0.016$\arcsec{} (smaller than the PSF, and hence can be considered point-like) and an exponential with scale length 0.05\arcsec{} (marginally resolved). The extended component is well fitted by a Gaussian with high ellipticity $e=0.73$, position
angle of $\theta=128^\circ$ and $\sigma=0.35$\arcsec{}. \citet{labbe2024} similarly fitted a Sérsic profile to this extended component in the F070W image, arguing that its surface brightness ($23.5~{\rm mag\,arcsec^{-2}}$) is much higher than would be expected if it were scattered light from the central source.

Since the F070W filter includes \lya{} emission, its morphology may be affected by this emission line. Therefore, we also analyze F090W.
The F090W data show a slightly more complex morphology. The core component is best-fitted by three Gaussians (unresolved components with $\sigma = 0.016\arcsec, 0.02\arcsec$, and 0.025\arcsec{}). The F090W image of \monster{} also shows a diffuse component, rather extended although less prominent than that of F070W. This extended component is best-fitted with an exponential with a similar position angle of $\theta=142^\circ$, ellipticity $e=0.36$ and scale length $h=0.1$\arcsec{}, aligned with the diffuse F070W component, although less extended, hinting that the diffuse component in F070W has bluer UV color---as confirmed by the spatial shift of the continuum emission centroid in the MUSE data (see Sect.~\ref{sec:muse_uv_continuum} below). See Appendix~\ref{sec:alternate_uv_morph_fit} for an alternative fitting procedure in which we attempt to fix the geometrical parameters within both filters.

\subsection{\muse{} spatially resolved UV spectroscopy}\label{sec:muse_uv_continuum}

Motivated by the complex rest-frame UV morphology seen in the NIRCam images, we examine the spatial distribution of the rest-frame UV continuum emission in the MUSE data.
These data cover the rest-frame UV of \monster{} up to $\lambda_0=1712$~\AA{}. We define two spectral intervals: 1235--1430~\AA{} (\contblue{}) and 1555--1712~\AA{} (\contred{}), deliberately avoiding the wavelengths of the prominent UV emission lines \niv{} and \ion{C}{iv}. After integrating the \muse{} datacube over these intervals, we observe noticeable offsets between the positions of the two continuum components: $\Delta ({\rm RA, DEC}) = (0.15\arcsec , 0.22\arcsec)$. We show the collapsed images for the corresponding intervals in Fig.~\ref{fig:continuum_centroids}.
The offset between \contblue{} and \contred{} suggests that the UV continuum of \monster{} arises from two distinct components, with the extended one exhibiting bluer colors.

ALMA observations presented in \citet{akins2025} revealed narrow (${\rm FWHM}=80^{+38}_{-22}$~km\,s$^{-1}$) [\ion{C}{i}](2--1) emission from \monster{}. The [\ion{C}{i}] emission appears 0.4\arcsec{} northeast of the spatial coordinates of the LRD. \citet{akins2025} argue that its position is consistent with emission from the phase center given the low S/N of their data. However, we note that this displacement aligns with our spatial centroids of the \contblue{} and \lya{} emission. Thus, the observed shift in [\ion{C}{i}] may be physical, suggesting that it originates from the extended associated galaxy and traces its cold gas.

\subsection{Extended {\lya{}} emission}\label{sec:extended_lya}

\begin{figure}
    \centering
    \includegraphics[width=\linewidth]{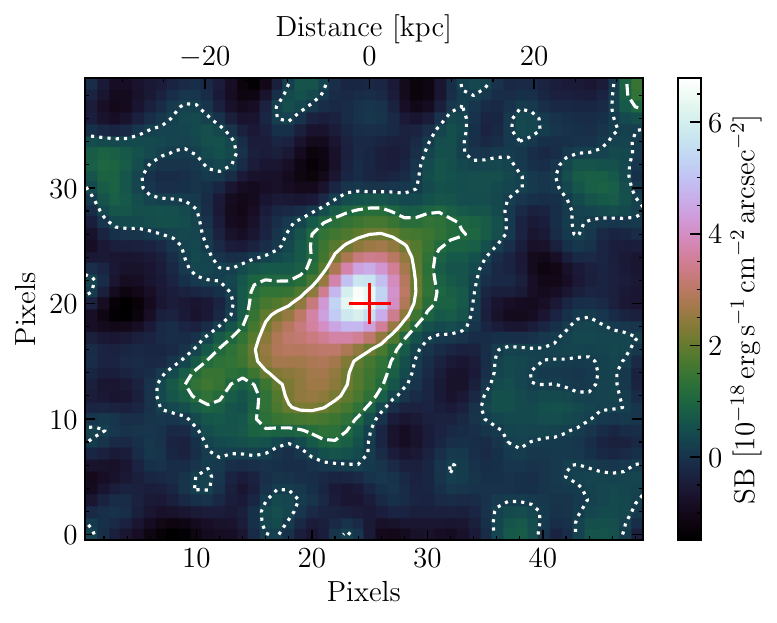}
    \caption{Surface brightness of the \lya{} halo in the wavelength interval 6638.75--6665.00~\AA{}. A 2D Gaussian smoothing kernel is applied across the spatial directions of the cube with $\sigma=1.5$~px.  The centroid of the NIRCam \Halpha{} emission is marked with a red cross. The dotted, dashed, and solid lines show the surface brightness contours corresponding to a $\rm S/N = 1$, 5, and~10, respectively.  The 1$\sigma$ SB limit corresponds to $2\times 10^{-19}$~erg\,s$^{-1}$\,$\rm cm^{-2}$\,$\rm arcsec^{-2}$.}
    \label{fig:SB_contour_map}
\end{figure}

We use the MUSE data to characterize extended Ly$\alpha$ emission from \monster{}. We produce a pseudo-narrowband by collapsing the \muse{} datacube along the observed vacuum wavelength axis between 6638.75 and 6665.00~\AA{}, covering the red \lya{} peak observed in a first optimal extraction (between -165.7~km\,s and 1019~km\,s$^{-1}$ with respect to the \Halpha{} systemic redshift). We subtract a continuum band, obtained in the same manner in the range 6750--7800~\AA{} (1235--1427 \AA{} in the rest-frame). We choose this wide interval assuming a roughly flat UV continuum \citep[$\beta_{\rm UV}\sim-2$, see][]{labbe2024}. The result is the continuum subtracted \lya{} image shown in the upper left panel of Fig.~\ref{fig:SB_contour_map}.
We fit this image using a two-component model, consisting of a 2D elliptical Gaussian (core) plus a 2D exponential (halo), convolved with the effective PSF of the \muse{} datacube (see Sect.~\ref{sec:data:muse}).
We also considered adding the fitted F090W morphology as a core component (since this filter does not include \lya{}), convolved with the MUSE PSF and appying a scaling factor; however, this component is completely unconstrained in our fits. We therefore conclude that there is no significant \lya{} emission from the compact components seen in the NIRCam images \citep[for a similar procedure see][]{leclercq2017}.

The observed \lya{} halo is well fitted by our two-component Gaussian model. We fit an exponential with scale length of $1.28\pm 0.18$~arcsec, and ellipticity of $e=0.66\pm 0.06$, with an angle of $\theta = (137\pm 3)^\circ$. For the core Gaussian component we fit ${\rm FWHM} = 0.43\arcsec$, marginally resolved given the PSF (Sect.~\ref{sec:data:muse}), and implying a size of $r_e \lesssim 2$~pkpc (magnification corrected). The decomposition of both components of the fit is shown in Fig.~\ref{fig:LyA_morph_model}.
To derive the physical geometrical parameters of the exponential halo in the source plane, we apply the radial and transverse magnification corrections (see Sect.~\ref{sec:data:jwst}). The halo is more strongly magnified along a direction approximately aligned with the major axis of the observed ellipse. However, even after applying these corrections, the halo remains significantly elliptical, with a scale length of $h = 5.7\pm 0.7$~pkpc and ellipticity $e = 0.56$. A schematic representation of the ellipses in the source and image planes is shown in Fig.~\ref{fig:source_plane_ellipse_lya}.

The total flux ratio between the halo and more compact Gaussian component is $\sim 5.4$, indicating that a large fraction of the \lya{} flux comes from an extended component. This is opposed to \Halpha{}, which presents a point-like morphology centered in the compact UV core \citep[see][]{labbe2024}. 
The \lya{} halo is consistent with the ones typically found around star-forming galaxies as we  discuss in Sect.~\ref{sec:discussion:halo}.

\subsection{\lya{} profile}\label{sec:lya_profile}

\begin{figure}
    \centering
    \includegraphics[width=\linewidth]{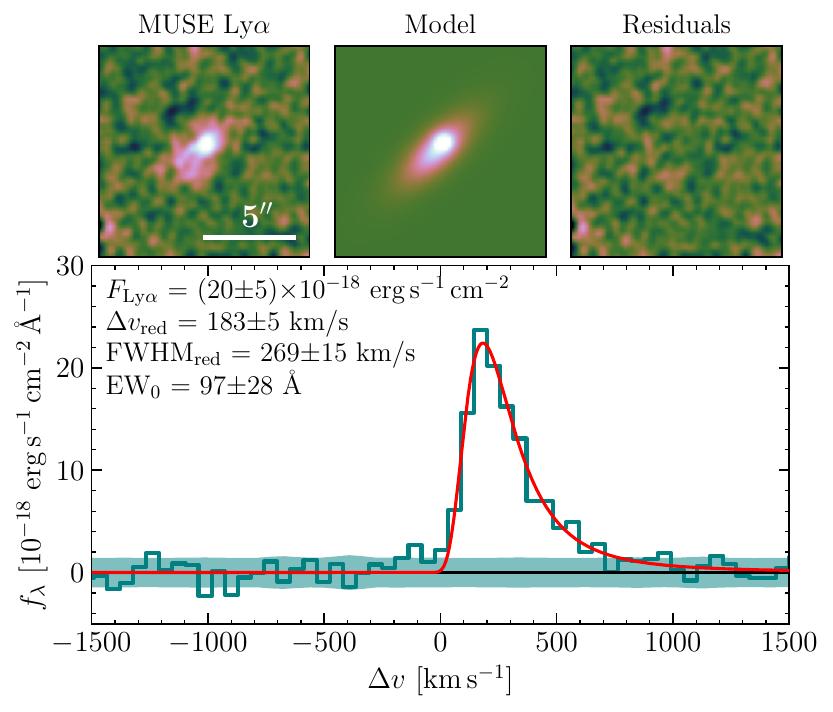}
    \caption{\lya{} spatial morphology and 1D spectrum. {\bf Top:} MUSE pseudo-narrowband of the \lya{} emission (left),  core+halo model (center), and  residuals of the fit (right). The morphology of the \lya{} halo is fitted to a two-component model composed of a 2D Gaussian and exponential. The scale length of the exponential component (halo) is $h=5.7\pm 0.7$~pkpc, correcting for magnification. {\bf Bottom:} Optimally extracted 1D spectrum of \lya{}. We show the fitted skewed Gaussian model (red line). The \lya{} line shows a narrow single-peaked profile, typical of star-forming galaxies. The bottom axis shows velocities with respect to the systemic redshift of the \Halpha{} emission.}
    \label{fig:monster_lya}
\end{figure}

We now turn to the spectral profile of the Ly$\alpha$ emission from \monster{}. We perform an optimal extraction \citep{horne1986} of \lya{} using the model fitted as described in Sect.~\ref{sec:extended_lya}. The extracted profile is shown in Fig.~\ref{fig:monster_lya}. We fit a skewed Gaussian to the \lya{} profile \citep[e.g.,][]{shibuya2014} obtaining best-fit parameters of $v_{\rm red}=183\pm 5$~km\,s$^{-1}$ (red peak velocity offset), $A=(22.7\pm0.7)\times10^{-18}$~\flambdaunits{} (amplitude), $a_{\rm asym.}=0.25\pm 0.02$ (asymetry parameter) and ${\rm FWHM_{\rm red}}=270\pm 15$~km\,s$^{-1}$. By integrating this best-fit curve, we obtain a total line flux of $F_\lya=(20\pm 5)\times 10^{-18}$~$\rm erg\,s^{-1}\,cm^{-2}$, corresponding to a luminosity of $10^{42.63\pm 0.11}$~erg\,s$^{-1}$ (magnification corrected flux and luminosity). We compute the EW of the \lya{}, first obtaining the continuum flux from the optimally extracted 1D spectrum in the range 1220-1450~\AA{}, after masking sky lines. We measure $\rm EW_\lya = 97\pm 28$~\AA{} from the MUSE spectrum.
We find moderate spatial variations of the fitted skewed Gaussian parameters, with peak velocities ranging from 100 to 200~km\,s$^{-1}$, and FWHM from 200 to 300~km\,s$^{-1}$ across the halo (Fig.~\ref{fig:monster_halo_lya_spatial_properties}). These spectral variations are typical in \lya{} halos of star-forming galaxies, and may be related to gas dynamics \citep[e.g.,][]{erb2018}.

\begin{figure}
    \centering
    \includegraphics[width=\linewidth]{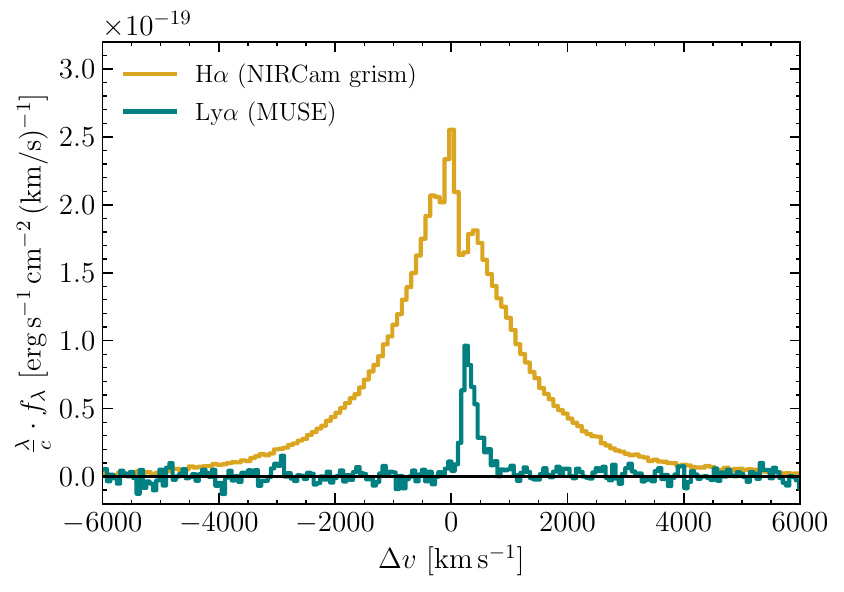}
    \caption{Flux density in velocity space of the \Halpha{} and \lya{} lines of \monster{}, centered on the systemic velocities of the two lines at $z=4.464$. The ratio of the line fluxes is $\lya{}/\Halpha{} = 0.066$. In addition, \Halpha{} has a prominently broad profile (${\rm FWHM} \approx 4500~\rm km\,s^{-1}$; \citealt{labbe2024}), while \lya{} presents the classical shape of a single narrow red peak (${\rm FWHM} = 270~\rm km\,s^{-1}$; see Fig.~\ref{fig:monster_lya}).}
    \label{fig:Monster_LyA_Halpha}
\end{figure}

In Fig.~\ref{fig:Monster_LyA_Halpha} we compare the absolute fluxes of the \lya{} and \Halpha{} lines of \monster{}. We measure a ratio of $\lya{}/\Halpha{} = 0.065$ (0.2 if we would consider only the narrow component of \Halpha{} fitted in \citealt{labbe2024}). This ratio is far from the theoretical $\lya{}/\Halpha{} = 8.7$ ratio for $\fesclya = 1$, under the traditional assumption of case B recombination and a gas temperature of $T=10^4$~K, typical of \ion{H}{ii} regions. 
This ratio is also very small in comparison to that measured in composite quasar spectra $\lya{}/\Halpha{}\sim 3$ \citep{vandenberk2001}.
The \lya{} EW and $v_{\lya{},\rm red}$ of \monster{} are comparable to typical star-forming galaxies in the field (Mascia et al. in prep.).

\subsection{Rest-frame UV emission lines}\label{sec:uv_lines}

\begin{figure}
    \centering
    \includegraphics[width=0.8\linewidth]{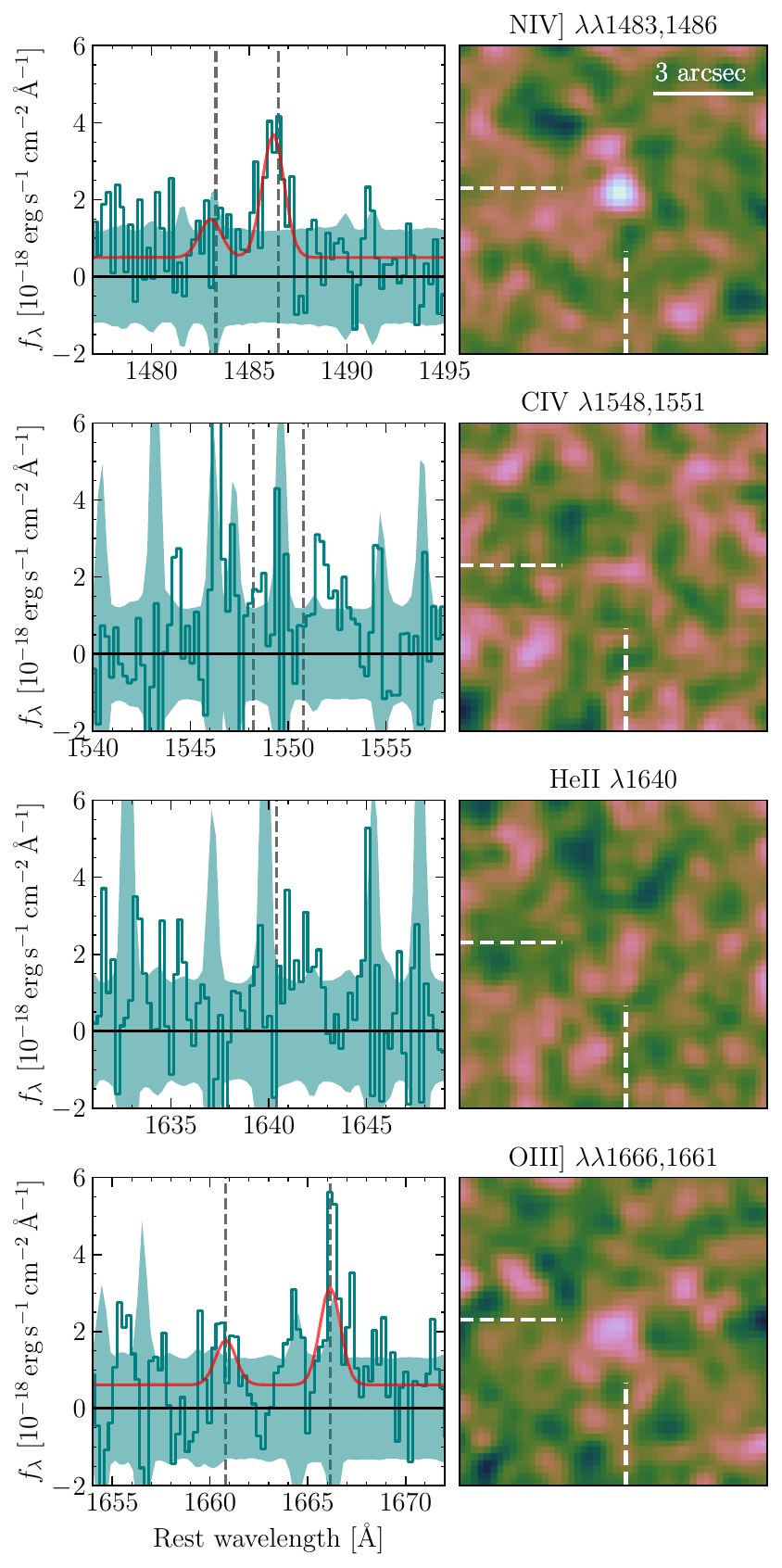}
    \caption{Rest-frame UV emission lines of \monster{}. The {\bf left column} shows the 1D spectrum around the wavelength of selected UV emission lines: \niv{} ($\lambda\lambda$1483,1486), \ion{C}{iv} ($\lambda\lambda$1548,1551), \ion{He}{ii} ($\lambda$1640), and \ion{O}{iii}] ($\lambda\lambda$1661,1666). In the case of \niv{} and \ion{O}{iii}]  the best-fitting two-component Gaussian model with fixed width is also shown. The {\bf right column} shows images obtained collapsing the \muse{} datacube in an interval of 10~\AA{} centered on the selected lines (for doublets, both components stacked), and subtracting the continuum, measured 1000~\AA{} around the position of the line (masking other emission lines and skylines).}
    \label{fig:uv_line_stamps}
\end{figure}

The rest-frame UV spectrum was already covered by the NIRspec/PRISM data presented in \cite{labbe2024}, but the very low spectral resolution ($R\sim 50$--$100$) of those data challenged the detection and characterization of various emission lines. In the UV spectrum of \monster{} in the MUSE data, which has a lower sensitivity but a much higher resolution, we also find a 4$\sigma$ detection of \niv{} ${\lambda1486}$ at a redshift consistent with the H$\alpha$ redshift. We fit a Gaussian profile to this emission line, and we measure a flux of $F_{\rm\niv{}\lambda 1486} = (2.5\pm 0.6)\times 10^{-18}~\rm erg\,s^{-1}\,cm^{-2}$ (${\rm EW}_{\rm\niv{}\lambda 1486} = 9 \pm 2$~\AA{}).
After Ly$\alpha$, this is the most prominent UV emission line in our data (see Sect.~\ref{sec:uv_lines}). By collapsing the \muse{} datacube in 10~\AA{} around the position of \niv{} $\lambda$1486 we obtain the continuum subtracted stamp shown in the top right panel of Fig.~\ref{fig:uv_line_stamps}.
The width of the Gaussian fit to \niv{} $\lambda$1486 is well resolved as ${\rm FWHM} = 270 \pm 50$~km\,s$^{-1}$.
Then, we fit a second Gaussian centered on the expected position of the other component of the \niv{} doublet, $\lambda$1483, fixing the width of the Gaussian to be the same. We get a tentative flux of $F_{\rm\niv{}\lambda 1483} = (0.8\pm 0.5)\times 10^{-18}~\rm erg\,s^{-1}\,cm^{-2}$ (${\rm EW}_{\rm\niv{}\lambda 1486} = 2.8 \pm 1.7$~\AA{}). We measure a doublet ratio of ${f_{1483}}/{f_{1486}} = 0.31 \pm 0.19$, which is suggestive of very high gas density \citep[$n_e > 10^5$ from Fig.~2 in][]{kewley2019}. Similarly low ratios were also observed in GN-z11 \citep{maiolino2024b} and CEERS-1019 \citep{marques-chaves2024}, both luminous high-redshift objects with suggested AGN activity.

We also marginally detect the \ion{O}{iii}]$\lambda\lambda$1661,1666 doublet (individual comopnents with ${\rm S/N}\approx 2$). The continuum-subtracted MUSE image suggests a clear detection (bottom right panel of Fig.~\ref{fig:uv_line_stamps}); however, the flux of \ion{O}{iii}] is affected by a skyline, and may be regarded as an upper limit.
The strong \niv{} emission, together with the \oiii{} flux reported in \citet{labbe2024}, suggests a moderate N/O abundance, with $\log_{10}({\rm N/O}) \gtrsim -1.1$ \citep[e.g.,][]{perez-montero2013} under the assumption of an electron temperature $T_e = 10^4$~K. However, deeper observations of both \niv{} and \ion{O}{iii}] are required to robustly determine the N/O ratio, as the \ion{O}{iii}] fluxes are sensitive to $T_e$.

Lastly, we neither find significant detection of \ion{He}{ii} ($\lambda$1640) or \ion{C}{iv} ($\lambda\lambda$1548,1551). \citet{labbe2024} measure a \ion{C}{iv} flux that is comparable to that of \niv{} in the NIRSpec prism spectrum of \monster{}, yet we do not detect \ion{C}{iv}. It is possible that the \ion{C}{iv} line is much broader than \niv{}, hence hindering the detectability in our spectrum. Moreover, several sky lines affect the measurement of this line, as evidenced by the uncertainty spikes in the left panel of the second row of Fig.~\ref{fig:uv_line_stamps}.

\section{Large-scale \lya{} emission}\label{sec:large_scale_lya}

\begin{figure*}
    \centering
    \includegraphics[width=0.8\linewidth]{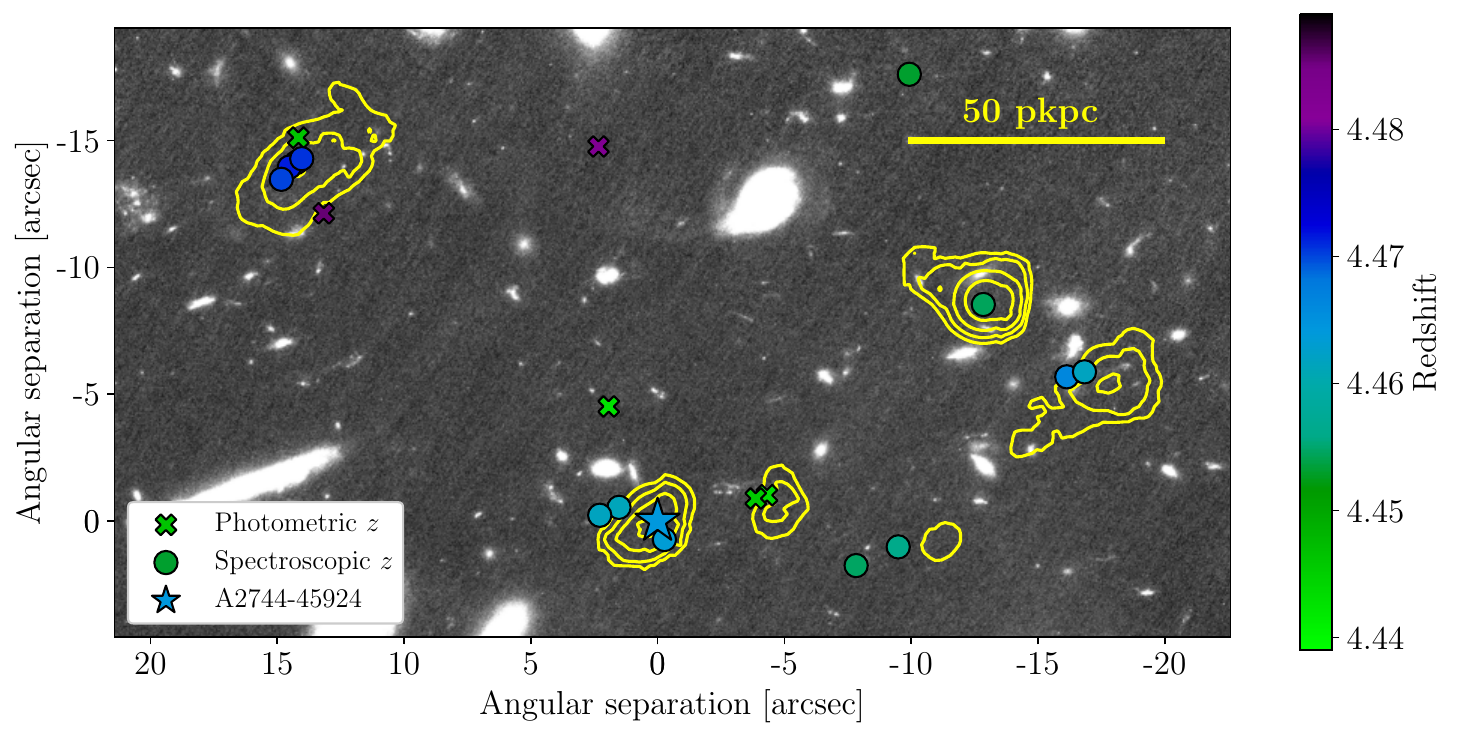}
    \caption{Detected \lya{} halos in the vicinity of \monster{}. The background image is the NIRCam F070W cutout. The search was performed in a box volume of $\approx 170$~cMpc$^3$ (correcting with the mean magnification of the sources in the field) defined by the FoV of the \muse{} data ($1.3\arcmin{}\times 1.3\arcmin$); however, we only show a cutout where the halos are detected. We show the contours of the \lya{} emission detected at $z=4.464\pm 0.02$. Each contour level corresponds to a surface brightness increase by a factor of 1.5 (baseline $1.8\times 10^{-18}$~$\rm erg\,s^{-1}\,cm^{-2}\,arcsec^{-2}$). Also shown are the positions of the \jwst{} objects with spectroscopic (circles) and photometric (crosses) redshift in the same interval (\protect\citetalias{naidu2024}, \protect\citealt{suess2024}). We detect an elevated number of \lya{} halos in the environment of \monster{}; however, this result is consistent with the galaxy overdensity \protect\citep{matthee2024b}. The total flux of \lya{} is moderate to low when compared with the total \Halpha{} emission of \monster{}'s neighbors (See Fig.~\ref{fig:lya_vs_ha_environment}).}
    \label{fig:f070w_with_LyA_contours}
\end{figure*}

We investigate the \lya{} emission of other sources in the vicinity of \monster{}. We identify \lya{} halos using the \textsc{shine}\footnote{\url{https://github.com/matteofox/SHINE}} tool, presented in \citet{tornotti2025}. We look for halos using this tool between observed wavelength 6600 and 6674~\AA{}, corresponding to $z_\lya = 4.43$--4.49~\AA{}. The field of view (FoV) of the \muse{} data corresponds to a box with side 1.3\arcmin{} roughly centered in \monster{}, hence we probe a volume of $\sim 172$~cMpc$^3$ (after correcting by the mean magnification of all the sources in the field, $\langle \mu \rangle = 1.8$, with values ranging from 1.5 to 2.9). We use a Gaussian smoothing kernel with $\sigma = 2$~px in the spatial directions, and impose a minimum spatial area of 50~px, 500 minimum connected voxels, and a size between 3 and 100 spectral pixels. After a visual inspection of the output 40 halo candidates produced using \textsc{shine}, we identify 4--5 extended \lya{} halos between $z\sim 4.45$ and 4.47. These halos are shown in Fig.~\ref{fig:f070w_with_LyA_contours}.
All the \lya{} halos have redshifts consistent with the spectroscopic redshifts of the sources they spatially overlap \citepalias{naidu2024}, with expected differences of a few hundred km\,s$^{-1}$, in line with the typical shift of \lya{} red peaks relative to the systemic redshift.  
Within the \muse{} FoV, we identify five additional sources with spectroscopic redshifts within the probed interval, but none of them exhibits significant \lya{} emission. Additionally, we detect one extra \lya{} emitter located $\sim 50\arcsec$ to the southeast, which shows prominent but compact \lya{} emission (not shown in Fig.~\ref{fig:f070w_with_LyA_contours}).
The discarded candidates were either clearly extended continuum sources or line emitters associated with sources with photometric or spectroscopic redshifts incompatible with \lya{}. We note that we use rather permissive search parameters in order to be able to spot even the fainter \lya{} halos.

\begin{figure}
    \centering
    \includegraphics[width=\linewidth]{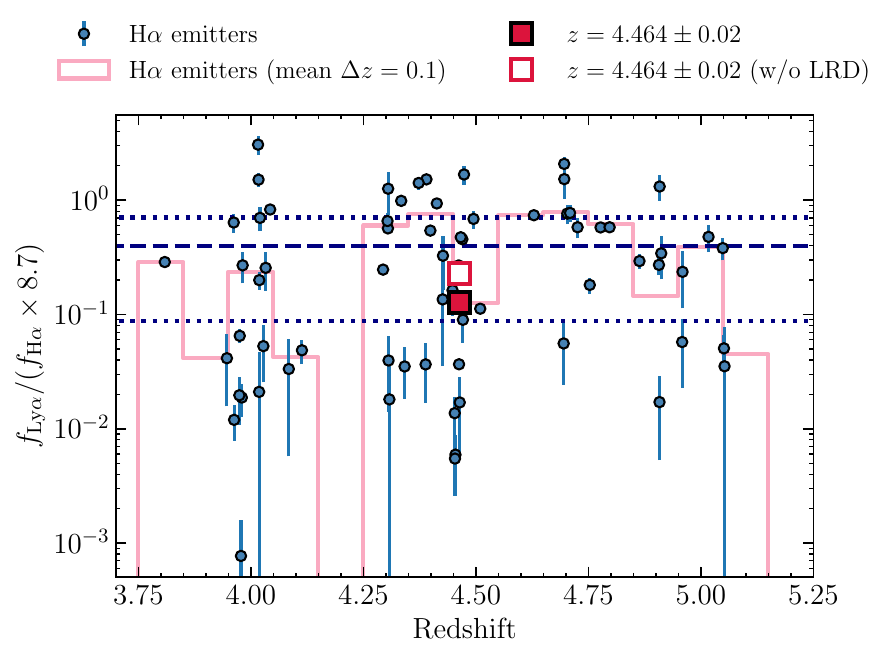}
    \caption{Ratio of total \lya{} to \Halpha{} flux for all the \Halpha{} emitters in the FoV of the \muse{} data presented in this work, and the data from \cite{richard2021} in the A2744 cluster. We include a $1/8.7$ factor for direct comparison with the Case B ($T=10^4$~K) ratio. The dashed and dotted blue lines show the mean and standard deviation in the field, respectively. We note that the fluxes are not dust-corrected. We show the total flux ratio variation in bins of redshift $\Delta z = 0.1$ (pink line), and within $\Delta z=0.02$ around the redshift of \monster{}, itself included (red filled square) and excluded (empty square). The $\lya{}/\Halpha{}$ ratio in the environment of \monster{} is moderate in comparison with other overdensities in our data, despite the large intrinsic overdensity of \Halpha{} emitters \citep{labbe2024, matthee2024b}.}
    \label{fig:lya_vs_ha_environment}
\end{figure}

We also study the strength of the \lya{} overdensity in relation with \Halpha{}. For this, we analyze the \lya{} emission of the spectroscopic sample of \Halpha{} emitters in the MUSE datacube centered in \monster{}, as well as in the data presented in \citet{richard2021} centered in the A2744 cluster. We note that the depth of the observations from \citet{richard2021} is comparable to that of our main dataset, with 3.5--7~h of on-source exposure, depending on the specific area; but, since we select all the sources from the homogeneous NIRCam images from their \Halpha{} emission, the differences between the MUSE datasets only have an impact on the accuracy of the individual \lya{} measurements, thus not affecting our results.
The FoV of both MUSE cubes are covered by NIRCam grism spectroscopy by the ALT survey that probes \Halpha{} at $3.8<z<5.1$.
\citet{matthee2024b} found evidence for a large overdensity ($1+\delta_{\rm 1 cMpc} \approx 30$) of galaxies (\Halpha{} emitters) around \monster{}. A discussion about the galaxy overdensities in the A2744 field is found in \citetalias{naidu2024}. Likewise, the detection of 4 prominent \lya{} halos in the neighborhood of \monster{} indicates an excess of \lya{} emission with respect to random.
To determine whether excess ionizing radiation---such as that produced by the AGN in \monster{}---has enhanced the photoionizing background and contributed to the excess \lya{} emission from halos around \monster{}, we must account for the underlying galaxy overdensity.  
The excess of \lya{} emission in $z=4.464\pm 0.02$ in the FoV of both \muse{} fields is $\delta + 1 \approx 20$ (comparing with the mean value in our data), while the excess of \Halpha{} is $\delta + 1 \approx 54$, when comparing the total flux of these lines with respect to the mean across $z=3.8$--5.5 in the same footprint.

We first extract MUSE spectra of all the \Halpha{} emitters in the ALT spectroscopic catalog using as extraction mask the morphology of the \Halpha{} emission, as described in Mascia et al. (in prep.). The \lya{} profile is then fitted in the 1D spectrum with a double skewed Gaussian to obtain priors for the line parameters, specifically the red peak wavelength and FWHM.
Then, we re-extract \lya{}, this time accounting for possible extended halos. To achieve this, we construct optimal weight masks by collapsing the MUSE datacubes along the spectral direction within a range defined by the FWHM and peak position of the red \lya{} component.
We mask pixels below a S/N threshold of 3 after applying a Gaussian smoothing kernel with $\sigma=2$~px. Finally, we re-extract the spectra of all \Halpha{} emitters and measure the \lya{} fluxes by performing a fits to double skewed Gaussian profiles.

Figure~\ref{fig:lya_vs_ha_environment} shows the ratio between the observed \lya{} and \Halpha{} emission from the \Halpha{} emitters within the two deep MUSE cubes in the A2744 field. We highlight the average Ly$\alpha$ to H$\alpha$ ratio at the redshift of \monster{}. Our main finding is that despite the large overdensity of \Halpha{} emitters, the \lya{} produced by these sources is moderate to low in comparison with other overdensities in the field, yet within the typical values (mean $\lya/\Halpha=0.4$, standard deviation 0.3). This result makes it unlikely that the observed \lya{} halos near \monster{} could be partly powered by its AGN, via scattering of \lya{} or photoionization \citep[e.g.,][]{cantalupo2014}. Moreover, \monster{} has three very close neighbors (angular separation $<3\arcsec$, $<15$~pkpc in projection, $\Delta z \lesssim 0.02$; $\Delta v\lesssim 1000\,\mathrm{km}\,\mathrm{s}^{-1}$), which totally lack detection of \lya{} emission despite being detected in \Halpha{} ($L_\Halpha{}\approx10^{42}$~erg\,s$^{-1}$), hence, there is no evidence that any escaping ionizing radiation from \monster{} is significantly affecting the CGM of these galaxies.

\section{What powers the \lya{} emission from \monster{}?}\label{sec:discussion_what_powers_lya}

\subsection{A weak halo for an extremely bright LRD}\label{sec:discussion:halo}

\begin{figure}
    \centering
    \includegraphics[width=\linewidth]{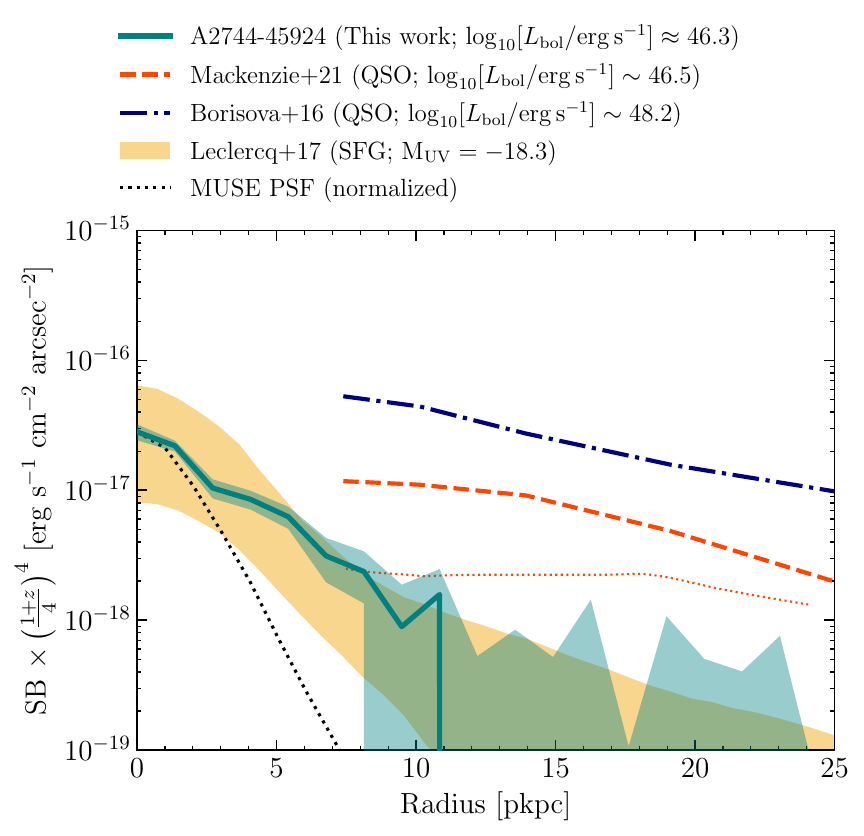}
    \caption{Radially average surface brightness (SB) profile of the \lya{} emission of \monster{}. Shown are the SB profile measured in the \muse{} image (teal solid line; the shaded area gives the 1$\sigma$ dispersion), and the SB produced by a point source for reference (black dotted line), normalized to the peak \lya{} emission of \monster{}. We compare our measurements to the halos measured around star-forming galaxies \protect\citep[$3<z<6$; $-22<{M}_{\rm UV}<-15$;][shaded orange region between 16th and 84th percentiles]{leclercq2017}, and also  to \lya{} halos of quasars at $z\sim 3.2$ \protect\citep[blue dot-dashed and orange dashed lines,][respectively]{borisova2016, mackenzie2021}. The \lya{} halo of \monster{} is comparable to that of SFGs both in terms of absolute SB and exponential scale length, despite having $L_{\rm bol}$ comparable to that of quasars from \protect\citet{mackenzie2021}. For reference, we also plot the faintest quasar halo in \citet{mackenzie2021} as a dotted orange line.}
    \label{fig:LyA_SB_profile}
\end{figure}

We begin by investigating the origin of the \lya{} emission by analyzing the properties of the extended \lya{} halo of \monster{}.
Extended \lya{} nebulae are ubiquitous around bright quasars \citep{borisova2016, arrigonibattaia2019}. \citet{mackenzie2021} reported a 100\% detection rate of \lya{} halos around fainter quasars,\footnote{However, see also \citet{herenz2015}, who reported 0\% detection rate in five radio-quiet quasars.} with extensions exceeding $60~{\rm pkpc}$. They also found that the extent of quasar \lya{} halos remains roughly independent of source luminosity, differing only by a normalization factor.
In contrast, we find a relatively compact \lya{} halo around \monster{}, whose surface brightness profile\footnote{The SB was measured in radii intervals in the source plane, considering radial and transverse magnification factors of  $\mu_t=1.68$, $\mu_r=1.07$, and a shear angle $\theta_\mu = 19.9^\circ$ \citep{furtak2025, price2025}.} is shown in Fig.~\ref{fig:LyA_SB_profile}. Fitting an exponential profile to this halo yields a scale length of $h=5.7\pm 0.7~{\rm pkpc}$ (see Sect.~\ref{fig:LyA_morph_model}).

Figure~\ref{fig:LyA_SB_profile} compares the \lya{} surface brightness profile of \monster{} with that around star-forming galaxies from \citet{leclercq2017}, as well as halos powered by luminous quasars (${M}_{\rm UV}\approx -29$) from \citet{borisova2016} and fainter quasars (${M}_{\rm UV}\approx -25$) from \citet{mackenzie2021}. The distinction between these populations is evident: quasar halos exhibit flatter profiles and extend over significantly larger distances.

As inferred from the H$\alpha$ luminosity and FWHM,
\monster{} shows a very high bolometric luminosity $\log_{10}(L_{\rm bol} / \rm erg\,s^{-1}) = 46.3\pm 0.2$ \citep[see][and references therein]{matthee2024b}, placing it in the regime of luminous quasars \citep{shen2020}.
For comparison, the average bolometric luminosities in \citet{borisova2016} and \citet{mackenzie2021} are $\log_{10}(L_{\rm bol} / \rm erg\,s^{-1}) \approx 48.2$ and $46.5$, respectively, estimated using the bolometric corrections from~\citet{richards2006}. The latter is comparable to the \Halpha{}-inferred bolometric luminosity of \monster{}, yet their \lya{} halos are $\sim$10 times brighter.
However, the validity of standard scaling relations for LRDs is uncertain, so $L_{\rm bol}$ could be overestimated. \citet{mackenzie2021} reported a tentative relation between $L_{\rm UV}$ and the \lya{} halo luminosity. Extrapolating their results, we obtaing that $L_{\rm bol}$ would need to be overestimated by $\gtrsim 2$~dex for the 1$\sigma$ limit to match the average SB in \citet{mackenzie2021}.
In turn, the halo of \monster{} is consistent with the halos found by \citet{leclercq2017} around star-forming galaxies with ${M}_{\rm UV}$ between $-15$ and $-22$ (\monster{} has ${M}_{\rm UV}=-19.39^{+0.12}_{-0.10}$), which have a median scale length of $h=4.5~{\rm pkpc}$ (ranging from 1 to 18~pkpc).
Overall, we find no evidence that the AGN in \monster{} is leaking enough ionizing photons to power a \lya{} halo like those observed around quasars with similar bolometric luminosities, either in terms of scale length or absolute surface brightness.

\subsection{Spatial shift of UV continuum and \lya{} emission}\label{sec:spatial_positions_components}

\begin{figure*}
    \centering
    \includegraphics[width=0.7\linewidth]{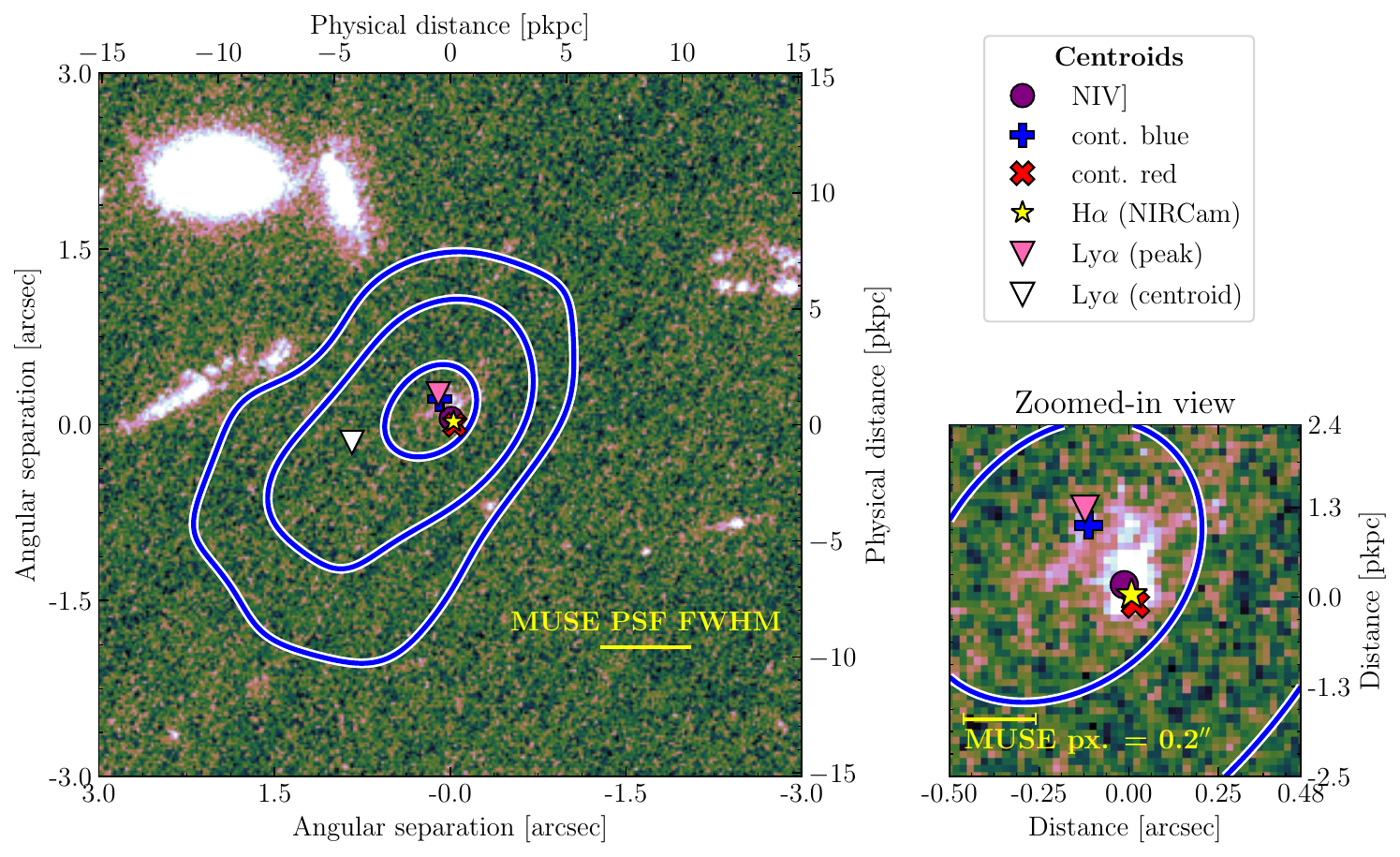}
    \caption{\lya{} contours and centroids of different spectral components of \monster{}, over the NIRCam F070W cutout. Shown are contours corresponding to 10\%, 50\%, and 90\% of the peak \lya{} flux (see Sect.~\ref{sec:extended_lya}). Also shown are the flux averaged centroids of the point-like \nircam{} H$\alpha$ emission (yellow star), the \muse{} collapsed continuum in the ranges 1235--1430 \AA{} (\contblue{}) and 1555--1712 \AA{} (\contred{}; blue and red crosses);  the centroid of the \niv{} (purple circle) and \lya{} lines (white triangle); as well as the position of the peak \lya{} (pink triangle). The bottom right panel shows a zoomed-in view of \monster{}. The spatial centroid of {\it cont. red} is consistent with the centroids of the \nircam{} and the position of \niv{}, while {\it cont. blue} is shifted toward the position of the peak \lya{} and the diffuse extended component in F070.}
    \label{fig:LyA_contour_f070w}
\end{figure*}

Our data revealed different spatially resolved components of \monster{}: extended \lya{} emission (Sect.~\ref{sec:extended_lya}), two rest-frame UV continuum components (Sect.~\ref{sec:muse_uv_continuum}), and a spatially compact \niv{} emission line (Sect.~\ref{sec:uv_lines}, Fig.~\ref{fig:uv_line_stamps}).
In Fig.~\ref{fig:LyA_contour_f070w} we show the contours of the \lya{} halo, overimposed on the F070W image, and we mark the spatial positions of the centroids of the \muse{} continuum components and \niv{}, peak \lya{}, and the coordinates of \Halpha{} from NIRCam. Although the PSF of NIRCam and \muse{} are very different, it is clear that the peak of the \lya{} emission is consistent with the center of the diffuse extended component seen in F070W (see Fig.~\ref{fig:NIRCam_UV_model}). Both these peaks are shifted $\sim 0.2\arcsec$ from the core component that marks the position of the AGN.

The unresolved morphology of \niv{} might suggest that it is originated close to the AGN component (see Sect.~\ref{sec:uv_lines}). We fit a 2D Gaussian with ${\rm FWHM} = 0.65\arcsec$ (for comparison, \muse{} average effective PSF of 0.75\arcsec{}), whose center is shown in Fig.~\ref{fig:LyA_contour_f070w}. The centroid of the \niv{} emission aligns with \contred{} and the \nircam{} centroids.
In addition, we compute the UV colors for the components fitted to the NIRCam F070W and F090W images, fixing the morphology of both components (see Appendix~\ref{sec:alternate_uv_morph_fit}). Although the \lya{} contamination in F070W and possible model degeneracies complicate the interpretation of these UV colors, we obtain $\rm F070W - F090W = 1.48$ for the compact component, aligned with \Halpha{} (AGN-associated); and 0.45 for the slighly more extended component (host galaxy-associated), respectively. Consistent with our findings from \muse{}, the AGN component appears redder and declines steeply from F090W to F070W, while the host galaxy component remains relatively blue. We note that the F070W band covers the rest-frame 1142--1429~\AA{}, extending blueward \lya{}, hence its flux is potentially affected by the \lya{} forest absorption.

From the spatial offset between the \lya{} peak and the core component (Fig.~\ref{fig:LyA_contour_f070w}) it is suggested that no significant \lya{} is emitted directly by the core along our line of sight. Additionally, the shift of the UV continuum centroid toward the \lya{} peak at bluer wavelengths implies that the \lya{} emission likely originates from this component. This is consistent with a scenario in which the core continuum is suppressed and a shifted component with bluer rest-UV colors dominates in the far-UV.
The alignment of the compact \niv{} emission with the \jwst{} coordinates of \Halpha{} ensures that our measured \lya{} to UV offset is robust against WCS uncertainties.  
The question to answer is whether the \lya{} emission is powered by star formation in a host galaxy or by the AGN through resonant scattering or photoionization.

\subsection{Can star formation power the \lya{} emission?}\label{sec:discussion:lya_vs_ha}

In star-forming galaxies at $z\approx0-2$, observations have shown that the observed EW of \lya{} correlates with \fesclya{} \citep{verhamme2017, sobral2017}.
\monster{} shows a fairly high \lya{} EW$_0 = 97\pm 28$~\AA{} with a narrow line-profile (FWHM $=270\pm 15$~km\,s$^{-1}$), suggesting a moderate $\fesclya = 0.48 \pm 0.18$ \citep{sobral2019}. If the \lya{} emission of the halo is produced by star formation, we should see the corresponding component of \Halpha{} in the grism data, with a total flux of $(2.4\pm 0.7)\times 10^{-18}~{\rm erg\,s^{-1}\,cm^{-2}}$ (magnification corrected and assuming a ratio of $\lya{}/\Halpha{}=8.7$). However, this component is likely outshined by the broad \Halpha{} emission from the central engine, with a flux of $\approx 40\times 10^{-18}~{\rm erg\,s^{-1}\,cm^{-2}}$, which we could not disentangle in the grism spectrum. \citet{chen2025} identified the extended component of \monster{} as a star-forming galaxy with a blue rest-frame UV slope ($\beta_{\rm UV}=-2.12$) and little dust attenuation. We estimate the star formation rate needed to power the \lya{} emission from \monster{} following Eq.~8 in \citet{sobral2019}, obtaining ${\rm SFR}_\lya = 4.5\pm 2.7~{\rm M_\sun\,yr^{-1}}$ (assuming $\fesclyc = 0$). We compare this value with the SFR obtained from the UV luminosity.
From the morphological model of F090W (Sect.~\ref{sec:uv_morphology}), we measure the UV luminosity subtracting the more compact component (assuming it is originated from the AGN), and obtain ${M}_{\rm UV} = -18.8 \pm 0.10$, whilst the value from the full morphological model is $M_{\rm UV}=-19.4\pm 0.1$.
For comparison, measuring the continuum from the MUSE optimally extracted spectrum, after masking emission lines, we likewise obtain $M_{\rm UV} = -19.39^{+0.12}_{-0.10}$.
The star formation rate associated with the UV luminosity minus the compact component is ${\rm SFR}_{\rm UV} = 3.6\pm0.3~{\rm M_\sun\,yr^{-1}}$ \citep{kennicutt2012}, fully compatible with the SFR inferred from \lya{}. This result suggests that the \lya{} emission of \monster{} can be fully powered by star formation from the extended host galaxy, without any contribution from AGN photoionization or scattering.

\section{Implications for the nature of A2744-45924}\label{sec:discussion:implications}

\subsection{A fully enshrouded SMBH}\label{subsec:fully_ensh_smbh}

Our results clearly indicate that Ly$\alpha$ emission in the A2744-45924 system is not centered on the main UV component (which almost certainly hosts an AGN, based on the co-spatial luminous and broad H$\alpha$ emission), but rather on the secondary UV component, probably a star-forming companion or satellite of the former. Furthermore, as discussed in Sect.~\ref{sec:discussion_what_powers_lya}, the extended Ly$\alpha$ emission appears consistent in luminosity, size and EW with being powered by star formation in the secondary UV component, with little or no influence from the AGN, despite its high \Halpha{}-inferred bolometric luminosity and close proximity ($\sim 1.7$~pkpc in projection, after correcting for magnification). 

Here we discuss how these findings can be reconciled with the fact that Type~1 AGN with a similar intrinsic luminosity are invariably surrounded by significantly more luminous and extended Ly$\alpha$ nebulae \citep[e.g.,][]{mackenzie2021}. One possibility is that the $L_\textrm{bol}$ of A2744-45924 derived from $L_{\rm H\alpha}$ is overestimated by more than 2 dex (see Sect.~\ref{sec:discussion:halo}), so that the true intrinsic bolometric luminosity of A2744-45924 is much lower, falling into a regime for which the ubiquity of luminous and extended Ly$\alpha$ nebulae has not been conclusively established yet. However, a simple explanation for the lack of a luminous Ly$\alpha$ nebula centered on it could be that the AGN photons usually responsible for powering Ly$\alpha$ emission do not reach the surrounding CGM due to some form of nuclear obscuration. 

An obscuration scenario is qualitatively consistent with the fact that A2744-45924 (as well as other LRDs) is a Type~2 AGN, thence obscured at least along our line of sight. Obscuration only along our line of sight (with a covering factor $f_\textrm{cov} \ll 1$), however, would not be sufficient to explain our findings, as AGN photons escaping in other directions would still be able to power a luminous Ly$\alpha$ nebula, as observed for instance around other Type~2 AGN \citep[e.g.,][]{brok2020}. Our findings therefore suggest that the obscuring material around the AGN in A2744-45924 (and possibly other LRDs) is likely to a have a large covering factor $f_\textrm{cov} \sim 1$.
This is consistent with the fact that the unobscured counterparts of LRDs has not yet been observed---That is, objects with similar broad H$\alpha$ luminosities and similar number density, but bluer colors.
A high $f_\textrm{cov}$ scenario is also consistent with the hypothesis, recently brought forward to explain LRDs, of a SMBH covered by a very dense envelope of gas (\citealt{inayoshi2025}; \citealt{ji2025}; \citealt{rusakov2025}; \citealt{naidu2025}; \citealt{graaff2025}), the so-called BH* model.

\subsection{Fate of AGN ionizing and Ly$\alpha$ photons}\label{subsec:fate}

To gain further insight on the implications of our observations for the properties of the obscuring gas, we need to make assumptions on the exact powering mechanism for AGN Ly$\alpha$ nebulae, thence the wavelength range of the photons that need to be obscured in the case of A2744-45924 and in turn the properties of the material that could be responsible for the obscuration. The dominant emission mechanism for AGN-powered Ly$\alpha$ nebulae is not fully clear and still subject to debate, the two most commonly invoked mechanisms being recombination and resonant scattering \citep[e.g.,][]{cantalupo2014}.\footnote{A third mechanism, collisional excitation, is also expected to play a role in extended Ly$\alpha$ emission around galaxies (e.g.,\ \citealt{dijkstra2009}) but is disfavored for AGN nebulae due to photo-ionization effects (e.g.,\ \citealt{pezzulli2019}).}

In the first scenario (recombination), ionizing photons reach and ionize the CGM, triggering the emission of recombination lines including Ly$\alpha$. If this interpretation is correct (as indicated for instance by the detection of nonresonant recombination lines from at least some AGN-powered Ly$\alpha$ nebulae; e.g., \citealt{langen2023}), then our main result---the {\it lack} of a luminous and extended Ly$\alpha$ nebula centered on the AGN---suggests a high covering factor of material that is optically thick to hydrogen-ionizing photons, requiring neutral hydrogen column density $N_{\rm H\,\text{\sc I}} >10^{17.2} \; \textrm{cm}^{-2}$. This is very consistent with (albeit not conclusive proof of) the BH* model, which predicts $N_{\rm H\,\text{\sc I}} \sim 10^{24} \; \textrm{cm}^{-2}$, fully sufficient to completely suppress the hydrogen ionizing flux, but is also subject to caveats, as discussed in Sect.~\ref{subsec:UVnature} below.

In the second scenario (scattering), part of the Ly$\alpha$ photons from the AGN broad line region (BLR) reach the CGM and are scattered by neutral hydrogen atoms therein into our line of sight.
The \lya{} to \Halpha{} ratio within $\sim 135$~km\,s$^{-1}$ around the \lya{} centroid is $\sim 0.07$, implying that a \lya{} escape fraction of $\sim 10\%$ (assuming case B recombination) would be sufficient, making this scenario energetically feasible ($\fesclya<1$), although it may be difficult to reconcile with the high gas densities inferred from other diagnostics.
Under the scattering hypothesis, the observed lack of a luminous \lya{} nebula implies a high $f_\textrm{cov}$ of obscuring material which is able to destroy AGN Ly$\alpha$ photons. Intriguingly, this is also in line with the observed lack of broad Ly$\alpha$ emission from A2744-45924. We note that this observation could in principle be explained by a proximate damped Ly$\alpha$ (PDLA) system \citep[e.g.,][]{marino2019}. This interpretation, however, is disfavored by the fact that a similar lack of broad Ly$\alpha$ emission is also observed in other LRDs \citep[e.g.,][]{ning2024}, suggesting, on statistical grounds, that the suppression of AGN Ly$\alpha$ photons is more likely due to an absorbing medium intrinsic to the system and surrounding the AGN with high $f_\textrm{cov}$.

One possible agent for the destruction of Ly$\alpha$ photons is dust. However, dust obscuration is  disfavored in \monster{} by the observed flat rest-IR slopes by NIRCam/MIRI and the far-IR and radio nondetections (Spitzer/Herschel, ALMA), which indicate no significant emission of reprocessed light by hot dust \citep{setton2024}. On the other hand, we   recall that resonant scattering of Ly$\alpha$ photons through a large column density of neutral hydrogen---as expected in the envelope of a BH*---increases the effective path length of Ly$\alpha$ photons, and therefore the efficiency of dust-induced suppression also for relatively low dust content \citep{scarlata2009}. For example, even for low values of dust attenuation such as $A_V\approx 0.1$,
one can obtain, for a static medium with $N_{\rm H\,\sc I}=10^{20}\,\mathrm{cm}^{-2}$ ($10^{21}\,\mathrm{cm}^{-2}$), escape fractions as low as $f_{\rm esc,Ly\alpha} \sim 0.03$ ($0.004$) \citep{neufeld1990,calzetti2000,verhamme2006}.
Overall, there is no evidence that Ly$\alpha$ photons from the AGN in A2744-45924 are being destroyed by dust, but this possibility cannot be fully discarded without more detailed modelling.

Another mechanism for Ly$\alpha$ suppression is efficient $l$\nobreakdash-changing (2p$\to$2s) collisions. A Ly$\alpha$ scattering event can be described as a succession of one Ly$\alpha$ absorption, inducing a transition of a neutral hydrogen atom from the 1s state to the 2p state, followed by spontaneous re-emission, accompanied by a decay back to the ground level (2p$\to$1s). It is  possible, however, that   between these two events, a collision---for instance with a proton---induces an $l$-changing transition from the 2p state to the 2s state, from which Ly$\alpha$ emission is not permitted. The transition back to the ground level will then occur through either two-photon decay (for $n \lesssim 10^4 \; \textrm{cm}^{-3}$) or collisional de-excitation (for $n \gtrsim10^4 \; \textrm{cm}^{-3}$; \citealt{spaans2006}; \citealt{neufeld1990}), in either case resulting in the effective destruction, rather than scattering, of the initial Ly$\alpha$ photon. For each scattering event, an $l$-changing (2p$\to$2s) collision, resulting in the destruction of a Ly$\alpha$ photon, is the most likely outcome only for extremely high densities $n_p \gtrsim n_{p, crit} = A_{Ly\alpha}/q_{2p2s} = 3.5 \times 10^{12} \; \textrm{cm}^{-3}$, where $A_{\textrm{Ly}\alpha} = 6.25 \times 10^8 \; \textrm{s}^{-1}$ is the Einstein coefficient for Ly$\alpha$ emission and $q_{\rm 2p2s} = 1.8 \times 10^{-4} \; \textrm{cm}^3 \; \textrm{s}^{-1}$ is the coefficient for a proton-induced 2p$\to$2s collisional transition (\citealt{dijkstra2016}). However, even for densities $n_p \ll n_{p,\rm crit}$, repeated scatterings across a path with high $N_{\rm H\,\text{\sc I}}$ boosts the probability of a Ly$\alpha$ photon being destroyed before escaping the system. Along similar lines as in \cite{dijkstra2016}, the number of required scattering events for Ly$\alpha$ destruction can be estimated as $N_\textrm{scatt} = n_{p,\rm crit}/n_p$. For the fiducial density $n_e \sim 10^9 \; \textrm{cm}^{-3}$ of the BH* model (\citealt{naidu2024}), a number of scatterings $N_\textrm{scatt} = 3.5 \times 10^3 $ would be sufficient, achievable (for a static medium and at line center) with a neutral hydrogen column density $N_{\rm H\,\text{\sc I}} \gtrsim 6 \times 10^{16} \; \textrm{cm}^{-2}$ (see equation 3 in~\citealt{dijkstra2016}) and abundantly met for $N_{\rm H\,\text{\sc I}} \sim 10^{24} \; \textrm{cm}^{-2}$, as in the BH* models. A more detailed analysis of this scenario would require taking into account the impact of kinematical effects, such as turbulence or possible inflow or outflow motions, on Ly$\alpha$ radiative transfer and is left for future investigation.\footnote{In the presence of differential motions with a constant velocity gradient, the required $N_\textrm{H\,\sc I}$ to obtain a given value of $N_\textrm{scatt}$ is only increased by a factor $\sim v_{\rm max}/v_{\rm thermal}$ \citep{bonilha1979,dijkstra2016}.}

Overall, our findings provide support for the presence of obscuring material with a large covering factor ($f_\textrm{cov} \sim 1$) of neutral hydrogen with at least a column density ($N_\textrm{HI} \gtrsim 10^{17.2} \; \textrm{cm}^{-2}$) around the AGN at the center of \monster{}. Furthermore, under the assumption that the dominant emission mechanism for AGN-powered Ly$\alpha$ nebulae is Ly$\alpha$ scattering, our observations appear fully consistent with the high gas density and hydrogen column densities expected in the envelope of a BH*.
We finally note that if LRDs have a higher than typical \Halpha{}/$L_{\rm bol}$
ratio \citep[e.g.,][]{graaff2025},
this could be in agreement with a scenario with $\fcov\sim 1$, where a large fraction of ionizing photons would be expected
to be reprocessed into recombination lines (including \Halpha{}) rather than
escaping to large distances.

\subsection{Origin of the compact UV emission and high-ionization emission lines}\label{subsec:UVnature}

Another important aspect to address, in the context of nuclear obscuration with $f_\textrm{cov} \sim 1$, is the nature of the main compact UV component in A2744-45924, as well as the presence of the  emission lines of highly ionized species, such as \ion{O}{III}], \ion{N}{IV}], and \ion{C}{IV}, in the spectrum (see Section \ref{sec:uv_lines} above, and \citealt{labbe2024}).
In one of the scenarios explored by~\citet{labbe2024}, the UV portion of the spectrum of A2744-45924 could represent AGN light escaping through unobscured channels and being scattered by free electrons into our line of sight. This scenario would imply $f_\textrm{cov} < 1$ and could also naturally explain the presence of high-ionization emission lines. Quantitatively, however, further work would be needed to determine the minimum fraction of the free (unobscured) solid angle ($f_\textrm{uncov} = 1-f_\textrm{cov}$) required to explain the observed UV and line fluxes in this scenario and verify whether it is consistent with $f_\textrm{uncov} \ll 1$, as indicated by our findings in this work. 

A second scenario considered by \cite{labbe2024} is that the UV spectrum may be dominated by a young, compact and metal-poor stellar population, which, in the context of the present discussion, could be interpreted as a compact nuclear star cluster surrounding the SMBH and its envelope. Because in this scenario the UV light originates outside the obscuring envelope, it would be consistent with $f_\textrm{cov} =1$, provided   that the high-ionization emission lines were powered by star formation rather than the AGN. This would be somewhat unusual, considering the high ionizing potential for producing \ion{N}{iv}] and \ion{C}{iv}, but perhaps not impossible considering the very blue observed slope ($-\beta_{UV} = 2.2-2.9$; \citealt{labbe2024}), plausibly accompanied by an ionizing spectrum correspondingly harder than usual \citep{topping2025}. This picture could also be supported by our finding that the \niv{}${\lambda1486}$ line is quite narrow (${\rm FWHM} \sim 270 \; \textrm{km} \; \textrm{s}^{-1}$; see Sect.~\ref{sec:uv_lines}), and therefore possibly arising from the interstellar medium (ISM) rather than the immediate proximity of the accretion disc. We note, in particular, that the relatively high density  $n_{\rm H} \sim 10^5 \; \textrm{cm}^{-3}$ inferred from the ${f_{1483}}/{f_{1486}}$ doublet ratio (Sect.~\ref{sec:uv_lines}) would in this case be indicative of a relatively dense ISM surrounding the even denser BH* envelope. For the implied ISM density $n \sim 10^5 \; \textrm{cm}^{-3}$, an ISM neutral hydrogen column density $N_{\rm H\,\text{\sc I}} \sim 6 \times 10^{20} \; \textrm{cm}^{-2}$ could already be sufficient to suppress also the Ly$\alpha$ photons from this hypothetical compact star-forming region through {$l$\nobreakdash-changing} collisions, as well as any hydrogen-ionizing photons. This could possibly explain why such a nuclear star cluster is not in turn powering an extended Ly$\alpha$ halo centered on the compact source.

One third possibility is that the compact UV component is due to AGN photons that traverse the dense gas envelope without being entirely absorbed. The BH* scenario is designed so that a non-negligible fraction of the hydrogen atoms in the envelope is collisionally excited to the $n=2$ level, capable of producing absorption features in Balmer lines, as well as producing a Balmer break by suppressing the Balmer continuum. Crucially, the photo-ionization cross-section for $n=2 \to \infty$ (i.e., the Balmer continuum) transitions decreases with decreasing wavelength (roughly as $\propto\lambda^3$). Therefore, we can expect that for high but not extreme column densities, UV photons could be able to traverse the envelope, possibly explaining the emergence of a blue UV spectrum and consistent with $f_\textrm{cov} = 1$. This scenario requires tailored modeling and will be the subject of future investigation. We note that, if produced \emph{within} the envelope (a few AU from the accretion disc in the model of \citealt{naidu2025}), UV high-ionization emission lines, such as \niv $\lambda 1486$, would also, in principle, be able to traverse the envelope, with attenuated flux but conserved doublet ratio and EW. It would be more difficult, however, to explain the narrow width. If instead the \niv{} $\lambda 1486$ line is produced outside the envelope (as suggested by the small FWHM), this scenario would need either a nonvanishing $f_\textrm{uncov}$ or some contribution from a compact young stellar population with a hard ionizing spectrum, as Lyman continuum photons encountering the envelope would be expected to be completely suppressed, in contrast to the Balmer continuum.

\section{Summary}\label{sec:summary}

In this paper we presented \muse{} observations of \monster{}, an extremely bright LRD ($L_\Halpha\approx10^{44}$~erg\,s$^{-1}$), revealing a narrow \lya{} emission line, and a moderately extended \lya{} halo. We summarize our main findings as follows:

\begin{enumerate}
    \item \monster{} shows a narrow (${\rm FWHM} = 270 \pm 15$~km\,s$^{-1}$) single-peaked \lya{} emission line, with ${\rm EW}_0 = 97\pm28$~\AA{} and a velocity separation of $\Delta v_{\rm red} = 183 \pm 5$km\,s$^{-1}$ with respect to the systemic redshift. The \lya{} emission has an extended spatial morphology, which can be well fitted by a combination of 2D Gaussian and exponential. The \lya{} nebula has an exponential scale length of $h=5.7\pm 0.7$~pkpc and ellipticity $e=0.46\pm 0.05$~(Sect.~\ref{sec:lya_profile}).

    \item We also report a significant (4$\sigma$) detection of \niv{}$\lambda$1486, which exhibits a very compact spatial morphology. This emission line is associated with AGN activity, and the low ratio with the marginally detected (1.6$\sigma$) \niv{}$\lambda$1483 suggest a high gas density (Sect.~\ref{sec:uv_lines}).
    
    \item The \lya{} emission peaks with a spatial offset of $\sim 0.2\arcsec{}$ relative to the NIRCam centroid of the compact core of \monster{}. This offset aligns with the position of the extended UV component seen in NIRCam. Additionally, the UV continuum detected with \muse{} shows a shift toward  shorter rest-frame wavelengths (1235--1430~\AA{}) compared to longer wavelengths (1555--1712~\AA{}). The spatial centroid of \niv{} is consistent with the compact core, supporting its AGN origin and confirming that the observed \lya{} and UV continuum shifts are not due to astrometry uncertainties (Sect.~\ref{sec:spatial_positions_components}).

    \item We investigated the \lya{} emission of sources in the neighborhood of \monster{}, in the \muse{} field of $\sim 1.3\arcmin\times 1.3\arcmin$. We find four extended \lya{} halos between 4.45 and 4.47 (in a volume corresponding to $\sim 172$~Mpc$^3$). We find an excess of total \lya{} emission in an interval of $\Delta z = 0.02$ around \monster{} ($1 + \delta_\lya \approx 20$). However, the ratio of the total \lya{} to  \Halpha{} flux is moderately low with respect to the average in the same FoV within $z=3.75$--5.25, but still within 1$\sigma$. This result implies that \monster{} does not significantly enhance the \lya{} luminosity of nearby halos through scattered \lya{} light (Sect.~\ref{sec:large_scale_lya}).

    \item The extended \lya{} halo of \monster{} resembles those of star-forming \lya{} emitters, more than quasars, in both scale length ($h=5.7 \pm 0.7$~pkpc) and surface brightness \citep{leclercq2017}. The narrow \lya{} line contrasts with the broad \Halpha{} line (${\rm FWHM} \approx 4500$~km\,s$^{-1}$), suggesting different origins. We show that the \lya{} emission can be fully explained by star formation in the host galaxy, without AGN contribution (Sects.~\ref{sec:discussion:halo} and~\ref{sec:discussion:lya_vs_ha}).

    \item We discuss the implications of our findings for a scenario in which the compact continuum emission and \Halpha{} are explained by a SMBH enshrouded in a dense cloud of gas (BH* models). We argue that the absence of a luminous Ly$\alpha$ halo, as seen in similarly bright Type I and II AGN, can be explained by obscuring neutral hydrogen gas with a covering factor $\fcov \sim 1$ and column density $N_{\rm H\,I} \gtrsim 10^{17.2}$~cm$^{-2}$. However, the main caveat for these interpretations is that $L_{\rm bol}$ could be largely overestimated (Sects.~\ref{subsec:fully_ensh_smbh} and~\ref{subsec:fate}).

    \item Finally, we consider the implications of the UV continuum and high-ionization emission lines for the dense obscuring gas hypothesis. The observed properties are consistent with a BH* model with $\fcov \sim 1$, although the exact interpretation depends on the gas distribution and geometry. Further modeling and observations are required to constrain the key parameters of the BH* model, i.e., $\fcov$, $N_{\rm H\,I}$, $n_e$ (Sect.~\ref{subsec:UVnature}).

\end{enumerate}

\begin{acknowledgements}

We thank the anonymous referee for constructive and useful comments.

We thank Sebastiano Cantalupo for comments on the draft.

Based on observations collected at the European Organisation for Astronomical Research in the Southern Hemisphere under ESO programme 114.27M6.001.

Funded by the European Union (ERC, AGENTS,  101076224). Views and opinions expressed are however those of the author(s) only and do not necessarily reflect those of the European Union or the European Research Council. Neither the European Union nor the granting authority can be held responsible for them. We acknowledge funding from {\it JWST} program GO-3516.

This work is based in part on observations made with the NASA/ESA/CSA James Webb Space Telescope. The data were obtained from the Mikulski Archive for Space Telescopes at the Space Telescope Science Institute, which is operated by the Association of Universities for Research in Astronomy, Inc., under NASA contract NAS 5-03127 for JWST. These observations are associated with program \#3516.

MG thanks the Max Planck Society for support through the MPRG.

FDE acknowledges support by the Science and Technology Facilities Council (STFC), by the ERC through Advanced Grant 695671 ``QUENCH'', and by the UKRI Frontier Research grant RISEandFALL.

TU acknowledges funding from the ERC-AdG grant SPECMAP-CGM, GA 101020943.

GK acknowledges support from the MERAC foundation.

\end{acknowledgements}

\bibliographystyle{aa}
\bibliography{my_bibliography}

@article{labbe2024,
	title        = {{An unambiguous AGN and a Balmer break in an Ultraluminous Little Red Dot at z=4.47 from Ultradeep UNCOVER and All the Little Things Spectroscopy}},
	author       = {{Labbe}, Ivo and {Greene}, Jenny E. and {Matthee}, Jorryt and {Treiber}, Helena and {Kokorev}, Vasily and {Miller}, Tim B. and {Kramarenko}, Ivan and {Setton}, David J. and {Ma}, Yilun and {Goulding}, Andy D. and {Bezanson}, Rachel and {Naidu}, Rohan P. and {Williams}, Christina C. and {Atek}, Hakim and {Brammer}, Gabriel and {Cutler}, Sam E. and {Chemerynska}, Iryna and {Cloonan}, Aidan P. and {Dayal}, Pratika and {de Graaff}, Anna and {Fudamoto}, Yoshinobu and {Fujimoto}, Seiji and {Furtak}, Lukas J. and {Glazebrook}, Karl and {Heintz}, Kasper E. and {Leja}, Joel and {Marchesini}, Danilo and {Nanayakkara}, Themiya and {Nelson}, Erica J. and {Oesch}, Pascal A. and {Pan}, Richard and {Price}, Sedona H. and {Shivaei}, Irene and {Sobral}, David and {Suess}, Katherine A. and {van Dokkum}, Pieter and {Wang}, Bingjie and {Weaver}, John R. and {Whitaker}, Katherine E. and {Zitrin}, Adi},
	year         = 2024,
	month        = {dec},
	journal      = {arXiv e-prints},
	pages        = {arXiv:2412.04557},
	doi          = {10.48550/arXiv.2412.04557},
	eid          = {arXiv:2412.04557},
	archiveprefix = {arXiv},
	eprint       = {2412.04557},
	primaryclass = {astro-ph.GA},
	adsurl       = {https://ui.adsabs.harvard.edu/abs/2024arXiv241204557L}
}

@ARTICLE{Arita25,
       author = {{Arita}, Junya and {Kashikawa}, Nobunari and {Onoue}, Masafusa and {Yoshioka}, Takehiro and {Takeda}, Yoshihiro and {Hoshi}, Hiroki and {Shimizu}, Shunta},
        title = "{The nature of low-luminosity AGNs discovered by JWST based on clustering analysis: progenitors of low-z quasars?}",
      journal = {\mnras},
         year = 2025,
        month = feb,
       volume = {536},
       number = {4},
        pages = {3677-3688},
          doi = {10.1093/mnras/stae2765},
archivePrefix = {arXiv},
       eprint = {2410.08707},
 primaryClass = {astro-ph.GA},
       adsurl = {https://ui.adsabs.harvard.edu/abs/2025MNRAS.536.3677A}
}

@ARTICLE{Lin25,
       author = {{Lin}, Xiaojing and {Fan}, Xiaohui and {Wang}, Feige and {Sun}, Fengwu and {Champagne}, Jaclyn B. and {Egami}, Eiichi and {Kakiichi}, Koki and {Lyu}, Jianwei and {Tee}, Wei Leong and {Yang}, Jinyi and {Bian}, Fuyan and {Bosman}, Sarah E.~I. and {Cai}, Zheng and {Casey}, Caitlin M. and {Decarli}, Roberto and {Faisst}, Andreas L. and {Fujimoto}, Seiji and {Harish}, Santosh and {Ilbert}, Olivier and {Inoue}, Akio K. and {Jin}, Xiangyu and {Kartaltepe}, Jeyhan S. and {Kocevski}, Dale D. and {Li}, Mingyu and {Liu}, Weizhe and {Liu}, Yichen and {Schindler}, Jan-Torge and {Shuntov}, Marko and {Tanaka}, Takumi S. and {Vestergaard}, Marianne and {Wu}, Yunjing and {Zhang}, Haowen and {Zhang}, Zijian},
        title = "{Bridging Quasars and Little Red Dots: Insights into Broad-Line AGNs at $z=5-8$ from the First JWST COSMOS-3D Dataset}",
      journal = {arXiv e-prints},
         year = 2025,
        month = apr,
          eid = {arXiv:2504.08039},
        pages = {arXiv:2504.08039},
          doi = {10.48550/arXiv.2504.08039},
archivePrefix = {arXiv},
       eprint = {2504.08039},
 primaryClass = {astro-ph.GA},
       adsurl = {https://ui.adsabs.harvard.edu/abs/2025arXiv250408039L}
}

@ARTICLE{Taylor25z9,
       author = {{Taylor}, Anthony J. and {Kokorev}, Vasily and {Kocevski}, Dale D. and {Akins}, Hollis B. and {Cullen}, Fergus and {Dickinson}, Mark and {Finkelstein}, Steven L. and {Arrabal Haro}, Pablo and {Bromm}, Volker and {Giavalisco}, Mauro and {Inayoshi}, Kohei and {Juneau}, St{\'e}phanie and {Leung}, Gene C.~K. and {P{\'e}rez-Gonz{\'a}lez}, Pablo G. and {Somerville}, Rachel S. and {Trump}, Jonathan R. and {Amor{\'\i}n}, Ricardo O. and {Barro}, Guillermo and {Burgarella}, Denis and {Brooks}, Madisyn and {Carnall}, Adam C. and {Casey}, Caitlin M. and {Cheng}, Yingjie and {Chisholm}, John and {Chworowsky}, Katherine and {Davis}, Kelcey and {Donnan}, Callum T. and {Dunlop}, James S. and {Ellis}, Richard S. and {Fern{\'a}ndez}, Vital and {Fujimoto}, Seiji and {Grogin}, Norman A. and {Gupta}, Ansh R. and {Hathi}, Nimish P. and {Jung}, Intae and {Hirschmann}, Michaela and {Kartaltepe}, Jeyhan S. and {Koekemoer}, Anton M. and {Larson}, Rebecca L. and {Leung}, Ho-Hin and {Llerena}, Mario and {Lucas}, Ray A. and {McLeod}, Derek J. and {McLure}, Ross and {Napolitano}, Lorenzo and {Papovich}, Casey and {Stanton}, Thomas M. and {Tripodi}, Roberta and {Wang}, Xin and {Wilkins}, Stephen M. and {Yung}, L.~Y. Aaron and {Zavala}, Jorge A.},
        title = "{CAPERS-LRD-z9: A Gas-enshrouded Little Red Dot Hosting a Broad-line Active Galactic Nucleus at z = 9.288}",
      journal = {\apjl},
         year = 2025,
        month = aug,
       volume = {989},
       number = {1},
          eid = {L7},
        pages = {L7},
          doi = {10.3847/2041-8213/ade789},
archivePrefix = {arXiv},
       eprint = {2505.04609},
 primaryClass = {astro-ph.GA},
       adsurl = {https://ui.adsabs.harvard.edu/abs/2025ApJ...989L...7T}
}

@ARTICLE{Bacon2017,
       author = {{Bacon}, Roland and {Conseil}, Simon and {Mary}, David and {Brinchmann}, Jarle and {Shepherd}, Martin and {Akhlaghi}, Mohammad and {Weilbacher}, Peter M. and {Piqueras}, Laure and {Wisotzki}, Lutz and {Lagattuta}, David and {Epinat}, Benoit and {Guerou}, Adrien and {Inami}, Hanae and {Cantalupo}, Sebastiano and {Courbot}, Jean Baptiste and {Contini}, Thierry and {Richard}, Johan and {Maseda}, Michael and {Bouwens}, Rychard and {Bouch{\'e}}, Nicolas and {Kollatschny}, Wolfram and {Schaye}, Joop and {Marino}, Raffaella Anna and {Pello}, Roser and {Herenz}, Christian and {Guiderdoni}, Bruno and {Carollo}, Marcella},
        title = "{The MUSE Hubble Ultra Deep Field Survey. I. Survey description, data reduction, and source detection}",
      journal = {\aap},
         year = 2017,
        month = dec,
       volume = {608},
          eid = {A1},
        pages = {A1},
          doi = {10.1051/0004-6361/201730833},
archivePrefix = {arXiv},
       eprint = {1710.03002},
 primaryClass = {astro-ph.GA},
       adsurl = {https://ui.adsabs.harvard.edu/abs/2017A&A...608A...1B}
}

@ARTICLE{Setton2025,
       author = {{Setton}, David J. and {Greene}, Jenny E. and {Spilker}, Justin S. and {Williams}, Christina C. and {Labb{\'e}}, Ivo and {Ma}, Yilun and {Wang}, Bingjie and {Whitaker}, Katherine E. and {Leja}, Joel and {de Graaff}, Anna and {Alberts}, Stacey and {Bezanson}, Rachel and {Boogaard}, Leindert A. and {Brammer}, Gabriel and {Cutler}, Sam E. and {Cleri}, Nikko J. and {Cooper}, Olivia R. and {Dayal}, Pratika and {Fujimoto}, Seiji and {Furtak}, Lukas J. and {Goulding}, Andy D. and {Hirschmann}, Michaela and {Kokorev}, Vasily and {Maseda}, Michael V. and {McConachie}, Ian and {Matthee}, Jorryt and {Miller}, Tim B. and {Naidu}, Rohan P. and {Oesch}, Pascal A. and {Pan}, Richard and {Price}, Sedona H. and {Suess}, Katherine A. and {Weaver}, John R. and {Xiao}, Mengyuan and {Zhang}, Yunchong and {Zitrin}, Adi},
        title = "{A Confirmed Deficit of Hot and Cold Dust Emission in the Most Luminous Little Red Dots}",
      journal = {\apjl},
         year = 2025,
        month = sep,
       volume = {991},
       number = {1},
          eid = {L10},
        pages = {L10},
          doi = {10.3847/2041-8213/ade78b},
archivePrefix = {arXiv},
       eprint = {2503.02059},
 primaryClass = {astro-ph.GA},
       adsurl = {https://ui.adsabs.harvard.edu/abs/2025ApJ...991L..10S}
}

@article{maiolino2024a,
	title        = {{JADES: The diverse population of infant black holes at 4 < z < 11: Merging, tiny, poor, but mighty}},
	author       = {{Maiolino}, Roberto and {Scholtz}, Jan and {Curtis-Lake}, Emma and {Carniani}, Stefano and {Baker}, William and {de Graaff}, Anna and {Tacchella}, Sandro and {{\"U}bler}, Hannah and {D'Eugenio}, Francesco and {Witstok}, Joris and {Curti}, Mirko and {Arribas}, Santiago and {Bunker}, Andrew J. and {Charlot}, St{\'e}phane and {Chevallard}, Jacopo and {Eisenstein}, Daniel J. and {Egami}, Eiichi and {Ji}, Zhiyuan and {Jones}, Gareth C. and {Lyu}, Jianwei and {Rawle}, Tim and {Robertson}, Brant and {Rujopakarn}, Wiphu and {Perna}, Michele and {Sun}, Fengwu and {Venturi}, Giacomo and {Williams}, Christina C. and {Willott}, Chris},
	year         = 2024,
	month        = {nov},
	journal      = {\aap},
	volume       = 691,
	pages        = {A145},
	doi          = {10.1051/0004-6361/202347640},
	eid          = {A145},
	archiveprefix = {arXiv},
	eprint       = {2308.01230},
	primaryclass = {astro-ph.GA},
	adsurl       = {https://ui.adsabs.harvard.edu/abs/2024A&A...691A.145M}
}

@article{harikane2023,
	title        = {{A JWST/NIRSpec First Census of Broad-line AGNs at z = 4-7: Detection of 10 Faint AGNs with M $_{BH}$ {}10$^{6}$-{}10$^{8}$ M $_{{\ensuremath{\odot}}}$ and Their Host Galaxy Properties}},
	author       = {{Harikane}, Yuichi and {Zhang}, Yechi and {Nakajima}, Kimihiko and {Ouchi}, Masami and {Isobe}, Yuki and {Ono}, Yoshiaki and {Hatano}, Shun and {Xu}, Yi and {Umeda}, Hiroya},
	year         = 2023,
	month        = {dec},
	journal      = {\apj},
	volume       = 959,
	number       = 1,
	pages        = 39,
	doi          = {10.3847/1538-4357/ad029e},
	eid          = 39,
	archiveprefix = {arXiv},
	eprint       = {2303.11946},
	primaryclass = {astro-ph.GA},
	adsurl       = {https://ui.adsabs.harvard.edu/abs/2023ApJ...959...39H}
}

@article{greene2024,
	title        = {{UNCOVER Spectroscopy Confirms the Surprising Ubiquity of Active Galactic Nuclei in Red Sources at z > 5}},
	author       = {{Greene}, Jenny E. and {Labbe}, Ivo and {Goulding}, Andy D. and {Furtak}, Lukas J. and {Chemerynska}, Iryna and {Kokorev}, Vasily and {Dayal}, Pratika and {Volonteri}, Marta and {Williams}, Christina C. and {Wang}, Bingjie and {Setton}, David J. and {Burgasser}, Adam J. and {Bezanson}, Rachel and {Atek}, Hakim and {Brammer}, Gabriel and {Cutler}, Sam E. and {Feldmann}, Robert and {Fujimoto}, Seiji and {Glazebrook}, Karl and {de Graaff}, Anna and {Khullar}, Gourav and {Leja}, Joel and {Marchesini}, Danilo and {Maseda}, Michael V. and {Matthee}, Jorryt and {Miller}, Tim B. and {Naidu}, Rohan P. and {Nanayakkara}, Themiya and {Oesch}, Pascal A. and {Pan}, Richard and {Papovich}, Casey and {Price}, Sedona H. and {van Dokkum}, Pieter and {Weaver}, John R. and {Whitaker}, Katherine E. and {Zitrin}, Adi},
	year         = 2024,
	month        = {mar},
	journal      = {\apj},
	volume       = 964,
	number       = 1,
	pages        = 39,
	doi          = {10.3847/1538-4357/ad1e5f},
	eid          = 39,
	archiveprefix = {arXiv},
	eprint       = {2309.05714},
	primaryclass = {astro-ph.GA},
	adsurl       = {https://ui.adsabs.harvard.edu/abs/2024ApJ...964...39G}
}

@article{kokorev2024a,
	title        = {{A Census of Photometrically Selected Little Red Dots at 4 < z < 9 in JWST Blank Fields}},
	author       = {{Kokorev}, Vasily and {Caputi}, Karina I. and {Greene}, Jenny E. and {Dayal}, Pratika and {Trebitsch}, Maxime and {Cutler}, Sam E. and {Fujimoto}, Seiji and {Labb{\'e}}, Ivo and {Miller}, Tim B. and {Iani}, Edoardo and {Navarro-Carrera}, Rafael and {Rinaldi}, Pierluigi},
	year         = 2024,
	month        = {jun},
	journal      = {\apj},
	volume       = 968,
	number       = 1,
	pages        = 38,
	doi          = {10.3847/1538-4357/ad4265},
	eid          = 38,
	archiveprefix = {arXiv},
	eprint       = {2401.09981},
	primaryclass = {astro-ph.GA},
	adsurl       = {https://ui.adsabs.harvard.edu/abs/2024ApJ...968...38K}
}

@article{matthee2024a,
	title        = {{Little Red Dots: An Abundant Population of Faint Active Galactic Nuclei at z {\ensuremath{\sim}} 5 Revealed by the EIGER and FRESCO JWST Surveys}},
	author       = {{Matthee}, Jorryt and {Naidu}, Rohan P. and {Brammer}, Gabriel and {Chisholm}, John and {Eilers}, Anna-Christina and {Goulding}, Andy and {Greene}, Jenny and {Kashino}, Daichi and {Labbe}, Ivo and {Lilly}, Simon J. and {Mackenzie}, Ruari and {Oesch}, Pascal A. and {Weibel}, Andrea and {Wuyts}, Stijn and {Xiao}, Mengyuan and {Bordoloi}, Rongmon and {Bouwens}, Rychard and {van Dokkum}, Pieter and {Illingworth}, Garth and {Kramarenko}, Ivan and {Maseda}, Michael V. and {Mason}, Charlotte and {Meyer}, Romain A. and {Nelson}, Erica J. and {Reddy}, Naveen A. and {Shivaei}, Irene and {Simcoe}, Robert A. and {Yue}, Minghao},
	year         = 2024,
	month        = {mar},
	journal      = {\apj},
	volume       = 963,
	number       = 2,
	pages        = 129,
	doi          = {10.3847/1538-4357/ad2345},
	eid          = 129,
	archiveprefix = {arXiv},
	eprint       = {2306.05448},
	primaryclass = {astro-ph.GA},
	adsurl       = {https://ui.adsabs.harvard.edu/abs/2024ApJ...963..129M}
}

@article{naidu2024,
	title        = {{All the Little Things in Abell 2744: $>$1000 Gravitationally Lensed Dwarf Galaxies at $z=0-9$ from JWST NIRCam Grism Spectroscopy}},
	author       = {{Naidu}, Rohan P. and {Matthee}, Jorryt and {Kramarenko}, Ivan and {Weibel}, Andrea and {Brammer}, Gabriel and {Oesch}, Pascal A. and {Lechner}, Peter and {Furtak}, Lukas J. and {Di Cesare}, Claudia and {Torralba}, Alberto and {Kotiwale}, Gauri and {Bezanson}, Rachel and {Bouwens}, Rychard J. and {Chandra}, Vedant and {Claeyssens}, Ad{\'e}la{\"\i}de and {Danhaive}, A. Lola and {Frebel}, Anna and {de Graaff}, Anna and {Greene}, Jenny E. and {Heintz}, Kasper E. and {Ji}, Alexander P. and {Kashino}, Daichi and {Katz}, Harley and {Labbe}, Ivo and {Leja}, Joel and {Li}, Yijia and {Maseda}, Michael V. and {Richard}, Johan and {Shivaei}, Irene and {Simcoe}, Robert A. and {Sobral}, David and {Suess}, Katherine A. and {Tacchella}, Sandro and {Williams}, Christina C.},
	year         = 2024,
	month        = {oct},
	journal      = {arXiv e-prints},
	pages        = {arXiv:2410.01874},
	doi          = {10.48550/arXiv.2410.01874},
	collaboration = {Naidu \& Matthee et al.},
	eid          = {arXiv:2410.01874},
	archiveprefix = {arXiv},
	eprint       = {2410.01874},
	primaryclass = {astro-ph.GA},
	adsurl       = {https://ui.adsabs.harvard.edu/abs/2024arXiv241001874N}
}

@inproceedings{bacon2010,
	title        = {{The MUSE second-generation VLT instrument}},
	author       = {{Bacon}, R. and {Accardo}, M. and {Adjali}, L. and {Anwand}, H. and {Bauer}, S. and {Biswas}, I. and {Blaizot}, J. and {Boudon}, D. and {Brau-Nogue}, S. and {Brinchmann}, J. and {Caillier}, P. and {Capoani}, L. and {Carollo}, C.~M. and {Contini}, T. and {Couderc}, P. and {Daguis{\'e}}, E. and {Deiries}, S. and {Delabre}, B. and {Dreizler}, S. and {Dubois}, J. and {Dupieux}, M. and {Dupuy}, C. and {Emsellem}, E. and {Fechner}, T. and {Fleischmann}, A. and {Fran{\c{c}}ois}, M. and {Gallou}, G. and {Gharsa}, T. and {Glindemann}, A. and {Gojak}, D. and {Guiderdoni}, B. and {Hansali}, G. and {Hahn}, T. and {Jarno}, A. and {Kelz}, A. and {Koehler}, C. and {Kosmalski}, J. and {Laurent}, F. and {Le Floch}, M. and {Lilly}, S.~J. and {Lizon}, J. -L. and {Loupias}, M. and {Manescau}, A. and {Monstein}, C. and {Nicklas}, H. and {Olaya}, J. -C. and {Pares}, L. and {Pasquini}, L. and {P{\'e}contal-Rousset}, A. and {Pell{\'o}}, R. and {Petit}, C. and {Popow}, E. and {Reiss}, R. and {Remillieux}, A. and {Renault}, E. and {Roth}, M. and {Rupprecht}, G. and {Serre}, D. and {Schaye}, J. and {Soucail}, G. and {Steinmetz}, M. and {Streicher}, O. and {Stuik}, R. and {Valentin},, H. and {Vernet}, J. and {Weilbacher}, P. and {Wisotzki}, L. and {Yerle}, N.},
	year         = 2010,
	month        = {jul},
	booktitle    = {Ground-based and Airborne Instrumentation for Astronomy III},
	series       = {Society of Photo-Optical Instrumentation Engineers (SPIE) Conference Series},
	volume       = 7735,
	pages        = 773508,
	doi          = {10.1117/12.856027},
	editor       = {{McLean}, Ian S. and {Ramsay}, Suzanne K. and {Takami}, Hideki},
	eid          = 773508,
	archiveprefix = {arXiv},
	eprint       = {2211.16795},
	primaryclass = {astro-ph.IM},
	adsurl       = {https://ui.adsabs.harvard.edu/abs/2010SPIE.7735E..08B}
}

@article{abell1989,
	title        = {{A Catalog of Rich Clusters of Galaxies}},
	author       = {{Abell}, George O. and {Corwin}, Jr., Harold G. and {Olowin}, Ronald P.},
	year         = 1989,
	month        = {may},
	journal      = {\apjs},
	volume       = 70,
	pages        = 1,
	doi          = {10.1086/191333},
	adsurl       = {https://ui.adsabs.harvard.edu/abs/1989ApJS...70....1A}
}

@ARTICLE{marino2019,
       author = {{Marino}, Raffaella Anna and {Cantalupo}, Sebastiano and {Pezzulli}, Gabriele and {Lilly}, Simon J. and {Gallego}, Sofia and {Mackenzie}, Ruari and {Matthee}, Jorryt and {Brinchmann}, Jarle and {Bouch{\'e}}, Nicolas and {Feltre}, Anna and {Muzahid}, Sowgat and {Schroetter}, Ilane and {Johnson}, Sean D. and {Nanayakkara}, Themiya},
        title = "{A Giant Ly{\ensuremath{\alpha}} Nebula and a Small-scale Clumpy Outflow in the System of the Exotic Quasar J0952+0114 Unveiled by MUSE}",
      journal = {\apj},
         year = 2019,
        month = jul,
       volume = {880},
       number = {1},
          eid = {47},
        pages = {47},
          doi = {10.3847/1538-4357/ab2881},
archivePrefix = {arXiv},
       eprint = {1906.06347},
 primaryClass = {astro-ph.GA},
       adsurl = {https://ui.adsabs.harvard.edu/abs/2019ApJ...880...47M}
}

@ARTICLE{langen2023,
       author = {{Langen}, Vivienne and {Cantalupo}, Sebastiano and {Steidel}, Charles C. and {Chen}, Yuguang and {Pezzulli}, Gabriele and {Gallego}, Sofia G.},
        title = "{Characterizing the circumgalactic medium of quasars at z   2.2 through H {\ensuremath{\alpha}} and Ly {\ensuremath{\alpha}} emission}",
      journal = {\mnras},
         year = 2023,
        month = mar,
       volume = {519},
       number = {4},
        pages = {5099-5113},
          doi = {10.1093/mnras/stac3205},
archivePrefix = {arXiv},
       eprint = {2303.05531},
 primaryClass = {astro-ph.GA},
       adsurl = {https://ui.adsabs.harvard.edu/abs/2023MNRAS.519.5099L}
}

@ARTICLE{scarlata2009,
       author = {{Scarlata}, C. and {Colbert}, J. and {Teplitz}, H.~I. and {Panagia}, N. and {Hayes}, M. and {Siana}, B. and {Rau}, A. and {Francis}, P. and {Caon}, A. and {Pizzella}, A. and {Bridge}, C.},
        title = "{The Effect of Dust Geometry on the Ly{\ensuremath{\alpha}} Output of Galaxies}",
      journal = {\apjl},
         year = 2009,
        month = oct,
       volume = {704},
       number = {2},
        pages = {L98-L102},
          doi = {10.1088/0004-637X/704/2/L98},
archivePrefix = {arXiv},
       eprint = {0909.3847},
 primaryClass = {astro-ph.CO},
       adsurl = {https://ui.adsabs.harvard.edu/abs/2009ApJ...704L..98S}
}

@article{bezanson2024,
	title        = {{The JWST UNCOVER Treasury Survey: Ultradeep NIRSpec and NIRCam Observations before the Epoch of Reionization}},
	author       = {{Bezanson}, Rachel and {Labbe}, Ivo and {Whitaker}, Katherine E. and {Leja}, Joel and {Price}, Sedona H. and {Franx}, Marijn and {Brammer}, Gabriel and {Marchesini}, Danilo and {Zitrin}, Adi and {Wang}, Bingjie and {Weaver}, John R. and {Furtak}, Lukas J. and {Atek}, Hakim and {Coe}, Dan and {Cutler}, Sam E. and {Dayal}, Pratika and {van Dokkum}, Pieter and {Feldmann}, Robert and {F{\"o}rster Schreiber}, Natascha M. and {Fujimoto}, Seiji and {Geha}, Marla and {Glazebrook}, Karl and {de Graaff}, Anna and {Greene}, Jenny E. and {Juneau}, St{\'e}phanie and {Kassin}, Susan and {Kriek}, Mariska and {Khullar}, Gourav and {Maseda}, Michael and {Mowla}, Lamiya A. and {Muzzin}, Adam and {Nanayakkara}, Themiya and {Nelson}, Erica J. and {Oesch}, Pascal A. and {Pacifici}, Camilla and {Pan}, Richard and {Papovich}, Casey and {Setton}, David J. and {Shapley}, Alice E. and {Smit}, Renske and {Stefanon}, Mauro and {Taylor}, Edward N. and {Williams}, Christina C.},
	year         = 2024,
	month        = {oct},
	journal      = {\apj},
	volume       = 974,
	number       = 1,
	pages        = 92,
	doi          = {10.3847/1538-4357/ad66cf},
	eid          = 92,
	archiveprefix = {arXiv},
	eprint       = {2212.04026},
	primaryclass = {astro-ph.GA},
	adsurl       = {https://ui.adsabs.harvard.edu/abs/2024ApJ...974...92B}
}

@article{labbe2025,
	title        = {{UNCOVER: Candidate Red Active Galactic Nuclei at 3 < z < 7 with JWST and ALMA}},
	author       = {{Labbe}, Ivo and {Greene}, Jenny E. and {Bezanson}, Rachel and {Fujimoto}, Seiji and {Furtak}, Lukas J. and {Goulding}, Andy D. and {Matthee}, Jorryt and {Naidu}, Rohan P. and {Oesch}, Pascal A. and {Atek}, Hakim and {Brammer}, Gabriel and {Chemerynska}, Iryna and {Coe}, Dan and {Cutler}, Sam E. and {Dayal}, Pratika and {Feldmann}, Robert and {Franx}, Marijn and {Glazebrook}, Karl and {Leja}, Joel and {Maseda}, Michael and {Marchesini}, Danilo and {Nanayakkara}, Themiya and {Nelson}, Erica J. and {Pan}, Richard and {Papovich}, Casey and {Price}, Sedona H. and {Suess}, Katherine A. and {Wang}, Bingjie and {Weaver}, John R. and {Whitaker}, Katherine E. and {Williams}, Christina C. and {Zitrin}, Adi},
	year         = 2025,
	month        = {jan},
	journal      = {\apj},
	volume       = 978,
	number       = 1,
	pages        = 92,
	doi          = {10.3847/1538-4357/ad3551},
	eid          = 92,
	archiveprefix = {arXiv},
	eprint       = {2306.07320},
	primaryclass = {astro-ph.GA},
	adsurl       = {https://ui.adsabs.harvard.edu/abs/2025ApJ...978...92L}
}

@article{horne1986,
	title        = {{An optimal extraction algorithm for CCD spectroscopy.}},
	author       = {{Horne}, K.},
	year         = 1986,
	month        = {jun},
	journal      = {\pasp},
	volume       = 98,
	pages        = {609--617},
	doi          = {10.1086/131801},
	adsurl       = {https://ui.adsabs.harvard.edu/abs/1986PASP...98..609H}
}

@article{matthee2020,
	title        = {{The nature of CR7 revealed with MUSE: a young starburst powering extended Ly {\ensuremath{\alpha}} emission at z = 6.6}},
	author       = {{Matthee}, Jorryt and {Pezzulli}, Gabriele and {Mackenzie}, Ruari and {Cantalupo}, Sebastiano and {Kusakabe}, Haruka and {Leclercq}, Floriane and {Sobral}, David and {Richard}, Johan and {Wisotzki}, Lutz and {Lilly}, Simon and {Boogaard}, Leindert and {Marino}, Raffaella and {Maseda}, Michael and {Nanayakkara}, Themiya},
	year         = 2020,
	month        = {oct},
	journal      = {\mnras},
	volume       = 498,
	number       = 2,
	pages        = {3043--3059},
	doi          = {10.1093/mnras/staa2550},
	archiveprefix = {arXiv},
	eprint       = {2008.01731},
	primaryclass = {astro-ph.GA},
	adsurl       = {https://ui.adsabs.harvard.edu/abs/2020MNRAS.498.3043M}
}

@article{matthee2024b,
       author = {{Matthee}, Jorryt and {Naidu}, Rohan P. and {Kotiwale}, Gauri and {Furtak}, Lukas J. and {Kramarenko}, Ivan and {Mackenzie}, Ruari and {Greene}, Jenny and {Adamo}, Angela and {Bouwens}, Rychard J. and {Di Cesare}, Claudia and {Eilers}, Anna-Christina and {de Graaff}, Anna and {Heintz}, Kasper E. and {Kashino}, Daichi and {Maseda}, Michael V. and {Tacchella}, Sandro and {Torralba}, Alberto},
        title = "{Environmental Evidence for Overly Massive Black Holes in Low-mass Galaxies and a Black Hole{\textendash}Halo Mass Relation at z {\ensuremath{\sim}} 5}",
      journal = {\apj},
         year = 2025,
        month = aug,
       volume = {988},
       number = {2},
          eid = {246},
        pages = {246},
          doi = {10.3847/1538-4357/ade886},
archivePrefix = {arXiv},
       eprint = {2412.02846},
 primaryClass = {astro-ph.GA},
       adsurl = {https://ui.adsabs.harvard.edu/abs/2025ApJ...988..246M}
}

@article{tornotti2025,
	title        = {{The MUSE Ultra Deep Field: A 5 Mpc Stretch of the z {\ensuremath{\approx}} 4 Cosmic Web Revealed in Emission}},
	author       = {{Tornotti}, Davide and {Fumagalli}, Michele and {Fossati}, Matteo and {Arrigoni Battaia}, Fabrizio and {Benitez-Llambay}, Alejandro and {Dayal}, Pratika and {Dutta}, Rajeshwari and {Peroux}, Celine and {Rafelski}, Marc and {Revalski}, Mitchell},
	year         = 2025,
	month        = {feb},
	journal      = {\apjl},
	volume       = 980,
	number       = 2,
	pages        = {L43},
	doi          = {10.3847/2041-8213/adb0ba},
	eid          = {L43},
	archiveprefix = {arXiv},
	eprint       = {2412.06895},
	primaryclass = {astro-ph.GA},
	adsurl       = {https://ui.adsabs.harvard.edu/abs/2025ApJ...980L..43T}
}

@article{vandenberk2001,
	title        = {{Composite Quasar Spectra from the Sloan Digital Sky Survey}},
	author       = {{Vanden Berk}, Daniel E. and {Richards}, Gordon T. and {Bauer}, Amanda and {Strauss}, Michael A. and {Schneider}, Donald P. and {Heckman}, Timothy M. and {York}, Donald G. and {Hall}, Patrick B. and {Fan}, Xiaohui and {Knapp}, G.~R. and {Anderson}, Scott F. and {Annis}, James and {Bahcall}, Neta A. and {Bernardi}, Mariangela and {Briggs}, John W. and {Brinkmann}, J. and {Brunner}, Robert and {Burles}, Scott and {Carey}, Larry and {Castander}, Francisco J. and {Connolly}, A.~J. and {Crocker}, J.~H. and {Csabai}, Istv{\'a}n and {Doi}, Mamoru and {Finkbeiner}, Douglas and {Friedman}, Scott and {Frieman}, Joshua A. and {Fukugita}, Masataka and {Gunn}, James E. and {Hennessy}, G.~S. and {Ivezi{\'c}}, {\v{Z}}eljko and {Kent}, Stephen and {Kunszt}, Peter Z. and {Lamb}, D.~Q. and {Leger}, R. French and {Long}, Daniel C. and {Loveday}, Jon and {Lupton}, Robert H. and {Meiksin}, Avery and {Merelli}, Aronne and {Munn}, Jeffrey A. and {Newberg}, Heidi Jo and {Newcomb}, Matt and {Nichol}, R.~C. and {Owen}, Russell and {Pier}, Jeffrey R. and {Pope}, Adrian and {Rockosi}, Constance M. and {Schlegel}, David J. and {Siegmund}, Walter A. and {Smee}, Stephen and {Snir}, Yehuda and {Stoughton}, Chris and {Stubbs}, Christopher and {SubbaRao}, Mark and {Szalay}, Alexander S. and {Szokoly}, Gyula P. and {Tremonti}, Christy and {Uomoto}, Alan and {Waddell}, Patrick and {Yanny}, Brian and {Zheng}, Wei},
	year         = 2001,
	month        = {aug},
	journal      = {\aj},
	volume       = 122,
	number       = 2,
	pages        = {549--564},
	doi          = {10.1086/321167},
	archiveprefix = {arXiv},
	eprint       = {astro-ph/0105231},
	primaryclass = {astro-ph},
	adsurl       = {https://ui.adsabs.harvard.edu/abs/2001AJ....122..549V}
}

@article{kewley2019,
	title        = {Theoretical {ISM} {Pressure} and {Electron} {Density} {Diagnostics} for {Local} and {High}-redshift {Galaxies}},
	author       = {Kewley, Lisa J. and Nicholls, David C. and Sutherland, Ralph and Rigby, Jane R. and Acharya, Ayan and Dopita, Michael A. and Bayliss, Matthew B.},
	year         = 2019,
	month        = {jul},
	journal      = {The Astrophysical Journal},
	volume       = 880,
	number       = 1,
	pages        = 16,
	doi          = {10.3847/1538-4357/ab16ed},
	issn         = {0004-637X, 1538-4357},
	url          = {https://iopscience.iop.org/article/10.3847/1538-4357/ab16ed},
	urldate      = {2025-03-15},
	language     = {en}
}

@article{maiolino2024b,
       author = {{Maiolino}, Roberto and {Scholtz}, Jan and {Witstok}, Joris and {Carniani}, Stefano and {D'Eugenio}, Francesco and {de Graaff}, Anna and {{\"U}bler}, Hannah and {Tacchella}, Sandro and {Curtis-Lake}, Emma and {Arribas}, Santiago and {Bunker}, Andrew and {Charlot}, St{\'e}phane and {Chevallard}, Jacopo and {Curti}, Mirko and {Looser}, Tobias J. and {Maseda}, Michael V. and {Rawle}, Timothy D. and {Rodr{\'\i}guez del Pino}, Bruno and {Willott}, Chris J. and {Egami}, Eiichi and {Eisenstein}, Daniel J. and {Hainline}, Kevin N. and {Robertson}, Brant and {Williams}, Christina C. and {Willmer}, Christopher N.~A. and {Baker}, William M. and {Boyett}, Kristan and {DeCoursey}, Christa and {Fabian}, Andrew C. and {Helton}, Jakob M. and {Ji}, Zhiyuan and {Jones}, Gareth C. and {Kumari}, Nimisha and {Laporte}, Nicolas and {Nelson}, Erica J. and {Perna}, Michele and {Sandles}, Lester and {Shivaei}, Irene and {Sun}, Fengwu},
        title = "{A small and vigorous black hole in the early Universe}",
      journal = {\nat},
         year = 2024,
        month = mar,
       volume = {627},
       number = {8002},
        pages = {59-63},
          doi = {10.1038/s41586-024-07052-5},
archivePrefix = {arXiv},
       eprint = {2305.12492},
 primaryClass = {astro-ph.GA},
       adsurl = {https://ui.adsabs.harvard.edu/abs/2024Natur.627...59M}
}

@article{marques-chaves2024,
	title        = {Extreme {N}-emitters at high redshift: {Possible} signatures of supermassive stars and globular cluster or black hole formation in action},
	shorttitle   = {Extreme {N}-emitters at high redshift},
	author       = {Marques-Chaves, R. and Schaerer, D. and Kuruvanthodi, A. and Korber, D. and Prantzos, N. and Charbonnel, C. and Weibel, A. and Izotov, Y. I. and Messa, M. and Brammer, G. and Dessauges-Zavadsky, M. and Oesch, P.},
	year         = 2024,
	month        = {jan},
	journal      = {Astronomy \& Astrophysics},
	volume       = 681,
	pages        = {A30},
	doi          = {10.1051/0004-6361/202347411},
	issn         = {0004-6361, 1432-0746},
	url          = {https://www.aanda.org/10.1051/0004-6361/202347411},
	urldate      = {2025-03-15},
	copyright    = {https://creativecommons.org/licenses/by/4.0},
	language     = {en}
}

@article{erwin2015,
	title        = {{IMFIT: A Fast, Flexible New Program for Astronomical Image Fitting}},
	author       = {{Erwin}, Peter},
	year         = 2015,
	month        = {feb},
	journal      = {\apj},
	volume       = 799,
	number       = 2,
	pages        = 226,
	doi          = {10.1088/0004-637X/799/2/226},
	eid          = 226,
	archiveprefix = {arXiv},
	eprint       = {1408.1097},
	primaryclass = {astro-ph.IM},
	adsurl       = {https://ui.adsabs.harvard.edu/abs/2015ApJ...799..226E}
}

@article{mackenzie2021,
	title        = {Revealing the impact of quasar luminosity on giant {Ly} α nebulae},
	author       = {Mackenzie, Ruari and Pezzulli, Gabriele and Cantalupo, Sebastiano and Marino, Raffaella A. and Lilly, Simon and Muzahid, Sowgat and Matthee, Jorryt and Schaye, Joop and Wisotzki, Lutz},
	year         = 2021,
	month        = {mar},
	journal      = {Monthly Notices of the Royal Astronomical Society},
	volume       = 502,
	number       = 1,
	pages        = {494--509},
	doi          = {10.1093/mnras/staa3277},
	issn         = {0035-8711},
	url          = {https://ui.adsabs.harvard.edu/abs/2021MNRAS.502..494M/abstract},
	urldate      = {2025-02-07},
	language     = {en}
}

@article{borisova2016,
	title        = {{Ubiquitous Giant Ly{\ensuremath{\alpha}} Nebulae around the Brightest Quasars at z {\ensuremath{\sim}} 3.5 Revealed with MUSE}},
	author       = {{Borisova}, Elena and {Cantalupo}, Sebastiano and {Lilly}, Simon J. and {Marino}, Raffaella A. and {Gallego}, Sofia G. and {Bacon}, Roland and {Blaizot}, Jeremy and {Bouch{\'e}}, Nicolas and {Brinchmann}, Jarle and {Carollo}, C. Marcella and {Caruana}, Joseph and {Finley}, Hayley and {Herenz}, Edmund C. and {Richard}, Johan and {Schaye}, Joop and {Straka}, Lorrie A. and {Turner}, Monica L. and {Urrutia}, Tanya and {Verhamme}, Anne and {Wisotzki}, Lutz},
	year         = 2016,
	month        = {nov},
	journal      = {\apj},
	volume       = 831,
	number       = 1,
	pages        = 39,
	doi          = {10.3847/0004-637X/831/1/39},
	eid          = 39,
	archiveprefix = {arXiv},
	eprint       = {1605.01422},
	primaryclass = {astro-ph.GA},
	adsurl       = {https://ui.adsabs.harvard.edu/abs/2016ApJ...831...39B}
}

@article{arrigonibattaia2019,
	title        = {{QSO MUSEUM I: a sample of 61 extended Ly {\ensuremath{\alpha}}-emission nebulae surrounding z {\ensuremath{\sim}} 3 quasars}},
	author       = {{Arrigoni Battaia}, Fabrizio and {Hennawi}, Joseph F. and {Prochaska}, J. Xavier and {O{\~n}orbe}, Jose and {Farina}, Emanuele P. and {Cantalupo}, Sebastiano and {Lusso}, Elisabeta},
	year         = 2019,
	month        = {jan},
	journal      = {\mnras},
	volume       = 482,
	number       = 3,
	pages        = {3162--3205},
	doi          = {10.1093/mnras/sty2827},
	archiveprefix = {arXiv},
	eprint       = {1808.10857},
	primaryclass = {astro-ph.GA},
	adsurl       = {https://ui.adsabs.harvard.edu/abs/2019MNRAS.482.3162A}
}

@article{naidu2025,
	title        = {{A ``Black Hole Star'' Reveals the Remarkable Gas-Enshrouded Hearts of the Little Red Dots}},
	author       = {{Naidu}, Rohan P. and {Matthee}, Jorryt and {Katz}, Harley and {de Graaff}, Anna and {Oesch}, Pascal and {Smith}, Aaron and {Greene}, Jenny E. and {Brammer}, Gabriel and {Weibel}, Andrea and {Hviding}, Raphael and {Chisholm}, John and {Labb\textbackslash'e}, Ivo and {Simcoe}, Robert A. and {Witten}, Callum and {Atek}, Hakim and {Baggen}, Josephine F.~W. and {Belli}, Sirio and {Bezanson}, Rachel and {Boogaard}, Leindert A. and {Bose}, Sownak and {Covelo-Paz}, Alba and {Dayal}, Pratika and {Fudamoto}, Yoshinobu and {Furtak}, Lukas J. and {Giovinazzo}, Emma and {Goulding}, Andy and {Gronke}, Max and {Heintz}, Kasper E. and {Hirschmann}, Michaela and {Illingworth}, Garth and {Inoue}, Akio K. and {Johnson}, Benjamin D. and {Leja}, Joel and {Leonova}, Ecaterina and {McConachie}, Ian and {Maseda}, Michael V. and {Natarajan}, Priyamvada and {Nelson}, Erica and {Setton}, David J. and {Shivaei}, Irene and {Sobral}, David and {Stefanon}, Mauro and {Tacchella}, Sandro and {Toft}, Sune and {Torralba}, Alberto and {van Dokkum}, Pieter and {van der Wel}, Arjen and {Volonteri}, Marta and {Walter}, Fabian and {Wang}, Bingjie and {Watson}, Darach},
	year         = 2025,
	month        = {mar},
	journal      = {arXiv e-prints},
	pages        = {arXiv:2503.16596},
	doi          = {10.48550/arXiv.2503.16596},
	eid          = {arXiv:2503.16596},
	archiveprefix = {arXiv},
	eprint       = {2503.16596},
	primaryclass = {astro-ph.GA},
	adsurl       = {https://ui.adsabs.harvard.edu/abs/2025arXiv250316596N}
}

@article{graaff2025,
       author = {{de Graaff}, Anna and {Rix}, Hans-Walter and {Naidu}, Rohan P. and {Labb{\'e}}, Ivo and {Wang}, Bingjie and {Leja}, Joel and {Matthee}, Jorryt and {Katz}, Harley and {Greene}, Jenny E. and {Hviding}, Raphael E. and {Baggen}, Josephine and {Bezanson}, Rachel and {Boogaard}, Leindert A. and {Brammer}, Gabriel and {Dayal}, Pratika and {van Dokkum}, Pieter and {Goulding}, Andy D. and {Hirschmann}, Michaela and {Maseda}, Michael V. and {McConachie}, Ian and {Miller}, Tim B. and {Nelson}, Erica and {Oesch}, Pascal A. and {Setton}, David J. and {Shivaei}, Irene and {Weibel}, Andrea and {Whitaker}, Katherine E. and {Williams}, Christina C.},
        title = "{A remarkable ruby: Absorption in dense gas, rather than evolved stars, drives the extreme Balmer break of a little red dot at z = 3.5}",
      journal = {\aap},
         year = 2025,
        month = sep,
       volume = {701},
          eid = {A168},
        pages = {A168},
          doi = {10.1051/0004-6361/202554681},
archivePrefix = {arXiv},
       eprint = {2503.16600},
 primaryClass = {astro-ph.GA},
       adsurl = {https://ui.adsabs.harvard.edu/abs/2025A&A...701A.168D}
}

@article{rusakov2025,
	title        = {{JWST's little red dots: an emerging population of young, low-mass AGN cocooned in dense ionized gas}},
	author       = {{Rusakov}, V. and {Watson}, D. and {Nikopoulos}, G.~P. and {Brammer}, G. and {Gottumukkala}, R. and {Harvey}, T. and {Heintz}, K.~E. and {Nielsen}, R.~D. and {Sim}, S.~A. and {Sneppen}, A. and {Vijayan}, A.~P. and {Adams}, N. and {Austin}, D. and {Conselice}, C.~J. and {Goolsby}, C.~M. and {Toft}, S.},
	year         = 2025,
	month        = {mar},
	journal      = {arXiv e-prints},
	pages        = {arXiv:2503.16595},
	doi          = {10.48550/arXiv.2503.16595},
	eid          = {arXiv:2503.16595},
	archiveprefix = {arXiv},
	eprint       = {2503.16595},
	primaryclass = {astro-ph.GA},
	adsurl       = {https://ui.adsabs.harvard.edu/abs/2025arXiv250316595R}
}

@article{ji2025,
	title        = {{BlackTHUNDER -- A non-stellar Balmer break in a black hole-dominated little red dot at $z=7.04$}},
	author       = {{Ji}, Xihan and {Maiolino}, Roberto and {{\"U}bler}, Hannah and {Scholtz}, Jan and {D'Eugenio}, Francesco and {Sun}, Fengwu and {Perna}, Michele and {Turner}, Hannah and {Arribas}, Santiago and {Bennett}, Jake S. and {Bunker}, Andrew and {Carniani}, Stefano and {Charlot}, St{\'e}phane and {Cresci}, Giovanni and {Curti}, Mirko and {Egami}, Eiichi and {Fabian}, Andy and {Inayoshi}, Kohei and {Isobe}, Yuki and {Jones}, Gareth and {Juod{\v{z}}balis}, Ignas and {Kumari}, Nimisha and {Lyu}, Jianwei and {Mazzolari}, Giovanni and {Parlanti}, Eleonora and {Robertson}, Brant and {Rodr{\'\i}guez Del Pino}, Bruno and {Schneider}, Raffaella and {Sijacki}, Debora and {Tacchella}, Sandro and {Trinca}, Alessandro and {Valiante}, Rosa and {Venturi}, Giacomo and {Volonteri}, Marta and {Willott}, Chris and {Witten}, Callum and {Witstok}, Joris},
	year         = 2025,
	month        = {jan},
	journal      = {arXiv e-prints},
	pages        = {arXiv:2501.13082},
	doi          = {10.48550/arXiv.2501.13082},
	eid          = {arXiv:2501.13082},
	archiveprefix = {arXiv},
	eprint       = {2501.13082},
	primaryclass = {astro-ph.GA},
	adsurl       = {https://ui.adsabs.harvard.edu/abs/2025arXiv250113082J}
}

@article{sobral2019,
       author = {{Sobral}, David and {Matthee}, Jorryt},
        title = "{Predicting Ly{\ensuremath{\alpha}} escape fractions with a simple observable. Ly{\ensuremath{\alpha}} in emission as an empirically calibrated star formation rate indicator}",
      journal = {\aap},
         year = 2019,
        month = mar,
       volume = {623},
          eid = {A157},
        pages = {A157},
          doi = {10.1051/0004-6361/201833075},
archivePrefix = {arXiv},
       eprint = {1803.08923},
 primaryClass = {astro-ph.GA},
       adsurl = {https://ui.adsabs.harvard.edu/abs/2019A&A...623A.157S}
}

@article{kennicutt2012,
	title        = {{Star Formation in the Milky Way and Nearby Galaxies}},
	author       = {{Kennicutt}, Robert C. and {Evans}, Neal J.},
	year         = 2012,
	month        = {sep},
	journal      = {\araa},
	volume       = 50,
	pages        = {531--608},
	doi          = {10.1146/annurev-astro-081811-125610},
	archiveprefix = {arXiv},
	eprint       = {1204.3552},
	primaryclass = {astro-ph.GA},
	adsurl       = {https://ui.adsabs.harvard.edu/abs/2012ARA&A..50..531K}
}

@article{leclercq2017,
	title        = {{The MUSE Hubble Ultra Deep Field Survey. VIII. Extended Lyman-{\ensuremath{\alpha}} haloes around high-z star-forming galaxies}},
	author       = {{Leclercq}, Floriane and {Bacon}, Roland and {Wisotzki}, Lutz and {Mitchell}, Peter and {Garel}, Thibault and {Verhamme}, Anne and {Blaizot}, J{\'e}r{\'e}my and {Hashimoto}, Takuya and {Herenz}, Edmund Christian and {Conseil}, Simon and {Cantalupo}, Sebastiano and {Inami}, Hanae and {Contini}, Thierry and {Richard}, Johan and {Maseda}, Michael and {Schaye}, Joop and {Marino}, Raffaella Anna and {Akhlaghi}, Mohammad and {Brinchmann}, Jarle and {Carollo}, Marcella},
	year         = 2017,
	month        = {dec},
	journal      = {\aap},
	volume       = 608,
	pages        = {A8},
	doi          = {10.1051/0004-6361/201731480},
	eid          = {A8},
	archiveprefix = {arXiv},
	eprint       = {1710.10271},
	primaryclass = {astro-ph.GA},
	adsurl       = {https://ui.adsabs.harvard.edu/abs/2017A&A...608A...8L}
}

@article{collaboration2020,
	title        = {{Planck 2018 results. VI. Cosmological parameters}},
	author       = {{Planck Collaboration} and {Aghanim}, N. and {Akrami}, Y. and {Ashdown}, M. and {Aumont}, J. and {Baccigalupi}, C. and {Ballardini}, M. and {Banday}, A.~J. and {Barreiro}, R.~B. and {Bartolo}, N. and {Basak}, S. and {Battye}, R. and {Benabed}, K. and {Bernard}, J. -P. and {Bersanelli}, M. and {Bielewicz}, P. and {Bock}, J.~J. and {Bond}, J.~R. and {Borrill}, J. and {Bouchet}, F.~R. and {Boulanger}, F. and {Bucher}, M. and {Burigana}, C. and {Butler}, R.~C. and {Calabrese}, E. and {Cardoso}, J. -F. and {Carron}, J. and {Challinor}, A. and {Chiang}, H.~C. and {Chluba}, J. and {Colombo}, L.~P.~L. and {Combet}, C. and {Contreras}, D. and {Crill}, B.~P. and {Cuttaia}, F. and {de Bernardis}, P. and {de Zotti}, G. and {Delabrouille}, J. and {Delouis}, J. -M. and {Di Valentino}, E. and {Diego}, J.~M. and {Dor{\'e}}, O. and {Douspis}, M. and {Ducout}, A. and {Dupac}, X. and {Dusini}, S. and {Efstathiou}, G. and {Elsner}, F. and {En{\ss}lin}, T.~A. and {Eriksen}, H.~K. and {Fantaye}, Y. and {Farhang}, M. and {Fergusson}, J. and {Fernandez-Cobos}, R. and {Finelli}, F. and {Forastieri}, F. and {Frailis}, M. and {Fraisse}, A.~A. and {Franceschi}, E. and {Frolov}, A. and {Galeotta}, S. and {Galli}, S. and {Ganga}, K. and {G{\'e}nova-Santos}, R.~T. and {Gerbino}, M. and {Ghosh}, T. and {Gonz{\'a}lez-Nuevo}, J. and {G{\'o}rski}, K.~M. and {Gratton}, S. and {Gruppuso}, A. and {Gudmundsson}, J.~E. and {Hamann}, J. and {Handley}, W. and {Hansen}, F.~K. and {Herranz}, D. and {Hildebrandt}, S.~R. and {Hivon}, E. and {Huang}, Z. and {Jaffe}, A.~H. and {Jones}, W.~C. and {Karakci}, A. and {Keih{\"a}nen}, E. and {Keskitalo}, R. and {Kiiveri}, K. and {Kim}, J. and {Kisner}, T.~S. and {Knox}, L. and {Krachmalnicoff}, N. and {Kunz}, M. and {Kurki-Suonio}, H. and {Lagache}, G. and {Lamarre}, J. -M. and {Lasenby}, A. and {Lattanzi}, M. and {Lawrence}, C.~R. and {Le Jeune}, M. and {Lemos}, P. and {Lesgourgues}, J. and {Levrier}, F. and {Lewis}, A. and {Liguori}, M. and {Lilje}, P.~B. and {Lilley}, M. and {Lindholm}, V. and {L{\'o}pez-Caniego}, M. and {Lubin}, P.~M. and {Ma}, Y. -Z. and {Mac{\'\i}as-P{\'e}rez}, J.~F. and {Maggio}, G. and {Maino}, D. and {Mandolesi}, N. and {Mangilli}, A. and {Marcos-Caballero}, A. and {Maris}, M. and {Martin}, P.~G. and {Martinelli}, M. and {Mart{\'\i}nez-Gonz{\'a}lez}, E. and {Matarrese}, S. and {Mauri}, N. and {McEwen}, J.~D. and {Meinhold}, P.~R. and {Melchiorri}, A. and {Mennella}, A. and {Migliaccio}, M. and {Millea}, M. and {Mitra}, S. and {Miville-Desch{\^e}nes}, M. -A. and {Molinari}, D. and {Montier}, L. and {Morgante}, G. and {Moss}, A. and {Natoli}, P. and {N{\o}rgaard-Nielsen}, H.~U. and {Pagano}, L. and {Paoletti}, D. and {Partridge}, B. and {Patanchon}, G. and {Peiris}, H.~V. and {Perrotta}, F. and {Pettorino}, V. and {Piacentini}, F. and {Polastri}, L. and {Polenta}, G. and {Puget}, J. -L. and {Rachen}, J.~P. and {Reinecke}, M. and {Remazeilles}, M. and {Renzi}, A. and {Rocha}, G. and {Rosset}, C. and {Roudier}, G. and {Rubi{\~n}o-Mart{\'\i}n}, J.~A. and {Ruiz-Granados}, B. and {Salvati}, L. and {Sandri}, M. and {Savelainen}, M. and {Scott}, D. and {Shellard}, E.~P.~S. and {Sirignano}, C. and {Sirri}, G. and {Spencer}, L.~D. and {Sunyaev}, R. and {Suur-Uski}, A. -S. and {Tauber}, J.~A. and {Tavagnacco}, D. and {Tenti}, M. and {Toffolatti}, L. and {Tomasi}, M. and {Trombetti}, T. and {Valenziano}, L. and {Valiviita}, J. and {Van Tent}, B. and {Vibert}, L. and {Vielva}, P. and {Villa}, F. and {Vittorio}, N. and {Wandelt}, B.~D. and {Wehus}, I.~K. and {White}, M. and {White}, S.~D.~M. and {Zacchei}, A. and {Zonca}, A.},
	year         = 2020,
	month        = {sep},
	journal      = {\aap},
	volume       = 641,
	pages        = {A6},
	doi          = {10.1051/0004-6361/201833910},
	eid          = {A6},
	archiveprefix = {arXiv},
	eprint       = {1807.06209},
	primaryclass = {astro-ph.CO},
	adsurl       = {https://ui.adsabs.harvard.edu/abs/2020A&A...641A...6P}
}

@article{oke1983,
	title        = {{Secondary standard stars for absolute spectrophotometry.}},
	author       = {{Oke}, J.~B. and {Gunn}, J.~E.},
	year         = 1983,
	month        = {mar},
	journal      = {\apj},
	volume       = 266,
	pages        = {713--717},
	doi          = {10.1086/160817},
	adsurl       = {https://ui.adsabs.harvard.edu/abs/1983ApJ...266..713O}
}

@article{verhamme2017,
	title        = {{Lyman-{\ensuremath{\alpha}} spectral properties of five newly discovered Lyman continuum emitters}},
	author       = {{Verhamme}, A. and {Orlitov{\'a}}, I. and {Schaerer}, D. and {Izotov}, Y. and {Worseck}, G. and {Thuan}, T.~X. and {Guseva}, N.},
	year         = 2017,
	month        = {jan},
	journal      = {\aap},
	volume       = 597,
	pages        = {A13},
	doi          = {10.1051/0004-6361/201629264},
	eid          = {A13},
	archiveprefix = {arXiv},
	eprint       = {1609.03477},
	primaryclass = {astro-ph.GA},
	adsurl       = {https://ui.adsabs.harvard.edu/abs/2017A&A...597A..13V}
}

@article{sobral2017,
	title        = {{The CALYMHA survey: Ly{\ensuremath{\alpha}} luminosity function and global escape fraction of Ly{\ensuremath{\alpha}} photons at z = 2.23}},
	author       = {{Sobral}, David and {Matthee}, Jorryt and {Best}, Philip and {Stroe}, Andra and {R{\"o}ttgering}, Huub and {Oteo}, Iv{\'a}n and {Smail}, Ian and {Morabito}, Leah and {Paulino-Afonso}, Ana},
	year         = 2017,
	month        = {apr},
	journal      = {\mnras},
	volume       = 466,
	number       = 1,
	pages        = {1242--1258},
	doi          = {10.1093/mnras/stw3090},
	archiveprefix = {arXiv},
	eprint       = {1609.05897},
	primaryclass = {astro-ph.GA},
	adsurl       = {https://ui.adsabs.harvard.edu/abs/2017MNRAS.466.1242S}
}

@article{richard2021,
	title        = {{An atlas of MUSE observations towards twelve massive lensing clusters}},
	author       = {{Richard}, Johan and {Claeyssens}, Ad{\'e}la{\"\i}de and {Lagattuta}, David and {Guaita}, Lucia and {Bauer}, Franz Erik and {Pello}, Roser and {Carton}, David and {Bacon}, Roland and {Soucail}, Genevi{\`e}ve and {Lyon}, Gonzalo Prieto and {Kneib}, Jean-Paul and {Mahler}, Guillaume and {Cl{\'e}ment}, Benjamin and {Mercier}, Wilfried and {Variu}, Andrei and {Tamone}, Am{\'e}lie and {Ebeling}, Harald and {Schmidt}, Kasper B. and {Nanayakkara}, Themiya and {Maseda}, Michael and {Weilbacher}, Peter M. and {Bouch{\'e}}, Nicolas and {Bouwens}, Rychard J. and {Wisotzki}, Lutz and {de la Vieuville}, Geoffroy and {Martinez}, Johany and {Patr{\'\i}cio}, Vera},
	year         = 2021,
	month        = {feb},
	journal      = {\aap},
	volume       = 646,
	pages        = {A83},
	doi          = {10.1051/0004-6361/202039462},
	eid          = {A83},
	archiveprefix = {arXiv},
	eprint       = {2009.09784},
	primaryclass = {astro-ph.GA},
	adsurl       = {https://ui.adsabs.harvard.edu/abs/2021A&A...646A..83R}
}

@article{kocevski2023,
       author = {{Kocevski}, Dale D. and {Onoue}, Masafusa and {Inayoshi}, Kohei and {Trump}, Jonathan R. and {Arrabal Haro}, Pablo and {Grazian}, Andrea and {Dickinson}, Mark and {Finkelstein}, Steven L. and {Kartaltepe}, Jeyhan S. and {Hirschmann}, Michaela and {Aird}, James and {Holwerda}, Benne W. and {Fujimoto}, Seiji and {Juneau}, St{\'e}phanie and {Amor{\'\i}n}, Ricardo O. and {Backhaus}, Bren E. and {Bagley}, Micaela B. and {Barro}, Guillermo and {Bell}, Eric F. and {Bisigello}, Laura and {Calabr{\`o}}, Antonello and {Cleri}, Nikko J. and {Cooper}, M.~C. and {Ding}, Xuheng and {Grogin}, Norman A. and {Ho}, Luis C. and {Hutchison}, Taylor A. and {Inoue}, Akio K. and {Jiang}, Linhua and {Jones}, Brenda and {Koekemoer}, Anton M. and {Li}, Wenxiu and {Li}, Zhengrong and {McGrath}, Elizabeth J. and {Molina}, Juan and {Papovich}, Casey and {P{\'e}rez-Gonz{\'a}lez}, Pablo G. and {Pirzkal}, Nor and {Wilkins}, Stephen M. and {Yang}, Guang and {Yung}, L.~Y. Aaron},
        title = "{Hidden Little Monsters: Spectroscopic Identification of Low-mass, Broad-line AGNs at z > 5 with CEERS}",
      journal = {\apjl},
         year = 2023,
        month = sep,
       volume = {954},
       number = {1},
          eid = {L4},
        pages = {L4},
          doi = {10.3847/2041-8213/ace5a0},
archivePrefix = {arXiv},
       eprint = {2302.00012},
 primaryClass = {astro-ph.GA},
       adsurl = {https://ui.adsabs.harvard.edu/abs/2023ApJ...954L...4K}
}

@article{barro2024,
	title        = {{A Comprehensive Photometric Selection of `Little Red Dots' in MIRI Fields: An IR-Bright LRD at $z=3.1386$ with Warm Dust Emission}},
	author       = {{Barro}, Guillermo and {Perez-Gonzalez}, Pablo G. and {Kocevski}, Dale D. and {McGrath}, Elizabeth J. and {Leung}, Gene C.~K. and {Cullen}, Fergus and {Dunlop}, James S. and {Ellis}, Richard S. and {Finkelstein}, Steven L. and {Grogin}, Norman A. and {Illingworth}, Garth and {Kartaltepe}, Jeyhan S. and {Koekemoer}, Anton M. and {Lucas}, Ray A. and {McLure}, Ross J. and {Yang}, Guang},
	year         = 2024,
	month        = {dec},
	journal      = {arXiv e-prints},
	pages        = {arXiv:2412.01887},
	doi          = {10.48550/arXiv.2412.01887},
	eid          = {arXiv:2412.01887},
	archiveprefix = {arXiv},
	eprint       = {2412.01887},
	primaryclass = {astro-ph.GA},
	adsurl       = {https://ui.adsabs.harvard.edu/abs/2024arXiv241201887B}
}

@article{akins2024,
       author = {{Akins}, Hollis B. and {Casey}, Caitlin M. and {Lambrides}, Erini and {Allen}, Natalie and {Andika}, Irham T. and {Brinch}, Malte and {Champagne}, Jaclyn B. and {Cooper}, Olivia and {Ding}, Xuheng and {Drakos}, Nicole E. and {Faisst}, Andreas and {Finkelstein}, Steven L. and {Franco}, Maximilien and {Fujimoto}, Seiji and {Gentile}, Fabrizio and {Gillman}, Steven and {Gozaliasl}, Ghassem and {Harish}, Santosh and {Hayward}, Christopher C. and {Hirschmann}, Michaela and {Ilbert}, Olivier and {Kartaltepe}, Jeyhan S. and {Kocevski}, Dale D. and {Koekemoer}, Anton M. and {Kokorev}, Vasily and {Liu}, Daizhong and {Long}, Arianna S. and {McCracken}, Henry Joy and {McKinney}, Jed and {Onoue}, Masafusa and {Paquereau}, Louise and {Renzini}, Alvio and {Rhodes}, Jason and {Robertson}, Brant E. and {Shuntov}, Marko and {Silverman}, John D. and {Tanaka}, Takumi S. and {Toft}, Sune and {Trakhtenbrot}, Benny and {Valentino}, Francesco and {Zavala}, Jorge},
        title = "{COSMOS-Web: The Overabundance and Physical Nature of ``Little Red Dots''{\textemdash}Implications for Early Galaxy and SMBH Assembly}",
      journal = {\apj},
         year = 2025,
        month = sep,
       volume = {991},
       number = {1},
          eid = {37},
        pages = {37},
          doi = {10.3847/1538-4357/ade984},
archivePrefix = {arXiv},
       eprint = {2406.10341},
 primaryClass = {astro-ph.GA},
       adsurl = {https://ui.adsabs.harvard.edu/abs/2025ApJ...991...37A}
}

@article{inayoshi2025,
	title        = {{Extremely Dense Gas around Little Red Dots and High-redshift Active Galactic Nuclei: A Nonstellar Origin of the Balmer Break and Absorption Features}},
	author       = {{Inayoshi}, Kohei and {Maiolino}, Roberto},
	year         = 2025,
	month        = {feb},
	journal      = {\apjl},
	volume       = 980,
	number       = 2,
	pages        = {L27},
	doi          = {10.3847/2041-8213/adaebd},
	eid          = {L27},
	archiveprefix = {arXiv},
	eprint       = {2409.07805},
	primaryclass = {astro-ph.GA},
	adsurl       = {https://ui.adsabs.harvard.edu/abs/2025ApJ...980L..27I}
}

@article{taylor2024,
       author = {{Taylor}, Anthony J. and {Finkelstein}, Steven L. and {Kocevski}, Dale D. and {Jeon}, Junehyoung and {Bromm}, Volker and {Amor{\'\i}n}, Ricardo O. and {Arrabal Haro}, Pablo and {Backhaus}, Bren E. and {Bagley}, Micaela B. and {Banados}, Eduardo and {Bhatawdekar}, Rachana and {Brooks}, Madisyn and {Calabr{\`o}}, Antonello and {Ch{\'a}vez Ortiz}, {\'O}scar A. and {Cheng}, Yingjie and {Cleri}, Nikko J. and {Cole}, Justin W. and {Davis}, Kelcey and {Dickinson}, Mark and {Donnan}, Callum and {Dunlop}, James S. and {Ellis}, Richard S. and {Fern{\'a}ndez}, Vital and {Fontana}, Adriano and {Fujimoto}, Seiji and {Giavalisco}, Mauro and {Grazian}, Andrea and {Guo}, Jingsong and {Hathi}, Nimish P. and {Holwerda}, Benne W. and {Hirschmann}, Michaela and {Inayoshi}, Kohei and {Kartaltepe}, Jeyhan S. and {Khusanova}, Yana and {Koekemoer}, Anton M. and {Kokorev}, Vasily and {Larson}, Rebecca L. and {Leung}, Gene C.~K. and {Lucas}, Ray A. and {McLeod}, Derek J. and {Napolitano}, Lorenzo and {Onoue}, Masafusa and {Pacucci}, Fabio and {Papovich}, Casey and {P{\'e}rez-Gonz{\'a}lez}, Pablo G. and {Pirzkal}, Nor and {Somerville}, Rachel S. and {Trump}, Jonathan R. and {Wilkins}, Stephen M. and {Yung}, L.~Y. Aaron and {Zhang}, Haowen},
        title = "{Broad-line AGNs at 3.5 < z < 6: The Black Hole Mass Function and a Connection with Little Red Dots}",
      journal = {\apj},
         year = 2025,
        month = jun,
       volume = {986},
       number = {2},
          eid = {165},
        pages = {165},
          doi = {10.3847/1538-4357/add15b},
archivePrefix = {arXiv},
       eprint = {2409.06772},
 primaryClass = {astro-ph.GA},
       adsurl = {https://ui.adsabs.harvard.edu/abs/2025ApJ...986..165T}
}

@article{perez-gonzalez2024,
       author = {{P{\'e}rez-Gonz{\'a}lez}, Pablo G. and {Barro}, Guillermo and {Rieke}, George H. and {Lyu}, Jianwei and {Rieke}, Marcia and {Alberts}, Stacey and {Williams}, Christina C. and {Hainline}, Kevin and {Sun}, Fengwu and {Pusk{\'a}s}, D{\'a}vid and {Annunziatella}, Marianna and {Baker}, William M. and {Bunker}, Andrew J. and {Egami}, Eiichi and {Ji}, Zhiyuan and {Johnson}, Benjamin D. and {Robertson}, Brant and {Rodr{\'\i}guez Del Pino}, Bruno and {Rujopakarn}, Wiphu and {Shivaei}, Irene and {Tacchella}, Sandro and {Willmer}, Christopher N.~A. and {Willott}, Chris},
        title = "{What Is the Nature of Little Red Dots and what Is Not, MIRI SMILES Edition}",
      journal = {\apj},
         year = 2024,
        month = jun,
       volume = {968},
       number = {1},
          eid = {4},
        pages = {4},
          doi = {10.3847/1538-4357/ad38bb},
archivePrefix = {arXiv},
       eprint = {2401.08782},
 primaryClass = {astro-ph.GA},
       adsurl = {https://ui.adsabs.harvard.edu/abs/2024ApJ...968....4P}
}

@article{leung2024,
       author = {{Leung}, Gene C.~K. and {Finkelstein}, Steven L. and {P{\'e}rez-Gonz{\'a}lez}, Pablo G. and {Morales}, Alexa M. and {Taylor}, Anthony J. and {Barro}, Guillermo and {Kocevski}, Dale D. and {Akins}, Hollis B. and {Carnall}, Adam C. and {Ch{\'a}vez Ortiz}, {\'O}scar A. and {Cleri}, Nikko J. and {Cullen}, Fergus and {Donnan}, Callum T. and {Dunlop}, James S. and {Ellis}, Richard S. and {Grogin}, Norman A. and {Hirschmann}, Michaela and {Koekemoer}, Anton M. and {Kokorev}, Vasily and {Lucas}, Ray A. and {McLeod}, Derek J. and {Papovich}, Casey and {Yung}, L.~Y. Aaron},
        title = "{Exploring the Nature of Little Red Dots: Constraints on Active Galactic Nucleus and Stellar Contributions from PRIMER MIRI Imaging}",
      journal = {\apj},
         year = 2025,
        month = oct,
       volume = {992},
       number = {1},
          eid = {26},
        pages = {26},
          doi = {10.3847/1538-4357/adfcce},
archivePrefix = {arXiv},
       eprint = {2411.12005},
 primaryClass = {astro-ph.GA},
       adsurl = {https://ui.adsabs.harvard.edu/abs/2025ApJ...992...26L}
}

@article{lin2024,
	title        = {{A SPectroscopic Survey of Biased Halos In the Reionization Era (ASPIRE): Broad-line AGN at z = 4‑5 Revealed by JWST/NIRCam WFSS}},
	author       = {{Lin}, Xiaojing and {Wang}, Feige and {Fan}, Xiaohui and {Cai}, Zheng and {Champagne}, Jaclyn B. and {Sun}, Fengwu and {Volonteri}, Marta and {Yang}, Jinyi and {Hennawi}, Joseph F. and {Ba{\~n}ados}, Eduardo and {Barth}, Aaron and {Eilers}, Anna-Christina and {Farina}, Emanuele Paolo and {Liu}, Weizhe and {Jin}, Xiangyu and {Jun}, Hyunsung D. and {Lupi}, Alessandro and {Kakiichi}, Koki and {Mazzucchelli}, Chiara and {Onoue}, Masafusa and {Pan}, Zhiwei and {Pizzati}, Elia and {Rojas-Ruiz}, Sof{\'\i}a and {Schindler}, Jan-Torge and {Trakhtenbrot}, Benny and {Shen}, Yue and {Trebitsch}, Maxime and {Zhuang}, Ming-Yang and {Endsley}, Ryan and {Meyer}, Romain A. and {Li}, Zihao and {Li}, Mingyu and {Pudoka}, Maria and {Tee}, Wei Leong and {Wu}, Yunjing and {Zhang}, Haowen},
	year         = 2024,
	month        = {oct},
	journal      = {\apj},
	volume       = 974,
	number       = 1,
	pages        = 147,
	doi          = {10.3847/1538-4357/ad6565},
	eid          = 147,
	archiveprefix = {arXiv},
	eprint       = {2407.17570},
	primaryclass = {astro-ph.GA},
	adsurl       = {https://ui.adsabs.harvard.edu/abs/2024ApJ...974..147L}
}

@article{volonteri2025,
	title        = {{Exploring active galactic nuclei and little red dots with the Obelisk simulation}},
	author       = {{Volonteri}, M. and {Trebitsch}, M. and {Greene}, J.~E. and {Dubois}, Y. and {Dong-Paez}, C. -A. and {Habouzit}, M. and {Lupi}, A. and {Ma}, Y. and {Beckmann}, R.~S. and {Dayal}, P. and {Schneider}, R.},
	year         = 2025,
	month        = {mar},
	journal      = {\aap},
	volume       = 695,
	pages        = {A33},
	doi          = {10.1051/0004-6361/202451963},
	eid          = {A33},
	archiveprefix = {arXiv},
	eprint       = {2408.12854},
	primaryclass = {astro-ph.GA},
	adsurl       = {https://ui.adsabs.harvard.edu/abs/2025A&A...695A..33V}
}

@article{kokorev2024b,
	title        = {{Silencing the Giant: Evidence of Active Galactic Nucleus Feedback and Quenching in a Little Red Dot at z = 4.13}},
	author       = {{Kokorev}, Vasily and {Chisholm}, John and {Endsley}, Ryan and {Finkelstein}, Steven L. and {Greene}, Jenny E. and {Akins}, Hollis B. and {Bromm}, Volker and {Casey}, Caitlin M. and {Fujimoto}, Seiji and {Labb{\'e}}, Ivo and {Larson}, Rebecca L.},
	year         = 2024,
	month        = {nov},
	journal      = {\apj},
	volume       = 975,
	number       = 2,
	pages        = 178,
	doi          = {10.3847/1538-4357/ad7d03},
	eid          = 178,
	archiveprefix = {arXiv},
	eprint       = {2407.20320},
	primaryclass = {astro-ph.GA},
	adsurl       = {https://ui.adsabs.harvard.edu/abs/2024ApJ...975..178K}
}

@article{setton2024,
	title        = {{Little Red Dots at an Inflection Point: Ubiquitous ``V-Shaped'' Turnover Consistently Occurs at the Balmer Limit}},
	author       = {{Setton}, David J. and {Greene}, Jenny E. and {de Graaff}, Anna and {Ma}, Yilun and {Leja}, Joel and {Matthee}, Jorryt and {Bezanson}, Rachel and {Boogaard}, Leindert A. and {Cleri}, Nikko J. and {Katz}, Harley and {Labbe}, Ivo and {Maseda}, Michael V. and {McConachie}, Ian and {Miller}, Tim B. and {Price}, Sedona H. and {Suess}, Katherine A. and {van Dokkum}, Pieter and {Wang}, Bingjie and {Weibel}, Andrea and {Whitaker}, Katherine E. and {Williams}, Christina C.},
	year         = 2024,
	month        = {nov},
	journal      = {arXiv e-prints},
	pages        = {arXiv:2411.03424},
	doi          = {10.48550/arXiv.2411.03424},
	eid          = {arXiv:2411.03424},
	archiveprefix = {arXiv},
	eprint       = {2411.03424},
	primaryclass = {astro-ph.GA},
	adsurl       = {https://ui.adsabs.harvard.edu/abs/2024arXiv241103424S}
}

@article{wang2024,
	title        = {{RUBIES: Evolved Stellar Populations with Extended Formation Histories at z {\ensuremath{\sim}} 7{\textendash}8 in Candidate Massive Galaxies Identified with JWST/NIRSpec}},
	author       = {{Wang}, Bingjie and {Leja}, Joel and {de Graaff}, Anna and {Brammer}, Gabriel B. and {Weibel}, Andrea and {van Dokkum}, Pieter and {Baggen}, Josephine F.~W. and {Suess}, Katherine A. and {Greene}, Jenny E. and {Bezanson}, Rachel and {Cleri}, Nikko J. and {Hirschmann}, Michaela and {Labb{\'e}}, Ivo and {Matthee}, Jorryt and {McConachie}, Ian and {Naidu}, Rohan P. and {Nelson}, Erica and {Oesch}, Pascal A. and {Setton}, David J. and {Williams}, Christina C.},
	year         = 2024,
	month        = {jul},
	journal      = {\apjl},
	volume       = 969,
	number       = 1,
	pages        = {L13},
	doi          = {10.3847/2041-8213/ad55f7},
	eid          = {L13},
	archiveprefix = {arXiv},
	eprint       = {2405.01473},
	primaryclass = {astro-ph.GA},
	adsurl       = {https://ui.adsabs.harvard.edu/abs/2024ApJ...969L..13W}
}

@article{latif2025,
	title        = {{Radio emission from little red dots may reveal their true nature}},
	author       = {{Latif}, Muhammad A. and {Aftab}, Ammara and {Whalen}, Daniel J. and {Mezcua}, Mar},
	year         = 2025,
	month        = {feb},
	journal      = {\aap},
	volume       = 694,
	pages        = {L14},
	doi          = {10.1051/0004-6361/202453194},
	eid          = {L14},
	archiveprefix = {arXiv},
	eprint       = {2502.03742},
	primaryclass = {astro-ph.GA},
	adsurl       = {https://ui.adsabs.harvard.edu/abs/2025A&A...694L..14L}
}

@article{gloudemans2025,
       author = {{Gloudemans}, Anniek J. and {Duncan}, Kenneth J. and {Eilers}, Anna-Christina and {Farina}, Emanuele Paolo and {Harikane}, Yuichi and {Inayoshi}, Kohei and {Lambrides}, Erini and {Vardoulaki}, Eleni},
        title = "{Another Piece to the Puzzle: Radio Detection of a JWST-detected Active Galactic Nucleus Candidate}",
      journal = {\apj},
         year = 2025,
        month = jun,
       volume = {986},
       number = {2},
          eid = {130},
        pages = {130},
          doi = {10.3847/1538-4357/adddb9},
archivePrefix = {arXiv},
       eprint = {2501.04912},
 primaryClass = {astro-ph.GA},
       adsurl = {https://ui.adsabs.harvard.edu/abs/2025ApJ...986..130G}
}

@article{perger2025,
	title        = {{Deep silence: Radio properties of little red dots}},
	author       = {{Perger}, K. and {Fogasy}, J. and {Frey}, S. and {Gab{\'a}nyi}, K. {\'E}.},
	year         = 2025,
	month        = {jan},
	journal      = {\aap},
	volume       = 693,
	pages        = {L2},
	doi          = {10.1051/0004-6361/202452422},
	eid          = {L2},
	archiveprefix = {arXiv},
	eprint       = {2411.19518},
	primaryclass = {astro-ph.GA},
	adsurl       = {https://ui.adsabs.harvard.edu/abs/2025A&A...693L...2P}
}

@article{ananna2024,
	title        = {{X-Ray View of Little Red Dots: Do They Host Supermassive Black Holes?}},
	author       = {{Ananna}, Tonima Tasnim and {Bogd{\'a}n}, {\'A}kos and {Kov{\'a}cs}, Orsolya E. and {Natarajan}, Priyamvada and {Hickox}, Ryan C.},
	year         = 2024,
	month        = {jul},
	journal      = {\apjl},
	volume       = 969,
	number       = 1,
	pages        = {L18},
	doi          = {10.3847/2041-8213/ad5669},
	eid          = {L18},
	archiveprefix = {arXiv},
	eprint       = {2404.19010},
	primaryclass = {astro-ph.GA},
	adsurl       = {https://ui.adsabs.harvard.edu/abs/2024ApJ...969L..18A}
}

@article{yue2024,
	title        = {{Stacking X-Ray Observations of ``Little Red Dots'': Implications for Their Active Galactic Nucleus Properties}},
	author       = {{Yue}, Minghao and {Eilers}, Anna-Christina and {Ananna}, Tonima Tasnim and {Panagiotou}, Christos and {Kara}, Erin and {Miyaji}, Takamitsu},
	year         = 2024,
	month        = {oct},
	journal      = {\apjl},
	volume       = 974,
	number       = 2,
	pages        = {L26},
	doi          = {10.3847/2041-8213/ad7eba},
	eid          = {L26},
	archiveprefix = {arXiv},
	eprint       = {2404.13290},
	primaryclass = {astro-ph.GA},
	adsurl       = {https://ui.adsabs.harvard.edu/abs/2024ApJ...974L..26Y}
}

@article{akins2025,
	title        = {{Tentative detection of neutral gas in a Little Red Dot at $z=4.46$}},
	author       = {{Akins}, Hollis B. and {Casey}, Caitlin M. and {Chisholm}, John and {Berg}, Danielle A. and {Cooper}, Olivia and {Franco}, Maximilien and {Fujimoto}, Seiji and {Lambrides}, Erini and {Long}, Arianna S. and {McKinney}, Jed},
	year         = 2025,
	month        = {mar},
	journal      = {arXiv e-prints},
	pages        = {arXiv:2503.00998},
	doi          = {10.48550/arXiv.2503.00998},
	eid          = {arXiv:2503.00998},
	archiveprefix = {arXiv},
	eprint       = {2503.00998},
	primaryclass = {astro-ph.GA},
	adsurl       = {https://ui.adsabs.harvard.edu/abs/2025arXiv250300998A}
}

@article{furtak2025,
author = {{Furtak}, Lukas J. and {Secunda}, Amy R. and {Greene}, Jenny E. and {Zitrin}, Adi and {Labb{\'e}}, Ivo and {Golubchik}, Miriam and {Bezanson}, Rachel and {Kokorev}, Vasily and {Atek}, Hakim and {Brammer}, Gabriel B. and {Chemerynska}, Iryna and {Cutler}, Sam E. and {Dayal}, Pratika and {Feldmann}, Robert and {Fujimoto}, Seiji and {Glazebrook}, Karl and {Leja}, Joel and {Ma}, Yilun and {Matthee}, Jorryt and {Naidu}, Rohan P. and {Nelson}, Erica J. and {Oesch}, Pascal A. and {Pan}, Richard and {Price}, Sedona H. and {Suess}, Katherine A. and {Wang}, Bingjie and {Weaver}, John R. and {Whitaker}, Katherine E.},
        title = "{Investigating photometric and spectroscopic variability in the multiply imaged little red dot A2744-QSO1}",
      journal = {\aap},
         year = 2025,
        month = jun,
       volume = {698},
          eid = {A227},
        pages = {A227},
          doi = {10.1051/0004-6361/202554110},
archivePrefix = {arXiv},
       eprint = {2502.07875},
 primaryClass = {astro-ph.GA},
       adsurl = {https://ui.adsabs.harvard.edu/abs/2025A&A...698A.227F}
}

@article{spaans2006,
	title        = {{Pregalactic Black Hole Formation with an Atomic Hydrogen Equation of State}},
	author       = {{Spaans}, Marco and {Silk}, Joseph},
	year         = 2006,
	month        = {dec},
	journal      = {\apj},
	volume       = 652,
	number       = 2,
	pages        = {902--906},
	doi          = {10.1086/508444},
	archiveprefix = {arXiv},
	eprint       = {astro-ph/0601714},
	primaryclass = {astro-ph},
	adsurl       = {https://ui.adsabs.harvard.edu/abs/2006ApJ...652..902S}
}

@article{cantalupo2014,
	title        = {{A cosmic web filament revealed in Lyman-{\ensuremath{\alpha}} emission around a luminous high-redshift quasar}},
	author       = {{Cantalupo}, Sebastiano and {Arrigoni-Battaia}, Fabrizio and {Prochaska}, J. Xavier and {Hennawi}, Joseph F. and {Madau}, Piero},
	year         = 2014,
	month        = feb,
	journal      = {\nat},
	volume       = 506,
	number       = 7486,
	pages        = {63--66},
	doi          = {10.1038/nature12898},
	archiveprefix = {arXiv},
	eprint       = {1401.4469},
	primaryclass = {astro-ph.CO},
	adsurl       = {https://ui.adsabs.harvard.edu/abs/2014Natur.506...63C}
}

@article{shibuya2014,
	title        = {{What is the Physical Origin of Strong Ly{\ensuremath{\alpha}} Emission? II. Gas Kinematics and Distribution of Ly{\ensuremath{\alpha}} Emitters}},
	author       = {{Shibuya}, Takatoshi and {Ouchi}, Masami and {Nakajima}, Kimihiko and {Hashimoto}, Takuya and {Ono}, Yoshiaki and {Rauch}, Michael and {Gauthier}, Jean-Rene and {Shimasaku}, Kazuhiro and {Goto}, Ryosuke and {Mori}, Masao and {Umemura.}, Masayuki},
	year         = 2014,
	month        = jun,
	journal      = {\apj},
	volume       = 788,
	number       = 1,
	pages        = 74,
	doi          = {10.1088/0004-637X/788/1/74},
	eid          = 74,
	archiveprefix = {arXiv},
	eprint       = {1402.1168},
	primaryclass = {astro-ph.CO},
	adsurl       = {https://ui.adsabs.harvard.edu/abs/2014ApJ...788...74S}
}

@article{furtak2023,
	title        = {{UNCOVERing the extended strong lensing structures of Abell 2744 with the deepest JWST imaging}},
	author       = {{Furtak}, Lukas J. and {Zitrin}, Adi and {Weaver}, John R. and {Atek}, Hakim and {Bezanson}, Rachel and {Labb{\'e}}, Ivo and {Whitaker}, Katherine E. and {Leja}, Joel and {Price}, Sedona H. and {Brammer}, Gabriel B. and {Wang}, Bingjie and {Marchesini}, Danilo and {Pan}, Richard and {Dayal}, Pratika and {van Dokkum}, Pieter and {Feldmann}, Robert and {Fujimoto}, Seiji and {Franx}, Marijn and {Khullar}, Gourav and {Nelson}, Erica J. and {Mowla}, Lamiya A.},
	year         = 2023,
	month        = aug,
	journal      = {\mnras},
	volume       = 523,
	number       = 3,
	pages        = {4568--4582},
	doi          = {10.1093/mnras/stad1627},
	archiveprefix = {arXiv},
	eprint       = {2212.04381},
	primaryclass = {astro-ph.GA},
	adsurl       = {https://ui.adsabs.harvard.edu/abs/2023MNRAS.523.4568F}
}

@article{price2025,
	title        = {{The UNCOVER Survey: First Release of Ultradeep JWST/NIRSpec PRISM Spectra for {\ensuremath{\sim}}700 Galaxies from z {\ensuremath{\sim}} 0.3{\textendash}13 in A2744}},
	author       = {{Price}, Sedona H. and {Bezanson}, Rachel and {Labbe}, Ivo and {Furtak}, Lukas J. and {de Graaff}, Anna and {Greene}, Jenny E. and {Kokorev}, Vasily and {Setton}, David J. and {Suess}, Katherine A. and {Brammer}, Gabriel and {Cutler}, Sam E. and {Leja}, Joel and {Pan}, Richard and {Wang}, Bingjie and {Weaver}, John R. and {Whitaker}, Katherine E. and {Atek}, Hakim and {Burgasser}, Adam J. and {Chemerynska}, Iryna and {Dayal}, Pratika and {Feldmann}, Robert and {F{\"o}rster Schreiber}, Natascha M. and {Fudamoto}, Yoshinobu and {Fujimoto}, Seiji and {Glazebrook}, Karl and {Goulding}, Andy D. and {Khullar}, Gourav and {Kriek}, Mariska and {Marchesini}, Danilo and {Maseda}, Michael V. and {Miller}, Tim B. and {Muzzin}, Adam and {Nanayakkara}, Themiya and {Nelson}, Erica and {Oesch}, Pascal A. and {Shipley}, Heath and {Smit}, Renske and {Taylor}, Edward N. and {Dokkum}, Pieter van and {Williams}, Christina C. and {Zitrin}, Adi},
	year         = 2025,
	month        = mar,
	journal      = {\apj},
	volume       = 982,
	number       = 1,
	pages        = 51,
	doi          = {10.3847/1538-4357/adaec1},
	eid          = 51,
	archiveprefix = {arXiv},
	eprint       = {2408.03920},
	primaryclass = {astro-ph.GA},
	adsurl       = {https://ui.adsabs.harvard.edu/abs/2025ApJ...982...51P}
}

@article{pizzati2024,
       author = {{Pizzati}, Elia and {Hennawi}, Joseph F. and {Schaye}, Joop and {Eilers}, Anna-Christina and {Huang}, Jiamu and {Schindler}, Jan-Torge and {Wang}, Feige},
        title = "{'Little red dots' cannot reside in the same dark matter haloes as comparably luminous unobscured quasars}",
      journal = {\mnras},
         year = 2025,
        month = jun,
       volume = {539},
       number = {4},
        pages = {2910-2925},
          doi = {10.1093/mnras/staf660},
archivePrefix = {arXiv},
       eprint = {2409.18208},
 primaryClass = {astro-ph.GA},
       adsurl = {https://ui.adsabs.harvard.edu/abs/2025MNRAS.539.2910P}
}

@ARTICLE{pezzulli2019,
       author = {{Pezzulli}, Gabriele and {Cantalupo}, Sebastiano},
        title = "{A high baryon fraction in massive haloes at z {\ensuremath{\sim}} 3}",
      journal = {\mnras},
         year = 2019,
        month = jun,
       volume = {486},
       number = {2},
        pages = {1489-1508},
          doi = {10.1093/mnras/stz906},
archivePrefix = {arXiv},
       eprint = {1903.11069},
 primaryClass = {astro-ph.GA},
       adsurl = {https://ui.adsabs.harvard.edu/abs/2019MNRAS.486.1489P}
}

@ARTICLE{dijkstra2009,
       author = {{Dijkstra}, Mark and {Loeb}, Abraham},
        title = "{Ly{\ensuremath{\alpha}} blobs as an observational signature of cold accretion streams into galaxies}",
      journal = {\mnras},
         year = 2009,
        month = dec,
       volume = {400},
       number = {2},
        pages = {1109-1120},
          doi = {10.1111/j.1365-2966.2009.15533.x},
archivePrefix = {arXiv},
       eprint = {0902.2999},
 primaryClass = {astro-ph.CO},
       adsurl = {https://ui.adsabs.harvard.edu/abs/2009MNRAS.400.1109D}
}

@ARTICLE{topping2025,
       author = {{Topping}, Michael W. and {Stark}, Daniel P. and {Senchyna}, Peter and {Chen}, Zuyi and {Zitrin}, Adi and {Endsley}, Ryan and {Charlot}, St{\'e}phane and {Furtak}, Lukas J. and {Maseda}, Michael V. and {Plat}, Adele and {Smit}, Renske and {Mainali}, Ramesh and {Chevallard}, Jacopo and {Molyneux}, Stephen and {Rigby}, Jane R.},
        title = "{Deep Rest-UV JWST/NIRSpec Spectroscopy of Early Galaxies: The Demographics of C IV and N-emitters in the Reionization Era}",
      journal = {\apj},
         year = 2025,
        month = feb,
       volume = {980},
       number = {2},
          eid = {225},
        pages = {225},
          doi = {10.3847/1538-4357/ada95c},
archivePrefix = {arXiv},
       eprint = {2407.19009},
 primaryClass = {astro-ph.GA},
       adsurl = {https://ui.adsabs.harvard.edu/abs/2025ApJ...980..225T}
}

@article{treiber2024,
       author = {{Treiber}, Helena and {Greene}, Jenny E. and {Weaver}, John R. and {Miller}, Tim B. and {Furtak}, Lukas J. and {Setton}, David J. and {Wang}, Bingjie and {de Graaff}, Anna and {Bezanson}, Rachel and {Brammer}, Gabriel and {Cutler}, Sam E. and {Dayal}, Pratika and {Feldmann}, Robert and {Fujimoto}, Seiji and {Goulding}, Andy D. and {Kokorev}, Vasily and {Labbe}, Ivo and {Leja}, Joel and {Marchesini}, Danilo and {Nanayakkara}, Themiya and {Nelson}, Erica and {Pan}, Richard and {Price}, Sedona H. and {Siegel}, Jared and {Suess}, Katherine A. and {Whitaker}, Katherine E.},
        title = "{UNCOVERing the High-redshift AGN Population among Extreme UV Line Emitters}",
      journal = {\apj},
         year = 2025,
        month = may,
       volume = {984},
       number = {1},
          eid = {93},
        pages = {93},
          doi = {10.3847/1538-4357/adc38f},
archivePrefix = {arXiv},
       eprint = {2409.12232},
 primaryClass = {astro-ph.GA},
       adsurl = {https://ui.adsabs.harvard.edu/abs/2025ApJ...984...93T}
}

@article{williams2024,
	title        = {{The Galaxies Missed by Hubble and ALMA: The Contribution of Extremely Red Galaxies to the Cosmic Census at 3 < z < 8}},
	author       = {{Williams}, Christina C. and {Alberts}, Stacey and {Ji}, Zhiyuan and {Hainline}, Kevin N. and {Lyu}, Jianwei and {Rieke}, George and {Endsley}, Ryan and {Suess}, Katherine A. and {Sun}, Fengwu and {Johnson}, Benjamin D. and {Florian}, Michael and {Shivaei}, Irene and {Rujopakarn}, Wiphu and {Baker}, William M. and {Bhatawdekar}, Rachana and {Boyett}, Kristan and {Bunker}, Andrew J. and {Cameron}, Alex J. and {Carniani}, Stefano and {Charlot}, Stephane and {Curtis-Lake}, Emma and {DeCoursey}, Christa and {de Graaff}, Anna and {Egami}, Eiichi and {Eisenstein}, Daniel J. and {Gibson}, Justus L. and {Hausen}, Ryan and {Helton}, Jakob M. and {Maiolino}, Roberto and {Maseda}, Michael V. and {Nelson}, Erica J. and {P{\'e}rez-Gonz{\'a}lez}, Pablo G. and {Rieke}, Marcia J. and {Robertson}, Brant E. and {Saxena}, Aayush and {Tacchella}, Sandro and {Willmer}, Christopher N.~A. and {Willott}, Chris J.},
	year         = 2024,
	month        = jun,
	journal      = {\apj},
	volume       = 968,
	number       = 1,
	pages        = 34,
	doi          = {10.3847/1538-4357/ad3f17},
	eid          = 34,
	archiveprefix = {arXiv},
	eprint       = {2311.07483},
	primaryclass = {astro-ph.GA},
	adsurl       = {https://ui.adsabs.harvard.edu/abs/2024ApJ...968...34W}
}

@article{deugenio2025,
	title        = {{BlackTHUNDER strikes twice: rest-frame Balmer-line absorption and high Eddington accretion rate in a Little Red Dot at $z=7.04$}},
	author       = {{D'Eugenio}, Francesco and {Maiolino}, Roberto and {Perna}, Michele and {Uebler}, Hannah and {Ji}, Xihan and {McClymont}, William and {Koudmani}, Sophie and {Sijacki}, Debora and {Juod{\v{z}}balis}, Ignas and {Scholtz}, Jan and {Bennett}, Jake and {Bunker}, Andrew J. and {Carniani}, Stefano and {Charlot}, St{\'e}phane and {Cresci}, Giovanni and {Curtis-Lake}, Emma and {Dalla Bont{\`a}}, Elena and {Jones}, Gareth C. and {Lyu}, Jianwei and {Marconi}, Alessandro and {Mazzolari}, Giovanni and {Nelson}, Erica J. and {Parlanti}, Eleonora and {Robertson}, Brant E. and {Schneider}, Raffaella and {Simmonds}, Charlotte and {Tacchella}, Sandro and {Venturi}, Giacomo and {Willott}, Chris and {Witstok}, Joris and {Witten}, Callum},
	year         = 2025,
	month        = mar,
	journal      = {arXiv e-prints},
	pages        = {arXiv:2503.11752},
	doi          = {10.48550/arXiv.2503.11752},
	eid          = {arXiv:2503.11752},
	archiveprefix = {arXiv},
	eprint       = {2503.11752},
	primaryclass = {astro-ph.GA},
	adsurl       = {https://ui.adsabs.harvard.edu/abs/2025arXiv250311752D}
}

@article{maiolino2025,
	title        = {{JWST meets Chandra: a large population of Compton thick, feedback-free, and intrinsically X-ray weak AGN, with a sprinkle of SNe}},
	author       = {{Maiolino}, Roberto and {Risaliti}, Guido and {Signorini}, Matilde and {Trefoloni}, Bartolomeo and {Juod{\v{z}}balis}, Ignas and {Scholtz}, Jan and {{\"U}bler}, Hannah and {D'Eugenio}, Francesco and {Carniani}, Stefano and {Fabian}, Andy and {Ji}, Xihan and {Mazzolari}, Giovanni and {Bertola}, Elena and {Brusa}, Marcella and {Bunker}, Andrew J. and {Charlot}, Stephane and {Comastri}, Andrea and {Cresci}, Giovanni and {DeCoursey}, Christa Noel and {Egami}, Eiichi and {Fiore}, Fabrizio and {Gilli}, Roberto and {Perna}, Michele and {Tacchella}, Sandro and {Venturi}, Giacomo},
	year         = 2025,
	month        = apr,
	journal      = {\mnras},
	volume       = 538,
	number       = 3,
	pages        = {1921--1943},
	doi          = {10.1093/mnras/staf359},
	adsurl       = {https://ui.adsabs.harvard.edu/abs/2025MNRAS.538.1921M}
}

@article{brok2020,
	title        = {{Probing the AGN unification model at redshift z {\ensuremath{\sim}} 3 with MUSE observations of giant Ly {\ensuremath{\alpha}} nebulae}},
	author       = {{den Brok}, J.~S. and {Cantalupo}, S. and {Mackenzie}, R. and {Marino}, R.~A. and {Pezzulli}, G. and {Matthee}, J. and {Johnson}, S.~D. and {Krumpe}, M. and {Urrutia}, T. and {Kollatschny}, W.},
	year         = 2020,
	month        = jun,
	journal      = {\mnras},
	volume       = 495,
	number       = 2,
	pages        = {1874--1887},
	doi          = {10.1093/mnras/staa1269},
	archiveprefix = {arXiv},
	eprint       = {2005.01732},
	primaryclass = {astro-ph.GA},
	adsurl       = {https://ui.adsabs.harvard.edu/abs/2020MNRAS.495.1874D}
}

@article{weibel2024,
	title        = {{Galaxy build-up in the first 1.5 Gyr of cosmic history: insights from the stellar mass function at z   4-9 from JWST NIRCam observations}},
	author       = {{Weibel}, Andrea and {Oesch}, Pascal A. and {Barrufet}, Laia and {Gottumukkala}, Rashmi and {Ellis}, Richard S. and {Santini}, Paola and {Weaver}, John R. and {Allen}, Natalie and {Bouwens}, Rychard and {Bowler}, Rebecca A.~A. and {Brammer}, Gabe and {Carnall}, Adam C. and {Cullen}, Fergus and {Dayal}, Pratika and {Dickinson}, Mark and {Donnan}, Callum T. and {Dunlop}, James S. and {Giavalisco}, Mauro and {Grogin}, Norman A. and {Illingworth}, Garth D. and {Koekemoer}, Anton M. and {Labbe}, Ivo and {Marchesini}, Danilo and {McLeod}, Derek J. and {McLure}, Ross J. and {Naidu}, Rohan P. and {P{\'e}rez-Gonz{\'a}lez}, Pablo G. and {Shuntov}, Marko and {Stefanon}, Mauro and {Toft}, Sune and {Xiao}, Mengyuan},
	year         = 2024,
	month        = sep,
	journal      = {\mnras},
	volume       = 533,
	number       = 2,
	pages        = {1808--1838},
	doi          = {10.1093/mnras/stae1891},
	archiveprefix = {arXiv},
	eprint       = {2403.08872},
	primaryclass = {astro-ph.GA},
	adsurl       = {https://ui.adsabs.harvard.edu/abs/2024MNRAS.533.1808W}
}

@article{suess2024,
	title        = {{Medium Bands, Mega Science: A JWST/NIRCam Medium-band Imaging Survey of A2744}},
	author       = {{Suess}, Katherine A. and {Weaver}, John R. and {Price}, Sedona H. and {Pan}, Richard and {Wang}, Bingjie and {Bezanson}, Rachel and {Brammer}, Gabriel and {Cutler}, Sam E. and {Labb{\'e}}, Ivo and {Leja}, Joel and {Williams}, Christina C. and {Whitaker}, Katherine E. and {Atek}, Hakim and {Dayal}, Pratika and {de Graaff}, Anna and {Feldmann}, Robert and {Franx}, Marijn and {Fudamoto}, Yoshinobu and {Fujimoto}, Seiji and {Furtak}, Lukas J. and {Goulding}, Andy D. and {Greene}, Jenny E. and {Khullar}, Gourav and {Kokorev}, Vasily and {Kriek}, Mariska and {Lorenz}, Brian and {Marchesini}, Danilo and {Maseda}, Michael V. and {Matthee}, Jorryt and {Miller}, Tim B. and {Mitsuhashi}, Ikki and {Mowla}, Lamiya A. and {Muzzin}, Adam and {Naidu}, Rohan P. and {Nanayakkara}, Themiya and {Nelson}, Erica J. and {Oesch}, Pascal A. and {Setton}, David J. and {Shipley}, Heath and {Smit}, Renske and {Spilker}, Justin S. and {van Dokkum}, Pieter and {Zitrin}, Adi},
	year         = 2024,
	month        = nov,
	journal      = {\apj},
	volume       = 976,
	number       = 1,
	pages        = 101,
	doi          = {10.3847/1538-4357/ad75fe},
	eid          = 101,
	archiveprefix = {arXiv},
	eprint       = {2404.13132},
	primaryclass = {astro-ph.GA},
	adsurl       = {https://ui.adsabs.harvard.edu/abs/2024ApJ...976..101S}
}

@article{erb2018,
	title        = {{The Kinematics of Extended Ly{\ensuremath{\alpha}} Emission in a Low-mass, Low-metallicity Galaxy at z = 2.3}},
	author       = {{Erb}, Dawn K. and {Steidel}, Charles C. and {Chen}, Yuguang},
	year         = 2018,
	month        = jul,
	journal      = {\apjl},
	volume       = 862,
	number       = 1,
	pages        = {L10},
	doi          = {10.3847/2041-8213/aacff6},
	eid          = {L10},
	archiveprefix = {arXiv},
	eprint       = {1807.00065},
	primaryclass = {astro-ph.GA},
	adsurl       = {https://ui.adsabs.harvard.edu/abs/2018ApJ...862L..10E}
}

@article{juodzbalis2024,
	title        = {{JADES - the Rosetta stone of JWST-discovered AGN: deciphering the intriguing nature of early AGN}},
	author       = {{Juod{\v{z}}balis}, Ignas and {Ji}, Xihan and {Maiolino}, Roberto and {D'Eugenio}, Francesco and {Scholtz}, Jan and {Risaliti}, Guido and {Fabian}, Andrew C. and {Mazzolari}, Giovanni and {Gilli}, Roberto and {Prandoni}, Isabella and {Arribas}, Santiago and {Bunker}, Andrew J. and {Carniani}, Stefano and {Charlot}, St{\'e}phane and {Curtis-Lake}, Emma and {de Graaff}, Anna and {Hainline}, Kevin and {Parlanti}, Eleonora and {Perna}, Michele and {P{\'e}rez-Gonz{\'a}lez}, Pablo G. and {Robertson}, Brant and {Tacchella}, Sandro and {{\"U}bler}, Hannah and {Williams}, Christina C. and {Willott}, Chris and {Witstok}, Joris},
	year         = 2024,
	month        = nov,
	journal      = {\mnras},
	volume       = 535,
	number       = 1,
	pages        = {853--873},
	doi          = {10.1093/mnras/stae2367},
	archiveprefix = {arXiv},
	eprint       = {2407.08643},
	primaryclass = {astro-ph.GA},
	adsurl       = {https://ui.adsabs.harvard.edu/abs/2024MNRAS.535..853J}
}

@article{herenz2015,
	title        = {{Where is the fuzz? Undetected Lyman {\ensuremath{\alpha}} nebulae around quasars at z \raisebox{-0.5ex}\textasciitilde 2.3}},
	author       = {{Herenz}, Edmund Christian and {Wisotzki}, Lutz and {Roth}, Martin and {Anders}, Friedrich},
	year         = 2015,
	month        = apr,
	journal      = {\aap},
	volume       = 576,
	pages        = {A115},
	doi          = {10.1051/0004-6361/201425580},
	eid          = {A115},
	archiveprefix = {arXiv},
	eprint       = {1502.05132},
	primaryclass = {astro-ph.GA},
	adsurl       = {https://ui.adsabs.harvard.edu/abs/2015A&A...576A.115H}
}

@article{neufeld1990,
	title        = {{The Transfer of Resonance-Line Radiation in Static Astrophysical Media}},
	author       = {{Neufeld}, David A.},
	year         = 1990,
	month        = feb,
	journal      = {\apj},
	volume       = 350,
	pages        = 216,
	doi          = {10.1086/168375},
	adsurl       = {https://ui.adsabs.harvard.edu/abs/1990ApJ...350..216N}
}

@article{dijkstra2016,
	title        = {{Ly{\ensuremath{\alpha}} Signatures from Direct Collapse Black Holes}},
	author       = {{Dijkstra}, Mark and {Gronke}, Max and {Sobral}, David},
	year         = 2016,
	month        = jun,
	journal      = {\apj},
	volume       = 823,
	number       = 2,
	pages        = 74,
	doi          = {10.3847/0004-637X/823/2/74},
	eid          = 74,
	archiveprefix = {arXiv},
	eprint       = {1602.07695},
	primaryclass = {astro-ph.GA},
	adsurl       = {https://ui.adsabs.harvard.edu/abs/2016ApJ...823...74D}
}

@article{calzetti2000,
	title        = {{The Dust Content and Opacity of Actively Star-forming Galaxies}},
	author       = {{Calzetti}, Daniela and {Armus}, Lee and {Bohlin}, Ralph C. and {Kinney}, Anne L. and {Koornneef}, Jan and {Storchi-Bergmann}, Thaisa},
	year         = 2000,
	month        = apr,
	journal      = {\apj},
	volume       = 533,
	number       = 2,
	pages        = {682--695},
	doi          = {10.1086/308692},
	archiveprefix = {arXiv},
	eprint       = {astro-ph/9911459},
	primaryclass = {astro-ph},
	adsurl       = {https://ui.adsabs.harvard.edu/abs/2000ApJ...533..682C}
}

@article{verhamme2006,
	title        = {{3D Ly{\ensuremath{\alpha}} radiation transfer. I. Understanding Ly{\ensuremath{\alpha}} line profile morphologies}},
	author       = {{Verhamme}, A. and {Schaerer}, D. and {Maselli}, A.},
	year         = 2006,
	month        = dec,
	journal      = {\aap},
	volume       = 460,
	number       = 2,
	pages        = {397--413},
	doi          = {10.1051/0004-6361:20065554},
	archiveprefix = {arXiv},
	eprint       = {astro-ph/0608075},
	primaryclass = {astro-ph},
	adsurl       = {https://ui.adsabs.harvard.edu/abs/2006A&A...460..397V}
}

@article{steidel2011,
	title        = {{Diffuse Ly{\ensuremath{\alpha}} Emitting Halos: A Generic Property of High-redshift Star-forming Galaxies}},
	author       = {{Steidel}, Charles C. and {Bogosavljevi{\'c}}, Milan and {Shapley}, Alice E. and {Kollmeier}, Juna A. and {Reddy}, Naveen A. and {Erb}, Dawn K. and {Pettini}, Max},
	year         = 2011,
	month        = aug,
	journal      = {\apj},
	volume       = 736,
	number       = 2,
	pages        = 160,
	doi          = {10.1088/0004-637X/736/2/160},
	eid          = 160,
	archiveprefix = {arXiv},
	eprint       = {1101.2204},
	primaryclass = {astro-ph.CO},
	adsurl       = {https://ui.adsabs.harvard.edu/abs/2011ApJ...736..160S}
}

@article{shen2020,
	title        = {{The bolometric quasar luminosity function at z = 0-7}},
	author       = {{Shen}, Xuejian and {Hopkins}, Philip F. and {Faucher-Gigu{\`e}re}, Claude-Andr{\'e} and {Alexander}, D.~M. and {Richards}, Gordon T. and {Ross}, Nicholas P. and {Hickox}, R.~C.},
	year         = 2020,
	month        = jan,
	journal      = {\mnras},
	volume       = 495,
	number       = 3,
	pages        = {3252--3275},
	doi          = {10.1093/mnras/staa1381},
	archiveprefix = {arXiv},
	eprint       = {2001.02696},
	primaryclass = {astro-ph.GA},
	adsurl       = {https://ui.adsabs.harvard.edu/abs/2020MNRAS.495.3252S}
}

@article{ning2024,
	title        = {{Unveiling Luminous Ly{\ensuremath{\alpha}} Emitters at z {\ensuremath{\approx}} 6 through JWST/NIRCam Imaging in the COSMOS Field}},
	author       = {{Ning}, Yuanhang and {Cai}, Zheng and {Lin}, Xiaojing and {Zheng}, Zhen-Ya and {Feng}, Xiaotong and {Li}, Mingyu and {Li}, Qiong and {Spinoso}, Daniele and {Wu}, Yunjing and {Zhang}, Haibin},
	year         = 2024,
	month        = mar,
	journal      = {\apjl},
	volume       = 963,
	number       = 2,
	pages        = {L38},
	doi          = {10.3847/2041-8213/ad292f},
	eid          = {L38},
	archiveprefix = {arXiv},
	eprint       = {2312.04841},
	primaryclass = {astro-ph.GA},
	adsurl       = {https://ui.adsabs.harvard.edu/abs/2024ApJ...963L..38N}
}

@article{richards2006,
	title        = {{Spectral Energy Distributions and Multiwavelength Selection of Type 1 Quasars}},
	author       = {{Richards}, Gordon T. and {Lacy}, Mark and {Storrie-Lombardi}, Lisa J. and {Hall}, Patrick B. and {Gallagher}, S.~C. and {Hines}, Dean C. and {Fan}, Xiaohui and {Papovich}, Casey and {Vanden Berk}, Daniel E. and {Trammell}, George B. and {Schneider}, Donald P. and {Vestergaard}, Marianne and {York}, Donald G. and {Jester}, Sebastian and {Anderson}, Scott F. and {Budav{\'a}ri}, Tam{\'a}s and {Szalay}, Alexander S.},
	year         = 2006,
	month        = oct,
	journal      = {\apjs},
	volume       = 166,
	number       = 2,
	pages        = {470--497},
	doi          = {10.1086/506525},
	archiveprefix = {arXiv},
	eprint       = {astro-ph/0601558},
	primaryclass = {astro-ph},
	adsurl       = {https://ui.adsabs.harvard.edu/abs/2006ApJS..166..470R}
}

@article{weilbacher2020,
	title        = {{The data processing pipeline for the MUSE instrument}},
	author       = {{Weilbacher}, Peter M. and {Palsa}, Ralf and {Streicher}, Ole and {Bacon}, Roland and {Urrutia}, Tanya and {Wisotzki}, Lutz and {Conseil}, Simon and {Husemann}, Bernd and {Jarno}, Aur{\'e}lien and {Kelz}, Andreas and {P{\'e}contal-Rousset}, Arlette and {Richard}, Johan and {Roth}, Martin M. and {Selman}, Fernando and {Vernet}, Jo{\"e}l},
	year         = 2020,
	month        = sep,
	journal      = {\aap},
	volume       = 641,
	pages        = {A28},
	doi          = {10.1051/0004-6361/202037855},
	eid          = {A28},
	archiveprefix = {arXiv},
	eprint       = {2006.08638},
	primaryclass = {astro-ph.IM},
	adsurl       = {https://ui.adsabs.harvard.edu/abs/2020A&A...641A..28W}
}

\appendix

\section{Morphology fits to NIRCam F070W and F090W}\label{sec:alternate_uv_morph_fit}

\begin{figure}
    \centering
    \includegraphics[width=0.5\linewidth]{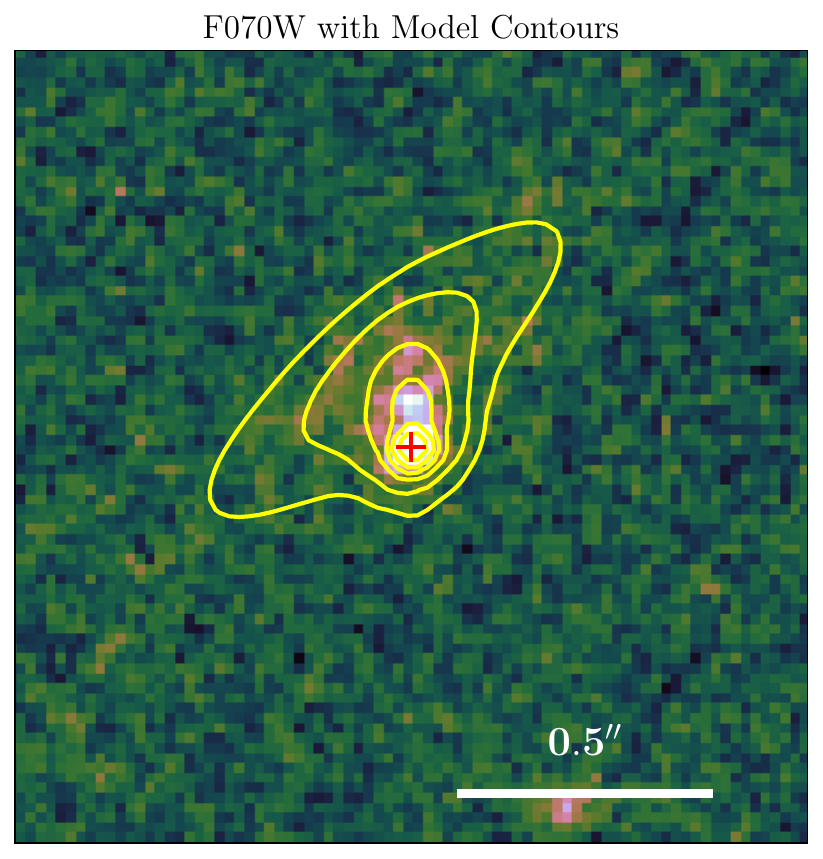}
    \includegraphics[width=0.5\linewidth]{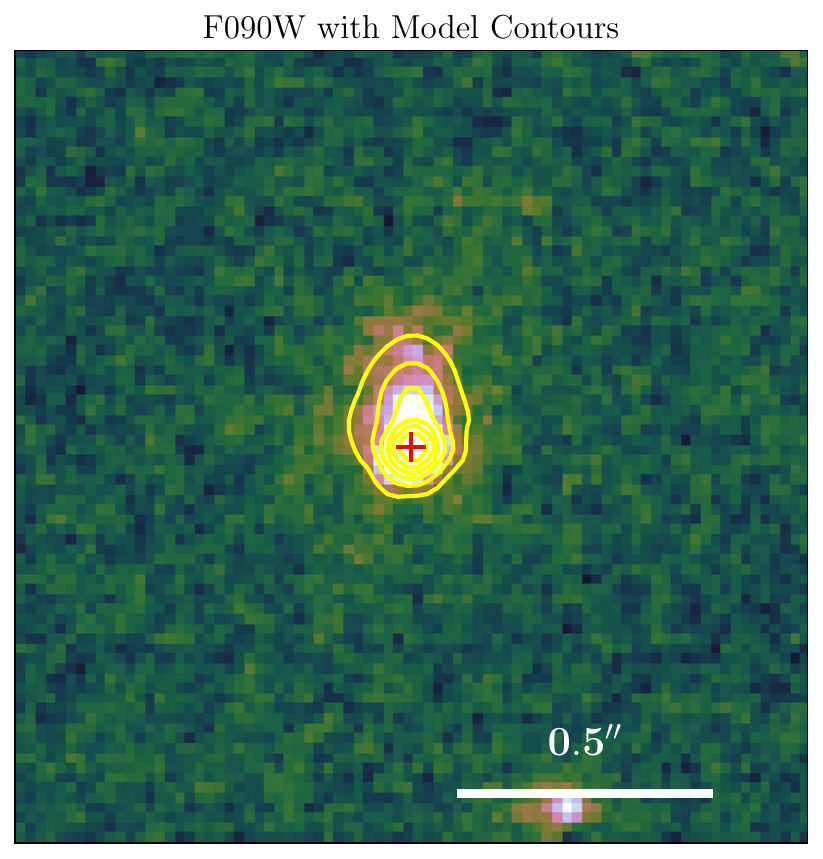}
    \caption{Same as in Fig.~\ref{fig:NIRCam_UV_model}, but shown as contours for visual clarity. We draw contours at every factor $2$ surface brightness decrease from the peak emission. A red cross marks the centroid of the \Halpha{} emission seen in NIRCam.}
    \label{fig:NIRCam_UV_model_contours}
\end{figure}

\begin{figure}
    \centering
    \includegraphics[width=\linewidth]{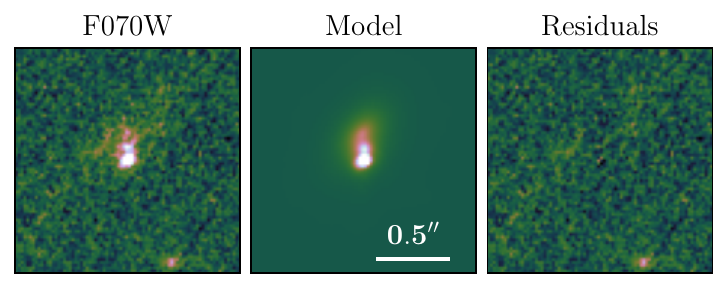}
    \includegraphics[width=\linewidth]{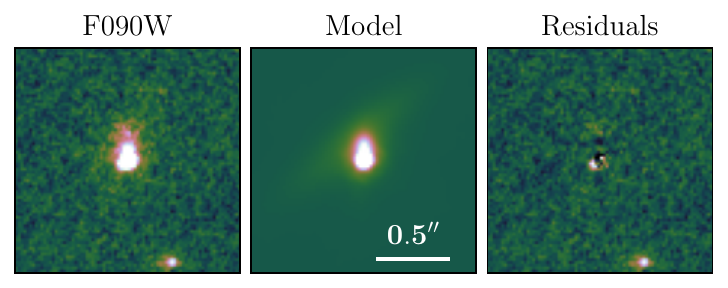}
    \caption{Alternative best-fit models of the UV morphology of \monster{} in the NIRCam filters F070W (top) and F090W (bottom). This figure is analogous to Fig.~\ref{fig:NIRCam_UV_model}, but  for each filter  we apply the best-fitting model from the other filter, keeping its geometrical parameters fixed and allowing only the intensities to vary. The residuals of the extended component reveal that its geometry differs between the two filters. The more extended morphology seen in F070W may reflect a significant contribution from \lya{} emission to the flux in this band.}
    \label{fig:uv_morph_fit_fixed_model}
\end{figure}

In Sect.~\ref{sec:uv_morphology}, we model the UV morphology of \monster{} in both F070W and F090W using two 2D elliptical Gaussians and one exponential component. The main difference between the fits in the two filters lies in the scale length of the exponential, which is significantly larger in F070W. In Fig.~\ref{fig:uv_morph_fit_fixed_model}, we present an alternative fitting approach in which the best-fitting model from F070W is used to fit the F090W image, and vice versa. In both cases, the positions and geometric parameters of all components are fixed to the best-fit values in one filter, allowing only the amplitudes to vary. The resulting residuals suggest that the geometry of the extended component is not consistent between filters, with F070W favoring a more extended exponential profile. This difference may point to a significant contribution from \lya{} emission to the F070W flux.

\section{Continuum centroids}

\begin{figure}
    \centering
    \includegraphics[width=\linewidth]{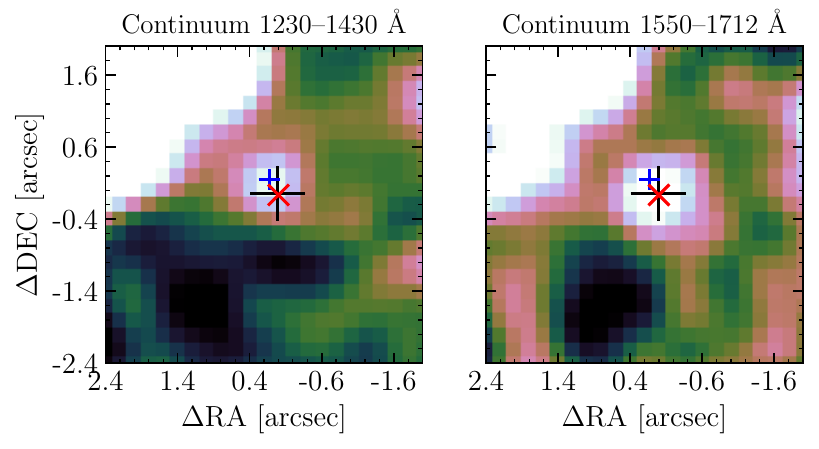}
    \caption{MUSE images of the continuum bands probing 1230--1430 \AA{} (\contblue{}) and 1550--1712 \AA{} (\contred{}). The centroid of the emission shows a clear shift of $\approx 0.2\arcsec{}$. The centroids of these two components are shown in Fig.~\ref{fig:LyA_contour_f070w} in comparison with the positions of other spatially resolved spectral features. The bright source in the northeast corner is the blend of two low-$z$ interlopers, clearly visible in Fig.~\ref{fig:f070w_with_LyA_contours}. The black cross marks the NIRCam coordinates of \monster{}. The blue and red crosses indicate the flux-weighted centroids of the blue and red continuum components, respectively.}
    \label{fig:continuum_centroids}
\end{figure}

In Fig.~\ref{fig:continuum_centroids} we show \muse{} images, result of collapsing the datacube in the specified wavelengths. We deliberately avoid the rest-frame wavelength range in which \niv{} and \ion{C}{iv} are located, to capture only continuum emission. We observe a clear shift in the spatial centroid of the continuum, as can be visually seen in the Fig.~\ref{fig:continuum_centroids}, with the black cross marking the NIRCam coordinates of \monster{} as reference. The flux weighted centroids of these components are shown in Fig.~\ref{fig:LyA_contour_f070w} ir relation to the positions of other spectral features (\niv{}, \lya{} and NIRCam rest-frame UV centroids).

\section{\lya{} morphology fit}

\begin{figure}
    \centering
    \includegraphics[width=\linewidth]{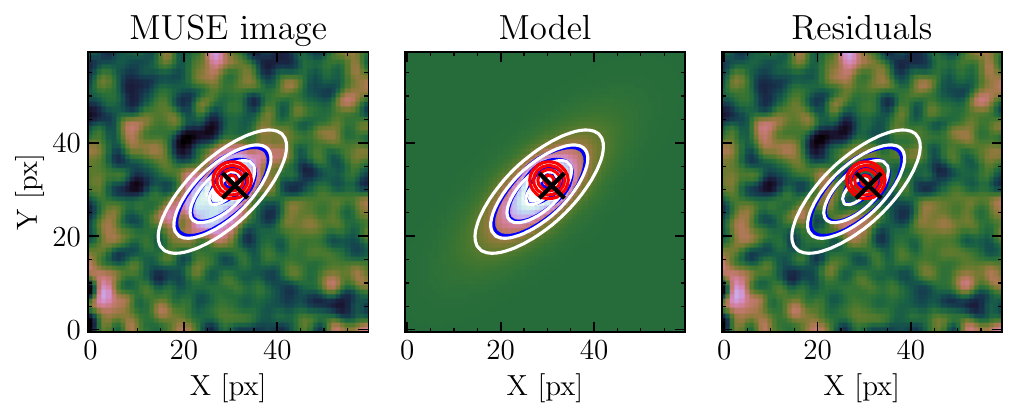}
    \caption{Morphological fit to the \lya{} emission of \monster{}. We fit a combination of the exponential (white), Gaussian (red), and a combination of the two (blue). The contours correspond to 80\%, 40\%, 20\%, and 10\% of the peak flux of each component. The \jwst{} coordinates of \monster{} are marked with a black cross.}
    \label{fig:LyA_morph_model}
\end{figure}

In Fig.~\ref{fig:LyA_morph_model}, we present the contours of the fitted model for the \lya{} halo of \monster{}. The fit was performed using the Python package \textsc{lmfit} with the \texttt{leastsq} method. We tested various combinations of exponential and Gaussian components and found that a model consisting of one exponential and one Gaussian provided the lowest $\chi^2$. We initially tested models including the morphology of NIRCam F090W (Sect.~\ref{sec:uv_morphology}) as a core component at the position of \monster{} with a small shift to allow for astrometric uncertainties \citep[see e.g.,][]{matthee2020}. However, we found that the fitting method could not satisfactorily constrain this component.

\begin{figure}
    \centering
    \includegraphics[width=0.8\linewidth]{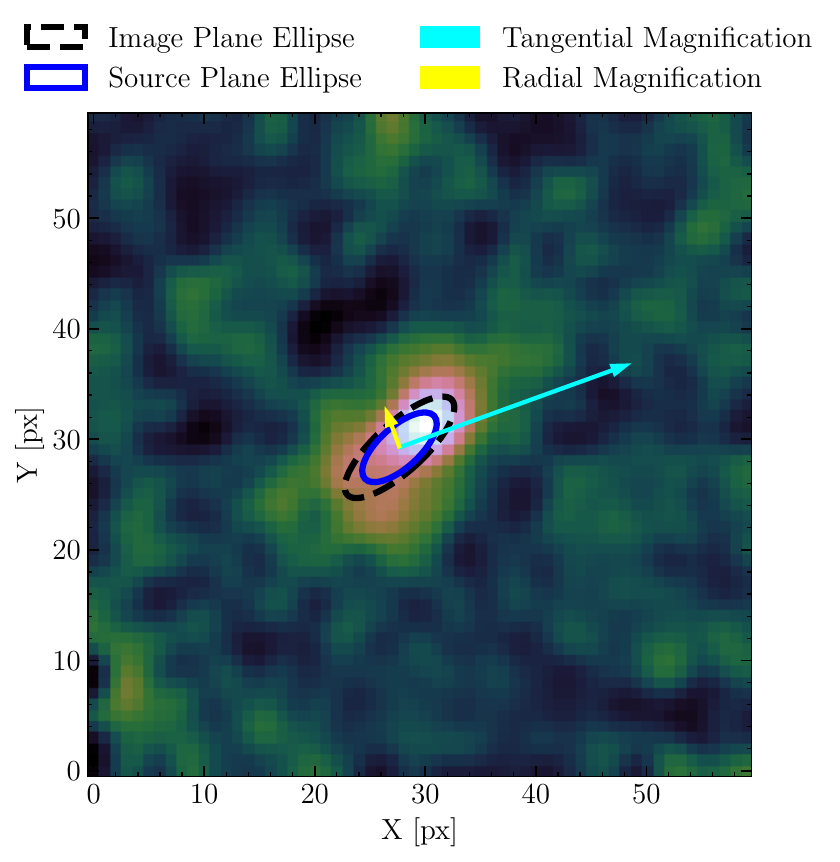}
    \caption{Ellipse defined by the fitted exponential at the scale length $h$ in the image plane (dashed black) and the transformation into the image plane (dark blue). These ellipses are plotted over the MUSE image of the \lya{} halo of \monster{} after applying a smoothing Gaussian filter of $\sigma=1$~px. Also show  are the directions of the radial and transverse magnification $\mu_r$ and $\mu_t$ (yellow and cyan arrows, respectively). The length of the arrows is proportional to $\mu_r-1$ and $\mu_t -1$ ($\mu_t = 1.68$, $\mu_r = 1.07$).}
    \label{fig:source_plane_ellipse_lya}
\end{figure}

In Fig.~\ref{fig:source_plane_ellipse_lya} we show the ellipses defined by the exponential fit to the \lya{} halo at the exponential scale length. The image is more strongly magnified in the direction of the semi-major axis of the image. However, after applying the transformation to the source plane, the fitted exponential remains fairly elliptical with $e=0.56$.

\section{Spatial variations of the \lya{} profile in the halo}

\begin{figure}
    \centering
    \includegraphics[width=0.8\linewidth]{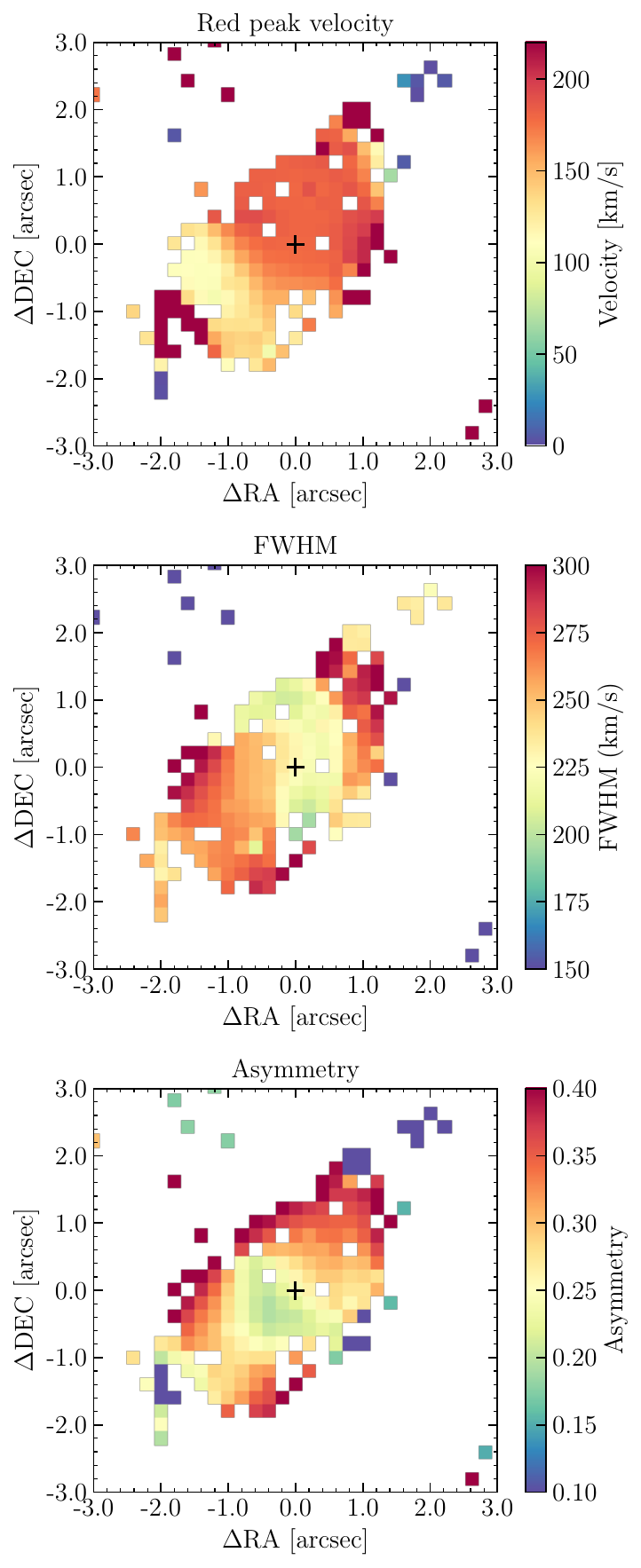}
    \caption{Spatial distribution of the fitted skewed Gaussian properties to the \lya{} halo of \monster{}. Shown are the velocity of the peak emission (top), the FWHM of the line in velocity space (middle), and the asymmetry parameter (see Eq.~\ref{eq:skew_gauss}; bottom). In these maps, we have masked out the pixels with a model amplitude ($A$ in Eq.~\ref{eq:skew_gauss}) with ${\rm S/N} < 3$. There are spatial variations of these parameters across the halo, in particular in a region with a slightly broader line profile and lower peak velocity toward the southwest. However, these differences are typical in \lya{} halos of star-forming galaxies in the literature \citep[e.g.,][]{erb2018}, and can be attributed to gas dynamics in the CGM.}
    \label{fig:monster_halo_lya_spatial_properties}
\end{figure}

We study the spatially resolved \lya{} line profile of \monster{}. For this, we fitted a skewed Gaussian defined as
\begin{equation}
    f(v) = A \exp\left(-\frac{(v - v_0)^2}{2 [a_{\rm asym}(v - v_0) + d]^2}\right)
    \label{eq:skew_gauss}
\end{equation}
\citep[e.g.,][]{shibuya2014} to every pixel in a box of $3\arcsec\times 3\arcsec$ around the coordinates of \monster{}, and after applying a Gaussian smoothing kernel with $\sigma=2$~px in the spatial directions. In Fig.~\ref{fig:monster_halo_lya_spatial_properties} we show the spatial maps of three parameters of the skewed Gaussian: the peak velocity offset with respect to the systemic redshift $z=4.464$ ($v_0$), the FWHM of the line profile,\footnote{With the definition of skewed Gaussian of Eq.~\ref{eq:skew_gauss}, the full width at half maximum is obtained as ${\rm FWHM} = \frac{2d\sqrt{2\ln 2}}{1 - (2\ln 2)a_{\rm asym}^2}$} and the asymmetry parameter ($a_{\rm asym}$). 

\end{document}